\documentclass[rmp,aps,nofootinbib,endfloats]{revtex4}

\usepackage{graphicx}
\usepackage{epsfig}
\usepackage{natbib}

\def\inbar{\,\vrule height1.5ex width.4pt depth0pt}
\def\IR{\relax{\rm I\kern-.18em R}}
\def\IC{\relax\hbox{$\inbar\kern-.3em{\rm C}$}}

\begin{document}
\title{Rotating massive stars through the ages,  with applications to WR stars, Pop III stars
and Gamma Ray Bursts}
\author{Andr\'e Maeder}
\email{Andre.Maeder@unige.ch}
\affiliation{Geneva Observatory, University of Geneva}
\author{Georges Meynet}
\email{Georges.Meynet@unige.ch}
\affiliation{Geneva Observatory, University of Geneva}

\begin{abstract}
This article first reviews the basic physics of rotating stars and their evolution. We examine in particular the changes of the mechanical and thermal equilibrium of rotating stars. An important  (predicted and observed)  effect is that rotating stars are hotter at the poles and cooler at the equator.  We  briefly discuss the mass loss by stellar winds, which are  influenced by the 
anisotropic temperature distribution.
These anisotropies in the interior are also driving circulation currents, which transports the chemical elements and the angular momentum in stars. Internal differential rotation,
if present, creates instabilities and mixing, 
 in particular the shear mixing, the horizontal
turbulence and their interactions.

A major check of the model predictions concerns the changes of the surface abundances, which are modified by mass loss in the very massive stars and  by 
rotational mixing in O- and B-type stars. We show that the observations confirm
the existence of rotational mixing, with much larger effects at lower metallicities. 

We discuss the predictions of stellar models concerning the evolution of the surface velocities, the evolutionary tracks
in the HR diagram and lifetimes,  the 
populations of blue, red supergiants  and Wolf-Rayet stars, and the progenitors of type Ibc supernovae. We  show, that in many aspects, rotating models provide a much better fit than non-rotating ones. Using the same physical ingredients as those which fit the best the observations
of stars at near solar metallicities, we explore the consequences of rotating models for the status of Be stars, the progenitors of Gamma Ray Bursts, the evolution
of Pop III stars and of very metal poor stars, the early chemical evolution of galaxies, the origin of the C-enhanced Metal Poor stars (CEMP) and of the chemical anomalies in globular clusters.

Rotation, together with mass loss are two key physical ingredients shaping the evolution of massive stars during the whole cosmic history.

\end{abstract}

\maketitle
\tableofcontents

\section{INTRODUCTION}
\label{sec:intro}

\begin{figure}[!th]
\centering
\includegraphics[angle=0,width=10.0cm]{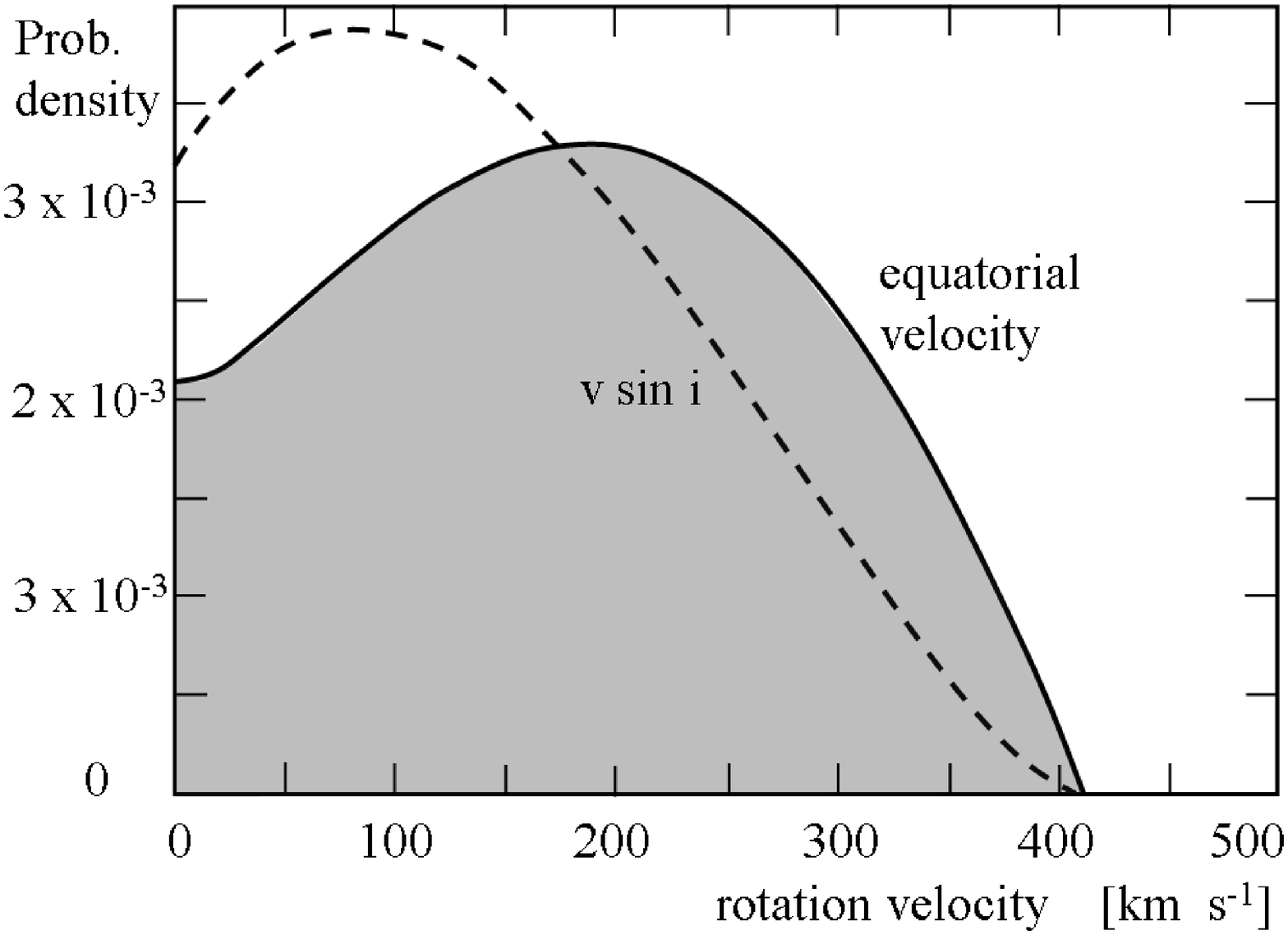}
         \caption{Probability density by km s$^{-1}$ of rotation velocities for  496  stars with types O9.5 to B8, i.e. masses between about 3 and 20 M$_{\odot}$ \citep{Huang2006a}. }        
\label{distrv}
\end{figure}

It may be surprising  that we emphasize the need of accounting for
the effects of rotation in stellar evolution, since in practice all stars are
rotating around their axis. However, it must be recalled that
the so-called "standard theory" of stellar  evolution generally ignores
the effects of stellar rotation and treats the stars as non--rotating bodies, despite the fact that  stars more massive than about 1.5 M$_{\odot}$ rotate fast on the average (Fig. \ref{distrv}). 

The recent progresses in astrophysical observations, particularly 
  in high resolution  spectroscopy and in asteroseismology, have shown many significant deviations from the standard models, for example the many large
   nitrogen enrichments resulting from mixing in massive stars
  (Sect. \ref{chimie}).
  These observations show the need to also account for the various effects of rotation in stellar modeling. All model outputs are finally modified by the proper account of
 rotation: the stellar luminosities and radii, the lifetimes, the chemical abundances
 at the surface, the helio- and asteroseismology responses, the nature of supernova explosions, the amounts of nucleosynthetic products, the nature of the final remnants, \ldots

 For practical purpose, it is often convenient to distinguish 4 main groups of
 rotational effects in stellar physics.
 
 1.- \textit{The  equilibrium configuration of rotating stars:} it results from the 
 centrifugal force on the stellar equilibrium. The equipotentials are modified
 and
 in particular the shape of the stellar surface, which affects the surface
  $T_{\mathrm{eff}}$ and gravity distributions.

  2.- \textit{The effects of rotation on mass loss or accretion:} in fast rotating stars, the isotropy of mass loss (or accretion when present) is
  destroyed and anisotropies appear and are effectively observed.

  3.- \textit{The rotational mixing:} the internal distortion induces circulation
  currents which transport the elements and angular momentum, while differential 
  rotation may produce several instabilities which also contribute to the transport
  processes.

  4.- \textit{The interactions with magnetic field:} the presence of an internal magnetic field may produce an internal  coupling of rotation, leading to solid body rotation,
  while external fields produce some magnetic braking.  
A major uncertainty concerns the
existence of a dynamo in radiative regions with  differential rotation. We examine some properties of such a dynamo.


  The above distinction is evidently a simplification since the various effects are related. For example, it is the modification of the internal equilibrium structure
  which drives the mixing. In turn, the mixing modifies  the internal distribution of the elements and this also influences the equilibrium structure.
  
  In this review, we focus on the rotational effects in the Main--Sequence (MS) and 
  post-MS phases, i.e. in the nuclear phases. The effects of rotation in star 
  formation are also a most important chapter of astrophysics, but they would deserve
  another specific review. We  consider here the case of single stars,
  the many effects of rotation in relation with tidal interactions in binaries 
  are also beyond the scope of this review. Most of the effects
   discussed here in the case of single stars evidently have their counterparts  in binaries,
  however their modeling is still in its infancy.
  
  Recent reviews on the observational and theoretical aspects of stellar rotation are
given in IAU symposium 215 \citep{MaederEenens2004} and the theoretical aspects have recently extensively been reviewed \citep{Maeder2009}.

\section{THE MECHANICAL AND THERMAL EQUILIBRIUM OF ROTATING STARS}

The equilibrium and stability configurations of rotating stars have been reviewed long ago \citep{Lebovitz1967}. In practice, except  for stars  with little internal 
density contrast, like white dwarfs or neutron stars, the approximation of the Roche model is  acceptable. It assumes that  the effective gravity results from  the matter  centrally condensed, with in addition  the  effect of the centrifugal force. 
 For all stellar masses, the rotational energy of the Roche model  represents at most about 1\% of the absolute value of the potential energy.

The properties of rotating stars depend on the  distribution  of the angular velocity $\Omega(r)$ inside the 
stars. The simplest case is that of solid body rotation, i.e. $\Omega=$ const., while
more elaborate models include differential rotation. 
  
\subsection{The mechanical equilibrium for uniform rotation}

  Let us first consider the case of a constant  angular velocity $\Omega$ in the Roche model in hydrostatic equilibrium. The gradient of  pressure $P$ is given by  
  \begin{eqnarray}
  \frac{1}{\varrho} \, \vec{\nabla}P \, = \, -\vec{\nabla}\Phi + \frac{1}{2} \, \Omega^2 \vec{\nabla}
  (r \, \sin\vartheta)^2  \; ,
  \label{fcentrif}
  \end{eqnarray} 
  \noindent
  $\varrho$ is the local  density, $\Phi$ is the gravitational
  potential, which is unmodified by rotation in the Roche approximation and which gives the gravitational acceleration
 $\vec{g} \, = -\, \vec{\nabla} \Phi \, = \,
- \frac{G\, M_r}{r^2} \frac{\vec{r}}{r}$, $r$ being  the distance to the center.
The components of $\vec{g}$ are $(-g,\;0,\; 0)$ and $g \, = \, \frac{\partial \Phi}{\partial r}$. 
  If $\Omega$ is constant or has a cylindrical symmetry, the centrifugal acceleration can also be derived from a potential, say $V$.
 In case of constant $\Omega$, one has  
  \begin{eqnarray}
  - \vec{\nabla}V \, = \, \Omega^2 \, \vec{\varpi} \quad \, \mathrm{and \; thus} \quad \,
  V \, = \, - \frac{1}{2} \Omega^2 \, \varpi^2  \; .
  \label{omegapot}
  \end{eqnarray}
  \noindent
   $\varpi= r \, \sin \vartheta $ is the distance to the rotation axis.
  The total potential $\Psi$ is 
 $ \Psi \, = \, \Phi + V $
  and with the usual Poisson equation one has
$ \nabla^2 \Psi \, = \, \nabla^2 \Phi + \nabla^2 V \quad \mathrm{with} \quad \nabla^2 \Phi = 4 \, \pi \, G \, \varrho  $.
 In a uniformly (or cylindrically)  rotating star, the centrifugal force derives from a potential and one has  
$ (\nabla^2 V)_{\varpi} =  \frac{1}{\varpi} \, \frac{\partial}{\partial \varpi}\left( -\varpi^2 \Omega^2
 \right) \,   = \, -2 \, \Omega^2 $.
 The Poisson equation \index{Poisson equation} becomes
 \begin{eqnarray}
 \nabla^2 \Psi \, = \, 4 \, \pi \, G \, \varrho - 2 \, \Omega^2  \; .
 \label{lapl}
 \end{eqnarray}
 \noindent
 and the equation of hydrostatic equilibrium 
   \begin{eqnarray}
  \frac{1}{\varrho} \vec{\nabla} P \, = \, -\vec{\nabla} \Psi  \, = 
  \, \vec{g}_{\mathrm{eff}} .
  \label{equpsi}
  \end{eqnarray} 
  \noindent
  The effective gravity\index{effective gravity} $\vec{g}_{\mathrm{eff}}$ results from both  gravitation and  centrifugal
  acceleration (take care  of the sign of $\Phi$ and $\Psi$).
  These expressions  imply that the pressure  is constant on an equipotential, i.e.  $P=P(\Psi)$.
  The equipotentials and isobars coincide  and the star is said {\emph{barotropic}},
  otherwise it is said {\emph{baroclinic}}.
   Eq.(\ref{equpsi}) shows that the density is also a function of 
  $\Psi$ only.
  Through the equation of state  $P \, = \, P(\varrho,T)$,  one also has
  $T=T(\Psi)$. The quantities $\varrho , \; P, \; T$ are constant on the equipotentials
  $\Psi =$ const.

The stellar surface is an equipotential. The total potential at a level $r$ and at colatitude
 $\vartheta$ ($\vartheta=0 $ at the pole) is 
 \begin{eqnarray}
  \Psi (r, \vartheta) \, = \,- \frac{G \,M_r}{r} - \frac{1}{2} \, \Omega^2 \, r^2 \sin^2 \vartheta \; .
 \label{psipot}
 \end{eqnarray}
 \noindent 
For a star of total mass $M$, let us  call $R(\vartheta)$ the  stellar radius  at colatitude 
 $\vartheta$. Since the centrifugal force is zero at the pole, the potential at the stellar pole is
just   ${G \, M}/R_{\mathrm{p}}$, where $R_{\mathrm{p}}$ is the polar radius. This fixes the constant value of the equipotential at the stellar surface, which  is given by
\begin{eqnarray}
  \frac{G \,M}{R} + \frac{1}{2} \, \Omega^2 \, R^2 \sin^2 \vartheta  \, = \, \frac{G \, M}{R_{\mathrm{p}}} \; .
  \label{surfpot}
  \end{eqnarray} 
 \noindent 
 This equation also applies to moderate differential rotation, if $\Omega$
 can be considered as constant on an equipotential surface.
 
 \begin{figure}[!t]
\centering
\includegraphics[width=8.5cm]{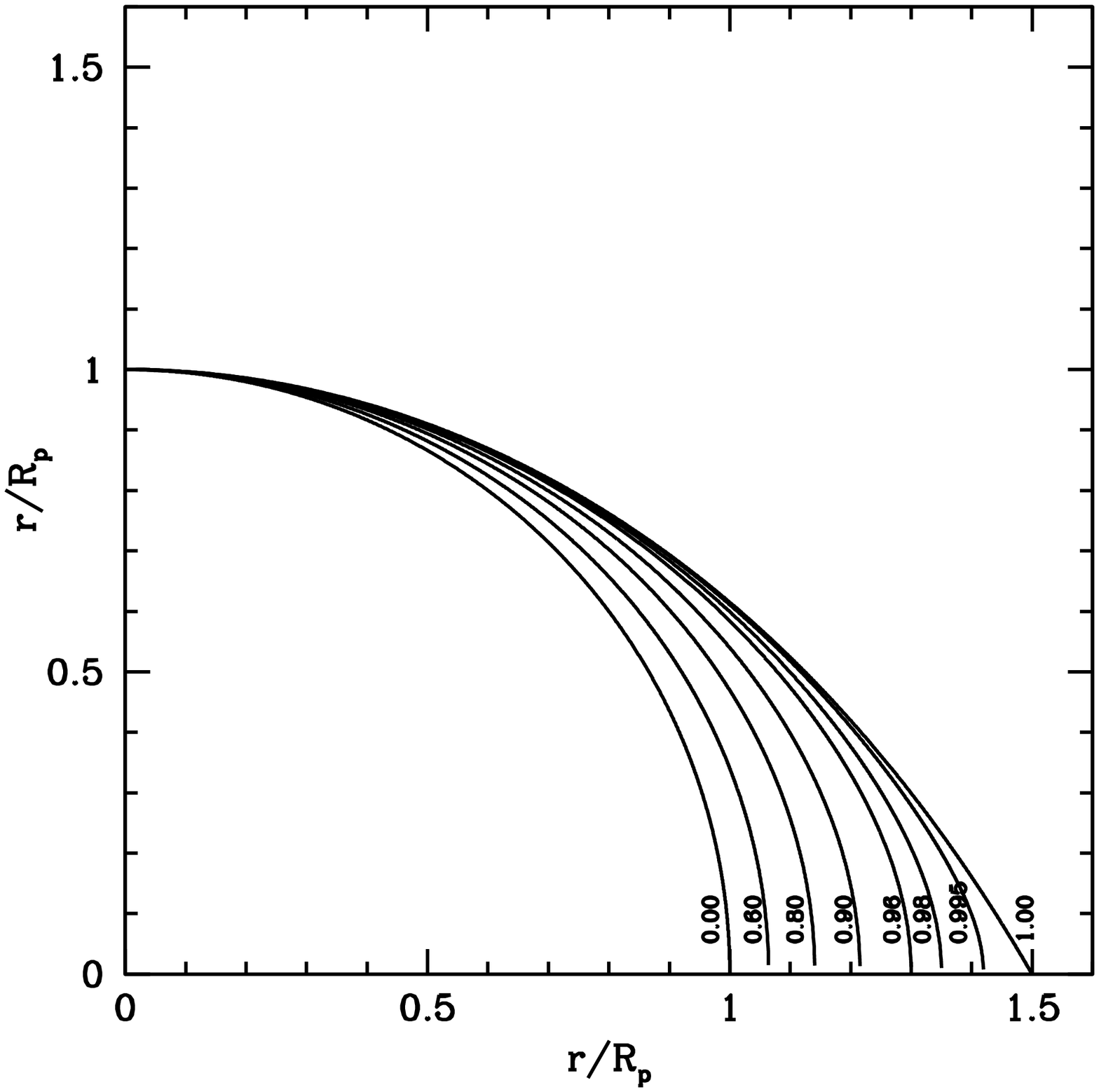}
\caption{Shape of the surface for various values of $\omega = \frac{\Omega}{\Omega_\mathrm{crit}}$ (labelled at the bottom of each curve). The x-axis is the equatorial radius, and the y-axis the polar one. Hence this is how we would see the star equator-on. 
$R_{\rm p}$ is the polar radius. Figure taken from \citet{Georgy2011}.}
\label{roche}
\end{figure}

  The shape of a Roche model\index{Roche model} is illustrated in Fig. \ref{roche} for different rotation velocities.
 The effective gravity resulting from the gravitational potential and from the centrifugal force  is given by
 (\ref{equpsi}).
 If $\vec{e}_r$ and $\vec{e}_{\vartheta}$ are the unity vectors in the radial and latitudinal directions, one has 
 \begin{eqnarray}
\vec{g}_{\mathrm{eff}} \, = \, \left[- \frac{G  M}{R^2(\vartheta)}+ \Omega^2 \, R(\vartheta) \sin^2 \vartheta \right]
\vec{e}_r + 
\left[  \Omega^2 \,  R(\vartheta) \sin \vartheta \cos \vartheta \right]
\vec{e}_{\vartheta}  \; . \quad \quad
\label{vecg}
\end{eqnarray} 
\noindent
Care has to be given for many applications that the gravity vector is not parallel to the vector radius, with an angle $\epsilon$ given by 
$\cos \epsilon \, = \, - {\vec{g_{\mathrm{eff}}} \cdot \vec{r}}/( {\left| \vec{g}_{\mathrm{eff}}\right| \cdot \left|\vec{r} \right|})$ between the two vectors.

\subsection{Differential rotation: the case of shellular rotation}
 
 The most studied  
case of differential rotation is that  
 of the so--called "shellular" rotation \cite{Zahn1992}, i.e. with a rotation law  $\Omega$
constant on isobaric shells and depending  to the first order   on the distance $r$ to the stellar center,
\begin{eqnarray}
 \Omega(r, \vartheta)= \overline{\Omega}(r)+ 
   \Omega_2(r) \,  P_2 (\cos \vartheta) \; ,
 \label{omegahat}
 \end{eqnarray} 
  \noindent
  where $P_2$ is the second Legendre polynomial.   
This rotation law results from the strong horizontal turbulence in differentially rotating stars, which imposes
a constancy of $\Omega$ on isobars, while  in the vertical direction, the turbulence is weaker due to
 the stable density stratification. The above writing of $\Omega(r, \vartheta)$
 implies some limitations, since  isobars are not identical to spherical surfaces,
 in particular 
     it only applies to low or moderate
  rotation velocities. 
 This case is non--conservative.  The surfaces defined 
 by $\Psi=$ const.  are  isobars \citep{paperI1997}, but they are not equipotentials and the star is  baroclinic. 
 
   The equations of stellar strucure
   for a star in shellular rotation
   can be written
   in a form  close  to that of the non--rotating case  in order to minimize
   the necessary modifications  \citep{paperI1997}.
  One  associates  a radius $r_{P}$ to  an isobar, it is defined  by
  $ V_P \, \equiv \, ({4 \, \pi}/{3}) \, r^3_P  \; $,
   where $V_P$ is the volume inside the  isobar. For any quantity like  $g_{\mathrm{eff}}$, which is not constant 
   over an isobaric surface, a mean value is defined by   
   \begin{eqnarray}
   <g_{\mathrm{eff}}> \, \equiv \, \frac{1}{S_P} \oint_{\Psi=\mathrm{const}} g_{\mathrm{eff}} \, d\sigma \; ,
   \label{qmoy}
   \end{eqnarray}   
   \noindent
   where $S_P$ is the total surface of the isobar and $d\sigma$ is an element 
  $d\sigma \, = \, {r^2 \, \sin  \vartheta \, d\varphi d\vartheta}/{\cos \epsilon}$ of this surface. In Lagrangian coordinates (where $M_P$ is the mass internal to an isobar), the equations of hydrostatic equilibrium, of mass conservation, of energy production and of energy transport can be written
  \cite{paperI1997},   
  \begin{eqnarray}
{d P \over d M_P}=-{G \, M_P \over 4 \, \pi \, r_P^4} \, f_P  \;, 
\end{eqnarray}
\begin{eqnarray}
 {d r_P \over
 d M_P}={1 \over 4\pi r_P^2 \overline\varrho}  \; 
\end{eqnarray}
\begin{eqnarray}
{d L_P \over d M_P}=\varepsilon_{nucl}-\varepsilon_{\nu}+
\varepsilon_{grav}  \; ,
\label{3rot}
\end{eqnarray}
\begin{eqnarray}
{d \ln T \over d M_P}  
=-{G \, M_P \over 4 \, \pi \, r_P^4} f_P  \, {\rm min}\left[\nabla_{\rm ad}, \nabla_{\rm rad}
{f_T \over f_P} \right]  \; ,
\label{4rot} 
\end{eqnarray}
\begin{eqnarray}
 \mathrm{whith} \; f_P={4 \pi   r_P^4 \over G M_P S_P} {1 \over <g_{\mathrm{eff}}^{-1}>}  
\quad 
 \mathrm{and} \quad f_T=\left( {4  \pi   r_P^2 \over S_P}\right)^2 {1 \over <g_{\mathrm{eff}}> <g_{\mathrm{eff}}^{-1}>}  \, .
 \label{ft}
\end{eqnarray}

 \noindent
 where $\varepsilon_{nucl}$, $\varepsilon_{\nu}$ and $\varepsilon_{grav}$
 are respectively the rate of nuclear energy production, of neutrino losses
 and of gravitational energy production or absorption by mass unit. 
 $\nabla_{\rm ad}$ and $ \nabla_{\rm rad}$ are the adiabatic and radiative
 temperature $T$ gradients ($d \ln T/d \ln P$). $\overline\varrho$
 is formally here the average density between two isobars. The above set of equations
 allows one to construct models of stars in differential rotation (shellular case)
 with little modifications with respect to the usual set of standard equations.
 The results are obtained as a function of $M_P$ the mass of an isobar of average
 coordinate $r_P$. In practice,  $r_P$ is the radius of the isobar
 at a colatitude  $\vartheta$ such that $P_2(\cos \vartheta)=0$, i.e. $\vartheta \approx 54 $ degrees.

\subsection{The critical velocities}

The critical  or break--up velocity is reached when the 
 centrifugal force becomes equal to the gravitational attraction at the equator, i.e. when
\begin{eqnarray}
\Omega^2_{\mathrm{crit}} \, = \, \frac{G \, M}{R^3_{\mathrm{e , crit}}}  \; ,
\end{eqnarray}
\noindent
where $R_{\mathrm{e , crit}}$ is the equatorial radius at break--up. With
$\Omega_{\mathrm{crit}}$ in the equation   of the surface  (\ref{surfpot}) at break--up, one gets
for the ratio of the equatorial to the polar radius at critical velocity,
$\frac{R_{\mathrm{e,  crit}}}{R_{\mathrm{p, crit}}} \, = \, \frac{3}{2}$. The equatorial
break--up velocity is thus
\begin{eqnarray}
v^2_{\mathrm{crit}} \, = \, \Omega^2_{\mathrm{crit}} \, R^2_{\mathrm{e,crit}} \, = \, 
\frac{G \, M}{R_{\mathrm{e,crit}}} \, = \frac{2 \,G \, M}{3 \, R_{\mathrm{p,crit}}}  \; .
\label{vrupture}
\end{eqnarray}
\noindent
Introducing
a non--dimensional rotation parameter $\omega$, defined as the ratio  of the angular velocity 
to the angular velocity at break--up,
$\omega \, = \, \frac{\Omega}{\Omega_{\mathrm{crit}}}$ i.e.
$\omega^2 \, = \, \frac{\Omega^2 \, R^3_{\mathrm{e , crit}}}{G \, M}$, 
$  \Omega^2 \, = \, \frac{8}{27} \, \frac{G \, M \omega^2 }{R^3_{\mathrm{p \,crit}}}$
 and the equation of the surface (\ref{surfpot}) becomes  with $x= R/{R_{\mathrm{p ,crit}}}$
  \begin{eqnarray}
 \frac{1}{x} + \frac{4}{27} \, \omega^2 \, x^2 \sin^2 \vartheta \, = \, \frac{R_{\mathrm{p ,crit}}}{R_{\mathrm{p}}(\omega)}
  \; .
\label{xsurf}
 \end{eqnarray}
 
       \begin{figure}[!t]
\centering
\includegraphics[angle=-90,width=9.0cm]{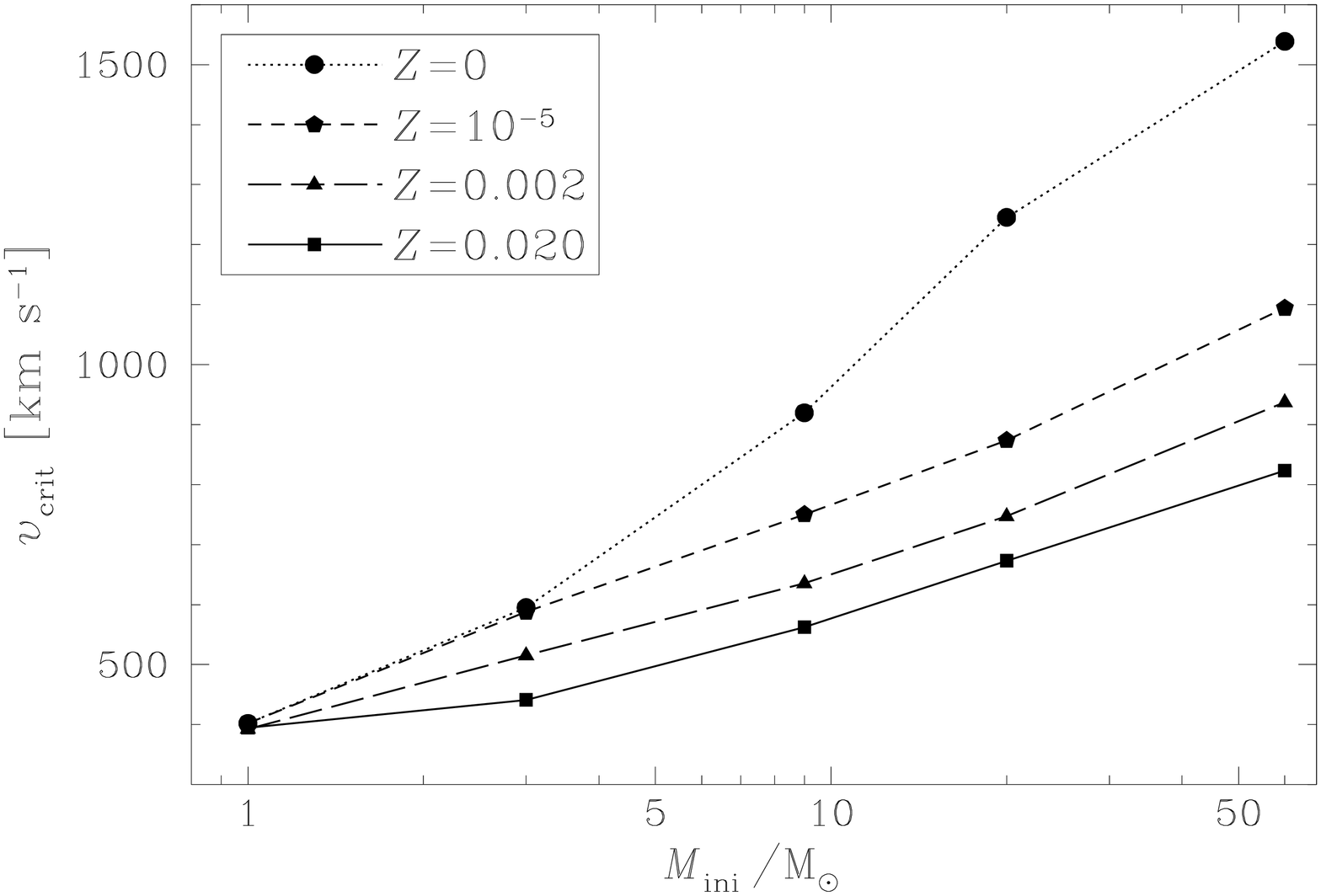}
\caption{The critical velocities $v_{\mathrm{crit}}$ as a function of stellar masses for different metallicities
$Z$ for stars on the ZAMS. The effects of the changes of the polar radius with rotation are accounted for. Figure taken from
\citet{Ekstrom2008b}.}
\label{vcrit}
\end{figure}

\noindent
If the polar radius would not change with $\omega$, the second term would be equal to 1;  Eq. ({\ref{xsurf})
is thus an algebraic equation of the 3$^{rd}$ degree, which gives the values of the radius as a function of $\omega$ and $\vartheta$. For low rotation ($\omega < 0.5$),  a linear approximation of $v$ in term of $\omega$ is valid
 \begin{eqnarray}
 v^2 \, = \, \frac{8}{27} \, \frac{G \, M \omega^2 }{R^3_{\mathrm{p ,crit}}} \, R^2_{\mathrm{e}} \approx
 \frac{8}{27} \, \frac{G \, M \omega^2 }{R_{\mathrm{p,crit }} } \, = \frac{4}{9} \omega^2 \,
  v^2_{\mathrm{crit}} \; .
  \label{v49}
  \end{eqnarray}  
Fig. \ref{vcrit} shows the critical velocities $v_{\mathrm{crit}}$ for stars of various masses and metallicities.  The critical velocities grow with stellar masses, because the stellar radii increase
only slowly with stellar masses. The critical velocities  are very large for 
low metallicity stars, since their radii are  much smaller as a result of their lower opacities. In the most massive stars, where radiation pressure nearly compensates gravity, the effects of radiation modify the expressions of the critical velocity making them lower. In this case, one speaks of the $\Omega$$\Gamma$ limit, since the stars reach their critical velocity, due to the combined effects of high radiation (the Eddington limit) and of centrifugal force \citep{paperVI2000}.

\subsection{The breakdown of thermal equilibrium in rotating stars} \label{breakd}

 Rotation deeply affects the internal  thermal equilibrium. 
 As seen above, the equipotentials are differently spaced as a function of the colatitude $\vartheta$, since the star is distorted.    
  The equipotentials  are closer in polar regions and more separated near the equator. 
 The radiative flux at some latitude is proportional to the local gradient 
 between the equipotentials (the effective gravity,  Sect. \ref{vz}). If there is no local energy source or sink, the total energy through an equipotential is conserved. Thus, there is  (in a simple picture of uniformly rotating stars) an excess of flux along the polar axis and a deficiency 
  near the equatorial plane. 
 This thermal imbalance generates global circulation motions in the meridian plane. 
   These currents then contribute
 to the mixing of chemical elements and transport of angular momentum (Sect. \ref{cm}).
 
\subsection{The von Zeipel Theorem and observations} \label{vz}

 The von Zeipel theorem relates the radiative flux at some colatitude $\vartheta$ 
 on the surface of a rotating star and the local effective gravity  $g_{\mathrm{eff}}(\Omega,\vartheta)$, function of the angular velocity $\Omega$ and $\vartheta$.  The flux is
 $\vec{F}(\Omega,\vartheta) \, = \, -\chi \, \vec{\nabla} T(\Omega,\vartheta) \, $ with $  
 \chi \, = \, \frac{4 \, a\, c\, T^3}{3 \, \kappa \, \varrho} $.
 We consider the case of a uniformly rotating star. The equipotentials and isobars  coincide, they are also surface of constant $T$ and $\varrho$.
 Thus, one has with (\ref{equpsi}) 
 \begin{eqnarray}
 \vec{F} (\Omega,\vartheta)\, = \,- \chi \, \frac{dT}{dP} \, \vec{\nabla} P(\Omega,\vartheta) \, = \, 
  -\varrho \, \chi \,  \frac{dT}{d P} \, \vec{g}_{\mathrm{eff}}(\Omega,\vartheta) \; .
 \end{eqnarray}
 \noindent
 The pressure gradient and effective gravity are parallel. 
 The term $\left(\varrho \, \chi \,  dT/{d P}\right)$ is constant on a given  equipotential, thus  the flux is proportional to the effective gravity on the equipotential. This coefficient of proportionality can easily be estimated 
 and one obtains
 \begin{eqnarray}
 \vec{F}(\Omega,\vartheta) \, = \, - \frac{L} {4 \, \pi \, G \, M^{*}} \; \vec{g}_{\mathrm{eff}} (\Omega,\vartheta)
 \label{vonzeipel}  \\[2mm]
 \mathrm{with} \quad \quad 
  M^{*} \,= \, M \left(1 - \frac{\Omega^2}{2 \, \pi \, G  \, \overline{\varrho}_M} \right) \; ,
  \label{mstar}
  \end{eqnarray} 
 \noindent
 where $\overline{\varrho}_M$  is the average internal density. The above relation applies to any equipotential, but in general the von Zeipel 
 theorem is considered at the  surface of  a star of total mass $M$ and luminosity $L$. 
    Eq. (\ref{vonzeipel}) is 
 the von Zeipel theorem: it says that the radiative flux at the surface of a rotating  star is proportional to the local effective gravity at the considered colatitude. 
 From $g_{\mathrm{eff}}$, one  obtains the radiative
 flux locally and thus $T_{\mathrm{eff}}$  
 \begin{eqnarray}
 T_{\mathrm{eff}}(\Omega,\vartheta) \, = \, \left(
 \frac{L} {4 \, \pi \,\sigma \,  G \, M^{*}}  
 \right)^{\frac{1}{4}} \;  \left[{g}_{\mathrm{eff}}(\Omega,\vartheta) \right]^{\frac{1}{4}}  \; .
 \label{tefftheta}
 \end{eqnarray}
 \noindent
 Both $g_{\mathrm{eff}}$  and $T_{\mathrm{eff}}$ vary over the  surface of a rotating star and  
 influence the emergent spectrum. The equatorial regions  are fainter  and cooler than the polar ones,
 which are brighter and hotter (the differences of $T_{\mathrm{eff}}$
 may reach a factor of 2). This effect is called the gravity--darkening.
 The von Zeipel theorem in a  star with shellular rotation shows  only minor differences \citep{Maeder1999a} with respect to (\ref{vonzeipel}). The differences, which essentially depend on the 
 $\Omega$ gradient close to the surface, may  slightly increase 
  the contrast between the pole and equator. 2--D models have tested the validity of the
  von Zeipel relation \citep{Lovekin2006}. Depending on the rotation laws, there are some small
  differences.

Interferometric observations  with the VLTI  of the Be star  Achernar   indicate a ratio   $R_{\mathrm{e}}/R_{\mathrm{p}} \approx 1.5$ \citep{Domiciano2003}. A recent
analysis of the data by  \citet{Carciofi2008} confirms that the observations well agree with a rigidly rotating star  at $\omega=0.992$ in the Roche model, provided it is surrounded by a small disk. 
VLTI 
 observations  \citep{Domiciano2005}  of the fast rotating star Altair ($M\approx 1.8$ M$_{\odot}$,  A7IV-V)   confirm a gravity--darkening as
 predicted by the von Zeipel theorem. This is also supported by further studies
  for Altair \citep{Peterson2006,Peterson2008},
which rotates at 90\% of its breakup angular velocity.
The effect of gravity--darkening receives further support \citep{Monnier2007}, but instead of an exponent $0.25$ as in Eq. (\ref{tefftheta}), an exponent $\beta=0.19$ is favored.
Also, these authors notice an  equatorial darkening stronger than predicted, which might result from faster equatorial rotation, of differences due to convection or opacity effects, etc. 
A  value of $\beta$ as small as $\beta=0.08$ is also found  \citep{vanBelle2006} in the case of  Alderamin ($\alpha$ Cep, type A7IV-V), which rotates at 83\% of its breakup velocity.
 Thus, the von Zeipel relation is confirmed, but there are some possible deviations.

   \section{MASS LOSS FROM MASSIVE STARS}
   
   Massive stars above $~ 10$ M$_{\odot}$ lose, at solar metallicity, a significant amount of mass during their evolution. As an example, a star with an initial mass of 60 M$_{\odot}$
   loses about  the half of its mass during its MS evolution. Due to mass loss in further stages, it is left with only about 10 M$_{\odot}$ at the time of  supernova explosion. The mass loss rates $\dot{M}$ of OB stars reach about $10^{-5}$
    M$_{\odot}$ yr$^{-1}$ with wind  velocities up to 3000 km s$^{-1}$. For these stars $\dot{M}$ is  a billion times larger than for the Sun, for which
  $\dot{M}$ is about  $10^{-14}$ M$_{\odot}$ yr$^{-1}$.The mass loss results from the stellar winds driven by the strong radiation pressure
of  luminous stars, which pushes the mass outside. The absorption of radiation by the
spectral  lines is the main effect transferring   momentum from radiation to matter.
In red supergiants, mass loss also occurs
due to the absorption and diffusion of radiation by dust. On the whole, mass loss is a dominant effect influencing all
the outputs of stellar evolution and nucleosynthesis.

 The   physics of stellar winds has been reviewed in several studies
  \citep{Lamers1999,Kudritzki2000}.   
Owing to the many uncertainties, 
stellar  model makers generally  apply expressions of $\dot{M}$ based on  observations, which are also not free from problems.
A major one is due to the  clumping effects of matter in the wind,
This has already lead to a reduction of the $\dot{M}$  by a factor 2 or 3 
over recent years and questions  arise about a further reduction.  For hot stars, there are several parametrizations of the  $\dot{M}$ rates deduced from theoretical models \cite{Kudritzki2000,Vink2000,Vink2001}
 for different $M$, $L$, $T_{\mathrm{eff}}$ and metallicities $Z$. For O stars, $\dot{M}$ (taken positive) behaves globally  like
 \begin{eqnarray}
  \dot{M} \sim L^{1.6}  \quad \quad \mathrm{and} \quad \quad  \dot{M} \sim Z^{0.7} \, .
  \label{mpointu}
  \end{eqnarray}
Let us note that \citet{Mokiem2006} find that for stars with log$L$/L$_\odot$ superior to about 5.4, the wind strengths are in excellent agreement with the theoretical predictions of \citet{Vink2001}.
 
Dust--enshrouded  red giants and  supergiants have strong mass loss rates due to the high dust  opacity. \cite{vanLoon2005} suggest the following
empirical formula:
\begin{eqnarray}
\log \dot{M}= -5.65+1.05  \log \left( \frac{L}{10^4 \mathrm{L}_{\odot}}\right) -6.3 \log 
\left(\frac{T_{\mathrm{eff}}}{3500 \;\mathrm{K}} \right) \,.
\end{eqnarray}

Wolf--Rayet (WR) stars are bare cores left over by mass loss from massive stars. Their $\dot{M}$ are  very high,  due mostly to their high $L/M$ ratios.
Models indicate that the $\dot{M}$ values  scale with the actual WR masses like \citep{Nugis2000,Nugis2002},
\begin{eqnarray}
 \log \dot{M} = A+B \log \frac{M}{\mathrm{M}_{\odot}} \;, 
\label{Mdotwr}
\end{eqnarray}   \index{mass--luminosity relation}
\noindent
with average values  $A=-5.73$, $B=0.88\pm0.14$ for 
WR stars in general. All these parametrizations are continuously revised \cite{Graefener2008}.

\begin{figure}[!t]
\centering
\includegraphics[angle=0,width=6cm]{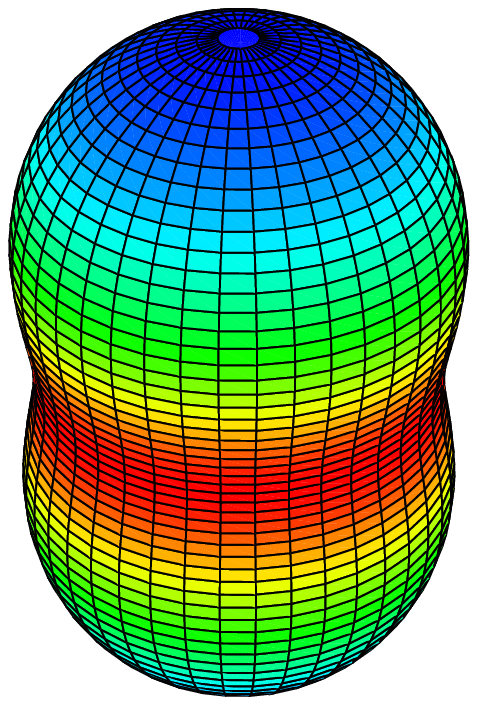}
         \caption{The mass fluxes around a rotating star of 100 M$_{\odot}$ with $10^{6.5}$ L$_{\odot}$ and 
a ratio of the angular velocity to the break-up angular velocity
$\omega = 0.80$, assuming a polar T$_{\mathrm{eff}} = 30000$ K 
\citep{MaederDesjacques2001}.}
\label{etacarFig12}
\end{figure}

\medskip
Rotation  affects  the mass loss in   two   ways \citep{Maeder2000}: - 1.  it makes the stellar winds anisotropic  because  the polar regions are hotter so that the radiation pressure is stronger, - 2. for  given luminosity $L$ and $T_{\mathrm{eff}}$ the average mass loss rates are increased.
Fig. \ref{etacarFig12} illustrates the distribution of the mass loss around a very massive stars. The higher $T$ in polar regions drives a stronger mass flux than 
in the equatorial regions. This characteristic peanut shape has been observed in several nebulae around very massive stars \citep{Nota1999}. For stars with an equatorial
$T_{\mathrm{eff}}$ lower than about 23000 K, the equatorial opacity  increases
enough to allow the radiative acceleration to  push  (despite the lower flux) the matter outward and to produce  an equatorial ring. Some stars, the so-called B[e] stars, simultaneously show a peanut--shaped nebula  and an  equatorial ring.

The anisotropic mass loss rates may be integrated over the stellar surface to give the overall $\dot{M}$ value. The result is that if we consider two stars with the
same $L$ and $T_{\mathrm{eff}}$, but one with rotation velocity $v$ and the other one
with $v=0$, the ratio of their $\dot{M}$--rates may reach very high values \cite{Maeder2000}, especially for stars close to the Eddington limit (i.~e. for the brightest stars where  the radiation pressure gradient becomes close to gravity).
 For  hot stars far enough from the
Eddington limit, the effects are limited, amounting to an increase of at most  70 \%.

 \section{THE CIRCULATION AND MAIN INSTABILITIES IN ROTATING STARS}  \label{cm}
 
 Rotation drives internal circulation currents and  several kinds of turbulent motions may arise from differential  rotation. All these motions may contribute to the internal transport
 of matter and angular momentum.
 
 \subsection{The transports of elements and angular momentum}
 
 Many effects of element transport in stars can be considered as a diffusion process
 characterized by a diffusion coefficient $D$.
 The kinetic theory of gases indicates that in an isotropic turbulent medium with an  average velocity $v$ and a mean free path $\ell$ the diffusion coefficient 
 is $D= (1/3) \;\ell \, v$. Let us call $X_i$, the abundance in mass fraction of particles $i$. If there is a gradient of composition $\vec{\nabla} X_i$,
 the diffusion velocity of particles $i$ is $\vec{v_i}= -(D/X_i) \, \vec{\nabla X_i}$. The diffusion equation of the chemical elements can be derived from 
 the continuity equation, this gives at a Lagrangian mass coordinate $M_r$,
 \begin{eqnarray}
	\varrho{\partial X_i \over \partial t}\bigg|_{M_r} = {1 \over r^2} \, {\partial \over \partial r}
\left (\varrho \, r^2 \, D \, {\partial X_i \over \partial r}\right )  \; .
\label{difx}
\end{eqnarray}
\noindent
The boundary  conditions at the center and surface are
$
	{\partial X_i \over \partial r }\bigg|_{M_r=0}\, = \, {\partial X_i \over \partial r }\bigg|_{M_r=M}\, = \, 0 $,
 where $M$ is the total mass. The diffusion coefficients may vary a lot and great care has to be given to their interpolation, if necessary \citep{MeynetMow2004}.

 \medskip
 
In a differentially rotating star, the evolution of the angular velocity $\Omega$ has to be followed at each level $r$ (for shellular rotation), so that a full description of $\Omega(r,\, t)$ is available. The values of $\Omega(r,\, t)$ influence the mixing of elements and in turn the evolution of $\Omega(r,\, t)$  also depends on the mixing processes and on the distribution of 
the elements. The derivation of the equation for the transport of angular momentum is not straightforward. In the case of shellular rotation, the equation in the Lagrangian form becomes \citep{Zahn1992, Maeder2009}
\begin{eqnarray}
\varrho \, {\partial \over \partial t}(r^2 \overline \Omega)_{M_r}={1 \over 5 \, r^2}{\partial \over \partial r}(\varrho \,  r^4 \overline \Omega \,U_2(r))
+{1 \over r^2}{\partial \over \partial r}\left(\varrho  \, D \,  r^4 \,  {\partial \overline \Omega \over \partial r} \right) \; .
\label{eqn7}
\end{eqnarray}
\noindent
There, $\overline \Omega$ is the average value of $\Omega$ on an isobar, or in approximation (\ref{omegahat}) the value at $P_2(\cos \vartheta)=0$. $U_2$ is the radial component of the meridional circulation velocity (see next subsection).
The second term on the right is a diffusion term, similar in its form  to
(\ref{difx}), while the first term on the right is  \emph{an advective term}, i.e.
the transport by a velocity current. We notice that Eq. (\ref{difx}) does not contain such an advective term. It could contain a term of that kind, however in general it is considered that the combined effect of turbulence and circulation currents
is equivalent  to a diffusion \emph{for the element transport} (see Eq. \ref{Deff}). 

Great care has to be brought to the solution of this equation, since $U_2$ given in the next subsection
contains terms up to the third derivative of $\overline{\Omega}$, so that this equation is of the fourth order. If there is no viscous momentum at the edges, one has two boundary conditions given by 
$
\frac{\partial \Omega}{\partial r} \, = \, 0  $ at the limits. The two other conditions are given by integrating the transport equations at the bottom and
top limits $r_b$ and $r_t$,
\begin{eqnarray}
\frac{1}{5} \, \varrho \, r^4 \, \overline{\Omega} \, U_{2} \bigg|_{\,r_{\mathrm{b}} }  = 
\,\frac{d}{dt} \left[\overline{\Omega} \,
\int^{r_b}_0 r^4 \, \varrho \, dr \right] \quad \mathrm{in} \; r=r_{\mathrm{b}} \, , \nonumber \\ [2mm]
 -\frac{1}{5} \, \varrho \, r^4 \, \overline{\Omega} \, U_2 \, \bigg|_{r_{\mathrm{t}}} = \,\frac{d}{dt} \left[\overline{\Omega} \,
 \int^{R}_{r_t} r^4 \, \varrho \, dr \right]   + \mathcal{M}_{\Omega} \quad 
 \mathrm{in} \; r=r_{\mathrm{t}} \; .
 \label{bounda}
 \end{eqnarray}
 \noindent
 The  radial components $U_2$ of the velocity  at the surface and center are zero.
 $\mathcal{M}_{\Omega}$ represents the momentum of force applied at the stellar surface,
 typically by magnetic field in solar type stars  or  by tidal 
 effects in binary systems. 
 
  \subsection{The meridional circulation}  \label{mcirc}
 The breakdown of the thermal stability  in rotating stars (Sect. \ref{breakd}) drives some circulation currents. For long there was a severe physical  problem: the solutions for meridional  circulation were not conserving the angular momentum. Thus the reality of the circulation was often
  questioned and solutions without circulation were envisaged.
  Zahn has made a great 
   step forward by showing that one must treat
 simultaneously the equation for
 energy conservation which expresses the thermal imbalance, the Poisson equation and the conservation of angular momentum, in order to have a self--consistent solution
 \citep{Zahn1992}.  
 
The derivation of the circulation velocity for shellular rotation is rather lengthy \citep{Zahn1992,Maeder1998}. The starting point is the equation of energy conservation,
 \begin{equation}
\rho \, T \frac{dS}{dt} =
\vec{\nabla} \cdot \left( \chi \vec{\nabla} T \right) + \varrho \,\varepsilon - \vec{\nabla} \cdot
\vec{F}_{\mathrm{h}} \, ,
\label{equdepart}
\end{equation} 
\noindent
where $S$ is the entropy per unit of mass and $\chi$ the thermal conductivity (Sect. \ref{vz}). The term $\varepsilon$ refers to the nuclear energy production rate 
only. $\vec{F}_{\mathrm{h}} = - D_{\mathrm{h}} \, \varrho\,  C_P \, \vec{\nabla}_{\mathrm{h}} T $ is the flux of thermal energy due
to the  horizontal turbulence with a diffusion  coefficient $D_{\mathrm{h}}$ (Sect. \ref{hzdiff}). $C_P$ is the specific heat at constant pressure $P$ and $\vec{\nabla}_{\mathrm{h}}$ expresses
the horizontal gradient. 
On an isobar, the variables are expanded around their average   up to the second Legendre polynomials $P_2(\cos \vartheta)$,
where $\vartheta$ is the colatitude, for example $T \left( P,\vartheta \right) = \overline{T} \left( P \right) + \widetilde{T}
\left(P\right) \, P_2 (\cos \vartheta) $.

  The velocity of meridional circulation is the main quantity characterizing this effect, it is expanded into 2 components    
\begin{eqnarray}
\vec{U} \, = \, U_2(r)\, P_2(\cos \vartheta) \, \vec{ e}_r + V_2(r) \, \frac{dP_2(\cos \vartheta)}
{d \vartheta} \, \vec{e}_\vartheta \; ,
\label{U22}
\end{eqnarray}
\noindent
$U_2(r)$ is the amplitude of the radial component of the meridional circulation velocity, $V_2(r)$ is the amplitude of the horizontal component. The general expression of $U_2(r)$ in a stationary situation is
\begin{eqnarray}
  U_2(r) =  \frac{P}{\overline{\varrho \,g}\, C_P  \overline{T}\, \left(\nabla_{\mathrm{ad}}-
  \nabla+ \frac{\varphi}{\delta}  \nabla_{\mu} \right)} \left[  \frac{L(r)}{M_{\ast}(r)}  \left(E_{\Omega}+E_{\mu} \right)
  \right] \, . \quad \; \; \; \; \; 
  \label{UU22}
  \end{eqnarray}
\noindent
 $\nabla_{\mu}=(d\ln \mu/d \ln P)$,
$\varphi= (\partial \ln \varrho/\partial \ln \mu)_{P,T}$,
$\delta= -(\partial \ln \varrho/\partial \ln T)_{P,\mu}$,
$E_{\Omega}$ contains dozens of terms up to the third derivative 
of $\Omega$ and $E_{\mu}$ up to the second derivative of  the mean molecular weight. An important term, also present in the case of uniform rotation in an homogeneous medium, is 
 \begin{eqnarray}
E_{\Omega}  \simeq  \frac{8}{3} \left[ 1 - \frac{{\Omega^2}}
{2\pi G\overline{\varrho}}\right] \left( \frac{\Omega^2r^3}{GM_r}  \right) .
 \end{eqnarray}
\noindent
This term is proportional to the ratio of the centrifugal 
to the gravitational force. The second term in the square bracket is the so-called
Gratton--\"{O}pik term. In the deep interior, it is generally negligible while it may become dominant in the outer layers, where the local average density $\overline{\varrho}$ is negligible. 
The sign of $U_2$ is however not  determined only by the sign of the Gratton--\"{O}pik term and the complete expression has to be considered.
Generally the inner cell is turning upward along the polar axis, while the external cell turns the opposite way (Fig. \ref{circu}).
 This means that the inner cell is bringing 
angular momentum inward, while the outer cell brings it outward.  However, the circulation patterns, which depends on the derivatives of $\Omega$ and $\mu$, may be different according to the stellar model considered.

\begin{figure}[!t]    
\centering
\includegraphics[angle=0,height=7.0cm,width=8.0cm]{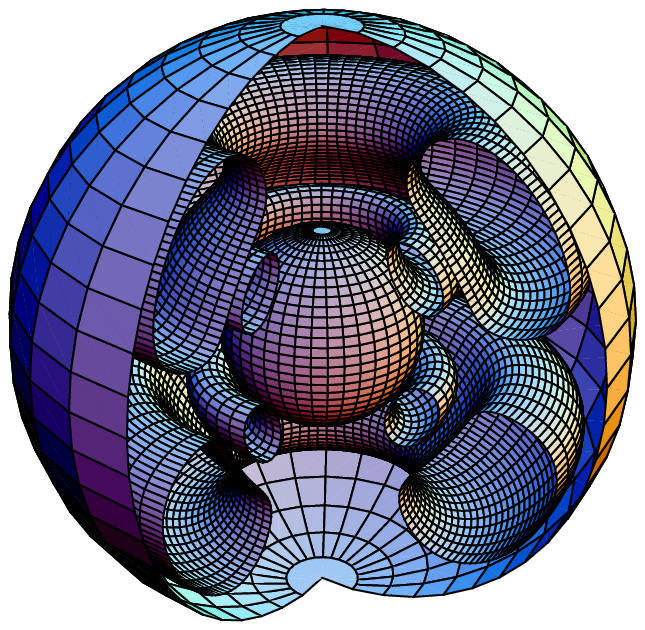}
\caption{Schematic  structure with stream lines of meridional circulation in a rotating 20 M$_\odot$
model of 5.2 R$_\odot$ with $Y=0.30$, $Z=0.02$ and
$v_{\rm ini}=300$ km s$^{-1}$ at the beginning of the MS phase. 
The figure  is made  as a function of $M_r$.
In the upper hemisphere on the right section, matter
is turning counterclockwise along the outer stream line and
clockwise along the inner one.  The inner sphere is the
convective core. It has a radius of 1.7 R$_\odot$  \citep{paperVIII2002}.}
\label{circu}
\end{figure}

The circulation velocities are typically of the order of a few cm s$^{-1}$ in upper 
MS stars. The velocities are larger at the beginning of the Main Sequence (MS) phase when rotation
is (supposedly) uniform, then in a few percents of the MS timescale the
 velocities become smaller reaching an equilibrium. At the end of the MS phase,
 they may again become large (negatively) due to the very high Gratton--\"{O}pik term,
 because of the very low density in the outer layers.
The circulation timescale $t_{\mathrm{circ}}$, also called the Eddington--Sweet 
  timescale $t_{\mathrm{ES}}$, is of the order of the stellar
  radius  divided by the velocity, 
 \begin{eqnarray}
 t_{\mathrm{ES}} \, \approx \, \frac{R}{U_2(R)} \, \approx \, \frac{GM^2}{L \, R} \, \frac{G \,M}{\Omega^2 R^3}
 \, \approx \,\frac{t_{\mathrm{KH}}}{\Omega^2 R^3/({GM})} \; .
 \label{tes}
 \end{eqnarray} 
\noindent
where $t_{\mathrm{KH}}$ is the Kelvin--Helmholtz timescale, i.e.
the  thermal timescale of the star (about $3 \times 10^7$ yr for the Sun).
The circulation timescale
is equal to about  $t_{\mathrm{KH}}$ divided by the ratio of the centrifugal force to the gravity  $\Omega^2 R^3/({GM})$, thus  
$t_{\mathrm{ES}}$ is some  multiple of $t_{\mathrm{KH}}$.


 Meridional circulation is important, because it transports both angular momentum and chemical elements. The presence of horizontal turbulence in differentially rotating stars affects these two kinds of transport differently.
  The interaction of the transport of chemical elements  by meridional circulation and horizontal turbulence results in a diffusion process \citep{Chaboyer1992}, with a diffusion coefficient 
  \begin{eqnarray}
D_{\mathrm{eff}} \, = \, \frac{(r \, U_2)^2}{30 \, D_{\mathrm{h}}} \, .
\label{Deff}
\end{eqnarray} 
\noindent 
where $D_{\mathrm{h}}$ is the diffusion coefficient by the horizontal turbulence as discussed in next Section. The above expression is valid only if $D_{\mathrm{h}}$
is large with respect to $r \, U_2$ and to the coefficient of vertical diffusion.

The above reasoning does not apply to the transport of angular momentum, thus Eq. (\ref{eqn7}) contains both  advective and diffusive terms. The reason is that angular momentum conservation limits the effects of horizontal motions on the transport of the angular momentum. Indeed, turbulence may be present without a net flux of angular momentum, however it maintains a shellular distribution of velocity,
as shown by 
numerical models \citep{Charbonneau1992}.

 \subsection{The horizontal turbulence in differentially rotating stars} \label{hzdiff}
 
 Rotation and especially differential rotation generates  turbulent motions.
 On the Earth, we have the example of west winds and  jet streams.
 In a  radiative zone, the turbulence is   stronger \citep{Zahn1992} in the horizontal 
than in the vertical direction, because in the vertical direction
the stable  thermal gradient opposes a strong force to the fluid motions. The horizontal turbulence is characterized by a diffusion coefficient
$D_{\mathrm{h}}$, which also expresses the horizontal viscosity. There are several observational indications supporting the existence of this turbulence \citep{Maeder2009}.
The expression of $D_{\mathrm{h}}$ is  uncertain
and three different forms have been proposed \citep{Zahn1992,Mathis2004a,Maeder2003}. That by Mathis et al.
is based on the Couette--Taylor experiment, while the other two are phenomenological. The expressions by Mathis et al. and Maeder  give 
similar results, larger   by orders of magnitude with respect to those obtained with Zahn's expression.

A  diffusion coefficient behaves like $ \ell^2/{t_{\mathrm{diff}}}$, where $\ell$ is a characteristic length (here typically $r$) and  $t_{\mathrm{diff}}$ a diffusion timescale. Thus, the question is to know which timescale is to be considered. One may see that the coefficient by Zahn
corresponds to about a value  $t_{\mathrm{diff}} \sim r/V_2$, where $V_2$ is defined
in Eq. (\ref{U22}), where one has accounted that $V_2 \sim U_2/3$.
The coefficient by Mathis et al. corresponds to a value  $t_{\mathrm{diff}} \sim 1/(\beta \Omega_2)$, where $\Omega_2$ is defined in Eq. (\ref{omegahat}) and
$\beta$ is a numerical coefficient of about $1.5 \cdot 10^{-5}$, determined empirically. The coefficient by Maeder corresponds to $t_{\mathrm{diff}} \sim
[r/(\Omega_2 V_2)]^{1/2}$, meaning that the timescale  considered is that necessary for the differential effects of the Coriolis 
force to cross the radius. At present, it is still uncertain which coefficient is to be chosen, even if the last two are in numerical agreement.

\subsection{Shear mixing and other instabilities}

A difference of horizontal velocity $V$ between two layers
 may generate a shear instability. The   vertical density 
 stratification (along axis $z$ or $r$) favors  stability and  the balance of the two effects
 has to be considered. 
 The criterion expressing the instability is known as the Richardson criterion, it is
 \begin{eqnarray}
\mathcal{R}i \, \equiv \, \frac{g}{\varrho} \, \frac{d \varrho/dz} {\left(dV/dz\right)^2} \, < \mathcal{R}i_{\mathrm{crit}}\; .
\label{Ri}
\end{eqnarray}
\noindent
The critical Richardson number is often taken equal to $1/4$, some studies suggest a value up to 1.0
\citep{Canuto2002}.
The criterion  says that an instability develops if the excess of energy $ \mathcal{R}i_{\mathrm{crit}}\varrho \,
(\delta V)^2$  in the differential motions (with respect to an average velocity)
is bigger than the work $g  \,\delta \varrho \, \, \delta z$ necessary to exchange the matter vertically. 
The criterion can be expressed in term of the Brunt--V\"{a}is\"{a}l\"{a} frequency $N$, i.e. the frequency of oscillatory motions under the restoring force of gravity,
\begin{eqnarray}
\mathcal{R}i \, \equiv \,\frac{N^2}{\left(dV/dz\right)^2} \, <  \mathcal{R}i_{\mathrm{crit}}  \label{Ri1} \, , \quad 
 \mathrm{with} \quad N^2 \, = \, \frac{g \, \delta}{H_P}
 \, \left( \, \nabla_{\mathrm{ad}}-\nabla + \frac{\varphi}{\delta}\nabla_{\mu} \right)  \, . 
 \label{Ri11}
 \end{eqnarray}

The density stratification, in particular in regions with a high $\mu$ gradient,
is generally able to prevent the shear instability.
However, the thermal diffusivity of a displaced fluid element weakens the stabilizing  effect of 
the  thermal stratification and  this may lead to a significant  mixing.
The left hand of Eq. (\ref{Ri11}) is then multiplied by $v \,\ell/(6 \,K)$, where 
$v$ and $\ell$ are the velocity and lengthscale of the vertical motions (the factor 6
applies to a spherical fluid element).
$K$ is the thermal diffusivity $K=4acT^3/(3 \kappa \varrho^2 C_P)$. The criterion is then \citep{Zahn1992, Maeder1995}
\begin{eqnarray}
\mathcal{R}i \, \equiv \,\frac{N^2}{\left(dV/dz\right)^2} \,\frac{v \ell}{6\, K} <  \mathcal{R}i_{\mathrm{crit}} \, .
 \label{Ri12}
 \end{eqnarray}
 \noindent
 For constant $\mu$, the coefficient for the vertical diffusion by 
shears is easily expressed. One supposes that the turbulent diffusion is dominated by the largest eddies 
satisfying  the Richardson's criterion 
\begin{eqnarray}
D_{\mathrm{shear}}  =  \frac{1}{3} v \, \ell \, =\, 2  \,
 \mathcal{R}i_{\mathrm{crit}} \, K \,\frac {\left(dV/dz\right)^2} {N^2} \; . \quad 
 \label{dkggg}
\end{eqnarray}
\noindent
This coeffcient expresses the transport of the elements and of the angular momentum
produced by shear turbulence in an homogeneous medium, with appropriate averages
on isobars \citep{Maeder2009}. In principle, the thermal gradient is slightly modified by the turbulence due to the shears, however this effect is often ignored.

In a zone with variable $\mu$, the situation is different.
The horizontal turbulence tends to reduce the composition 
difference between the turbulent eddies and their surroundings. Thus, the restoring force of buoyancy is smaller   and the 
shear instability is favored. The resulting coefficient becomes \citep{Talon1997}
\begin{eqnarray}
D_{\mathrm{shear}}  =\, 2  \,
 \mathcal{R}i_{\mathrm{crit}} \,\frac {\left(dV/dz\right)^2}
 {\frac{N_{T,\, \mathrm{ad}}^2}{K+D_{\mathrm{h}}}+\frac{N^2_{\mu}}{D_{\mathrm{h}}}} \; , \\[2mm]
 \mathrm{with} \quad N_{T,\, \mathrm{ad}}^2 \, = \, \frac{g \, \delta}{H_P}
 \, \left( \, \nabla_{\mathrm{ad}}-\nabla  \right) \quad \mathrm{and} \quad 
  N_{\mu}^2 \, = \, \frac{g }{H_P}
 \,  {\varphi}\nabla_{\mu} \,.
 \label{dkg}
\end{eqnarray}
\noindent
The horizontal turbulence favors the diffusion by the shear, reducing the inhibiting
effect of the $\mu$--gradient.

There are other possible instabilities, which may play a role in differentially rotating stars (baroclinic instabilities).
The Goldreich-Schubert-Fricke (GSF) instability \citep{Goldreich1967} which says that instability  arises if the $T$ gradient, with account for thermal and viscous diffusivities, is
 insufficient to compensate for the growth of the centrifugal force during an arbitrary small displacement. If there is a significant $\mu$-gradient, the instability is killed, thus it is generally not important.
 Another possible instability is the so-called  Axisymmetric-BaroClinic-Diffusive (ABCD) instability \citep{Knobloch1983,Talon1997}. If a fluid element
   moves over an arbitrary displacement from point A to C, it is brought back to C due to the conservation of angular momentum. 
   However, in C the fluid eddy may  be hotter than the surroundings
   (since the lines of constant $\Omega$ and $T$ do  not necessarily coincide in a baroclinic star). Due to thermal diffusivity,  the fluid element
     loses some energy on its way. Thus, when back to A, the eddy is  cooler and  it overpasses 
    its equilibrium position and the oscillation amplitude grows. The viscosity due 
    to the horizontal turbulence  tends to suppress this instability. On the whole, one must say that the real importance of the GSF and ABCD instabilities is still uncertain.

\section{ROTATION AND MAGNETIC FIELDS}

All the way from star formation to the endpoints, rotation and magnetic fields may be interacting. 
From the initial  molecular clouds to the present Sun, the specific angular momentum
 decreases by  a factor $\sim 10^6$. During the pre-MS phase, disk locking is a major effect for this reduction \citep{Hartmann1998}.   Fields of $\sim 1$ kG are sufficient for  coupling a solar type  star with an extended  accretion disk. The contracting star is bound to its disk and it keeps the same angular velocity during contraction, thus losing  lots of angular momentum.
 In solar type stars,
the magnetic field creates a strong coupling between the star and the 
solar wind, which leads to further losses of angular momentum during the
end of the pre-MS phase and  the MS phase \citep{Kawaler1988,Krishnamurthi1997}.

Spectropolarimetric surveys have obtained evidences for the presence of magnetic field at the surface 
of OB stars \citep[see e.g. the recent review by ][and references therein]{Walder2011}. 
The origin of these magnetic fields is still unknown. It might be fossil fields \citep[e.g. the
spectral characteristics of  Of?p stars are indicative of organized magnetic fields, most likely 
of a fossil origin according to][]{WadeGrun2010}, or fields produced through a dynamo mechanism. Recent simulations by  \citet{Cantiello2010}
of subsurface convective zone in massive stars show dynamo-generated magnetic fields of the order of one kG. According to these authors, these 
magnetic fields might reach the surface of OB stars.

A big question 
is whether there is  a dynamo working in internal radiative zones. This could have  far reaching consequences
concerning the mixing of the  elements and the loss of  angular momentum.
This question is also essential regarding the rotation periods of pulsars and the origin of GRBs (Sect. \ref{pulsar} and \ref{grb}). It may also be important in order to explain the flat profile of the internal rotation of the Sun \citep{Eggenberger2005a}. Note that an alternative process has been invoked to explain
this feature, the action of internal gravity waves \citep{CT2005} .

 \subsection{Some properties of a dynamo in a radiative zone}
 
 We  examine  some properties of  a dynamo in a radiative region. 
 A  particular case is  the Tayler--Spruit  dynamo   \citep{Spruit2002} and we are using many  equations of this dynamo \citep{Maeder2005a}.
Some  numerical simulations  \citep{Braithwaite2006,Zahn2007}   confirm the existence of a magnetic instability, however the existence of the dynamo is still controversial.

The two quantities we are mainly interested in for stellar evolution are the magnetic viscosity $\nu$, which expresses the mechanical coupling due to the  magnetic field $\vec{B}$, and the magnetic diffusivity $\eta$ which expresses 
the transport by a magnetic instability and thus also the damping of the instability.
Some general expressions for these two quantities may be derived.
Let us  first examine the  energy conservation.
The rate of magnetic energy production 
  $W_{\mathrm{B}}$ per unit of time and volume is of the same order of magnitude as  the rate
  $W_{\nu}$ of  dissipation of the excess of rotational energy by the magnetic viscosity
  $\nu$.
  The differential motions in shellular rotation give velocity differences  $dv= r \, d\Omega$.
  The amount of energy  dissipated  during a time $dt$  for an element of matter 
 $ dm$ in a volume $dV$ is 
  \begin{eqnarray}
W_\nu=\frac{1}{2} \, dm \,(dv)^2 \frac{1}{dV} \, \frac{1}{dt}= \frac{1}{2} {\varrho} \, \nu
\left(\frac{dv}{dr}\right)^2 = {1\over 2}\varrho\nu \Omega^2q^2 \quad \quad
\mathrm{with} \quad q = r \left|\nabla \Omega\right|/\Omega,
\label{Wnu}
\end{eqnarray}
\noindent
because the viscous time is $dt=(dr)^2/\nu$.
The magnetic energy density is $u_{\mathrm{B}}={B^2}/(8 \pi)$, it
is produced within the characteristic growth time of the magnetic field $\sigma_{\mathrm{B}}^{-1}$, 
thus the rate $W_{\mathrm{B}}$ of magnetic energy creation by unit of volume and time is
\begin{equation}
W_{\mathrm{B}}= \frac{B^2}{8 \pi} \sigma_{\mathrm{B}}= \frac{1}{2} \omega^2_{\mathrm{A}}r^2 \sigma_{\mathrm{B}}\varrho \;,
\label{WB}
\end{equation}

\noindent
where we have used   the expression of the Alfv\'en frequency 
$\omega_{\mathrm{A}} \, = \, \frac{B}{r (4 \, \pi \varrho)^{1/2}} $.
Let us  assume $W_{\nu}  =  W_{\mathrm{B}}$, i.e. that the excess of energy in the differential rotation
is converted to magnetic energy
by unit of time. This gives for the viscosity coefficient of magnetic coupling
\begin{eqnarray}
\nu = \frac{\omega^2_{\mathrm{A}}\, r^2 \, \sigma_{\mathrm{B}}}{\Omega^2 \,q^2} \; .
\label{nu1}
\end{eqnarray}
\noindent
This  coefficient  intervenes in the expression for the transport
of angular momentum, in the Lagrangian form as given by Eq. (\ref{eqn7}).
Compared to the  energy available for the solar dynamo, the amount of energy available from differential rotation is very limited.

\medskip
If due to an instability, some  vertical motions
 with an  amplitude $l_r/2$ occur around an average  position,
 the restoring buoyancy force produces  vertical oscillations at
 the Brunt--V\H{a}is\H{a}l\H{a} frequency $N$. 
The restoring oscillations will have an average density of kinetic energy 
$u_{\mathrm{N}}  =   f \, \rho \, l_r^2 \, N^2$,
where $f$ is of the order of unity.
The magnetic field must overcome the buoyancy effect, which implies 
$ u_{\mathrm{B}} >  u_{\mathrm{N}} $,
 otherwise the restoring force of gravity
 would  counteract the magnetic instability  at the dynamical timescale. From this
 condition, one obtains 
$l_r^2 <  \frac{1}{2f} \,r^2 \, \frac{\omega^2_{\mathrm{A}}}{N^2}$.
 If, $f= {1 \over 2}$, we  have the condition \cite{Spruit2002}
 \begin{equation}
 l_r \; < \,r \, \frac{\omega_{\mathrm{A}}}{N}  \; .
 \label{lr}
 \end{equation}

An  unstable vertical displacement of size $l_r$ of the azimuthal field of  lengthscale  $r$
and intensity $B_{\varphi}$ also feeds a radial field component $B_r$. The relative sizes of these two field components are
defined by the induction equation,
 which gives the following scaling over the time $\delta t$ characteristic 
of the unstable displacement,
$
B_r \, \approx \,  \delta B  \, \approx \, \frac{1}{r} \,\frac{l_r}{\delta t} \, B_{\varphi} \,   \delta t  \; .
$
For the maximum displacement $l_r$ given by Eq. (\ref{lr}), this  gives
\begin{eqnarray}
\frac{B_r}{B_{\varphi}} \approx \frac{l_r}{r}  \; ,
\label{lrr}
\end{eqnarray}
\noindent
which provides  an estimate of the ratio of the radial to azimuthal fields.

\medskip

 Let us now consider the magnetic diffusivity $\eta$, which  damps the  instability.
 The thermal diffusivity $K$ produces heat losses from the unstable fluid elements
and thus  reduces the buoyancy forces opposed to the magnetic instability.  Both effects
have to be accounted for.
 If the radial scale of the vertical instability is   small, the perturbation is quickly damped by the magnetic diffusivity  
  $\eta$ (in cm$^2$ s$^{-1}$). The radial amplitude  must satisfy the relation,  
  \begin{eqnarray}
 l_r^2 >  \frac{\eta}{\sigma_{\mathrm{B}}} \; , 
 \label{lmin}
 \end{eqnarray} 
 \noindent
 where, as seen above, 
 $\sigma_{\mathrm{B}}$ is the characteristic frequency for the growth  of the  instability. 
The combination of the two limits   (\ref{lr}) and (\ref{lmin}) gives for the case of marginal stability,
\begin{equation}
\eta \, = \, \frac{r^2 \, \omega^2_{\mathrm{A}} \, \sigma_{\mathrm{B}}}{N^2}   \; .
\label{premier}
\end{equation} 

\noindent
For given $\eta$ and $\sigma_{\mathrm{B}}$, this provides the minimum value of $\omega_{\mathrm{A}}$, and thus of the magnetic field $B$, for the instability to 
occur.
The instability is confined within a domain, limited on the large side 
by the stable stratification (\ref{lr}) and on the small scales by magnetic diffusion (\ref{lmin}). For the case of marginal stability, which is likely reached in evolution, this equation relates the magnetic diffusivity $\eta$ and the Alfv\'en frequency $\omega_{\mathrm{A}}$.

 The Brunt--V\"{a}is\"{a}l\"{a} frequency $N$  of a fluid element displaced
in a medium with account of both the magnetic and thermal diffusivities $\eta$ and
$K$ is  \citep{magnII2004},
\begin{eqnarray}
N^2 = \frac{\frac{\eta}{  K}} {\frac{\eta}{  K}  + \, 2} \, \; N^2_{T, \, \mathrm{ad}}+ N^2_{\mu}  \; , \label{N2} 
\end{eqnarray}
\noindent
with $N^2_{T, \, \mathrm{ad}}$ and  $N^2_{\mu}$ defined in Eq. ({\ref{dkg}). 
The ratio $\eta/K$ of the magnetic to thermal diffusivities determines the
  heat losses. The factor of 2 is determined by the geometry of the instability, a factor of 2 applies to a thin slab, for a spherical element a factor of 6 is appropriate
 \cite{magnIII2005}. 
In the interior of a 15 M$_{\odot}$ star, $\eta/K$ lies between $10^{-6}$ and $10^{-2}$. Thus the $T$--stratification only has a  little stabilizing effect, compared to the $\mu$--gradient.

\medskip
Let us now examine the magnetic coupling, 
The  momentum of force $\vec{S}$ by volume unity due to the magnetic field  is obtained 
  by writing  the momentum of  the Lorentz force $\vec{F}_{\mathrm{L}}$.
  The current density $\vec{j}$ is given by the Maxwell equation $\frac{4 \, \pi}{c} \,\vec{j}= \vec{\nabla } \times {\vec{B }}$. Thus, one has 
  \begin{eqnarray}
  \vec{S}=\vec{r} \times \vec{F}_{\mathrm{L}}=\frac{1}{c} \vec{r} \times(\vec{j} \times \vec{B})
  =\frac{1}{4  \, \pi} \vec{r} \times \left((\vec{\nabla} \times \vec{B}) \times \vec{B} \right) \, , \; \;\\[2mm]
  \mathrm{in \; modulus} \quad S \, \approx \,  \frac{1}{4 \; \pi} \; B_{\mathrm{r}} B_{\varphi} \; = \;
  \frac{1}{4 \; \pi} \;  \left(\frac{l_{\mathrm{r}}}
  {r}\right) B_{\varphi}^2 = 
  \; \rho \; r^2 \; \left(\frac{\omega_{\mathrm{A}}^3}{N}\right)  \; .
  \label{S}
  \end{eqnarray}
 
 \noindent
 The units of $S$ are g s$^{-2}$ cm$^{-1}$, the same as for $B^2$ in the Gauss system.
 The kinematic viscosity $\nu$ (in cm$^2$ s$^{-1}$) 
 for the vertical transport of angular momentum is
 \begin{eqnarray}
 \nu=\frac{\eta}{\varrho} = \frac{1}{\varrho}  \, F  \, \frac{dr}{dv}=
 \frac{1}{\varrho} \, F \, \frac{dr}{r d\Omega} = 
 \frac{1}{\varrho} \, F \, \frac{d \ln r}{\Omega \,d \ln \Omega} \;,
 \end{eqnarray}
  
 \noindent
where  F   is a force by surface unity, which also
 corresponds to a momentum of force by volume unity in
   g s$^{-2}$ cm$^{-1}$. $\vec{F}$ is  applied horizontally to a slab of velocity $v$ in
   a direction perpendicular to $r$. 
 Considering only positive quantities, with  $q= \left|d\ln \Omega/d \ln r\right|$,
 one has
 \begin{equation}
 \nu = \frac{S}{\rho \; q \; \Omega} = \; \frac{\omega^3_{\mathrm{A}} \, r^2}{N\, q \, \Omega} \;.
 \label{nu2}
 \end{equation}
 
 Now, we can compare this expression for $\nu$ to Eq. (\ref{nu1}) and  get
 \begin{eqnarray}
 \sigma_{\mathrm{B}} \; = \; \frac{\omega_{\mathrm{A}} \, \Omega \, q}{N} \;,
 \label{sigfdt} 
 \end{eqnarray}
 \noindent
 which relates the growth rate of the magnetic field to 
 its amplitude (through $\omega_{\mathrm{A}}$).

\subsection{The basic equations}

Introducing the expression (\ref{sigfdt}) of $\sigma_{\mathrm{B}}$ in Eq. (\ref{premier}), we get
 \begin{eqnarray}
 \eta \,=\, r^2 \, \Omega \, q \left(\frac{\omega_{\mathrm{A}}}{N}\right)^3 \, \, .
 \label {1etaN}
 \end{eqnarray}
 \noindent
 Also, with the expression (\ref{N2}), we can write for $\sigma^2_{\mathrm{B}}$
 \begin{eqnarray}
 \sigma^2_{\mathrm{B}} \, = \,\frac{\omega^2_{\mathrm{A}} \, \Omega^2 \, q^2}{\frac{\frac{\eta}{  K}} {\frac{\eta}{  K}  + \, 2} \, \; N^2_{T, \, \mathrm{ad}}+ N^2_{\mu}  } \;.
 \label{2sigmaN}
 \end{eqnarray} 
\noindent
These equations are quite general. If the growth rate $\sigma_{\mathrm{B}}$ of the 
instability is known, the two equations (\ref{1etaN}) and (\ref{2sigmaN}) form a system of 2 equations with 2 unknowns $\eta$  and $\omega_{\mathrm{A}}$.

\subsection{The case of Tayler--Spruit dynamo}

\begin{figure}[!t]
\centering
\includegraphics[height=4.2cm,width=10cm]{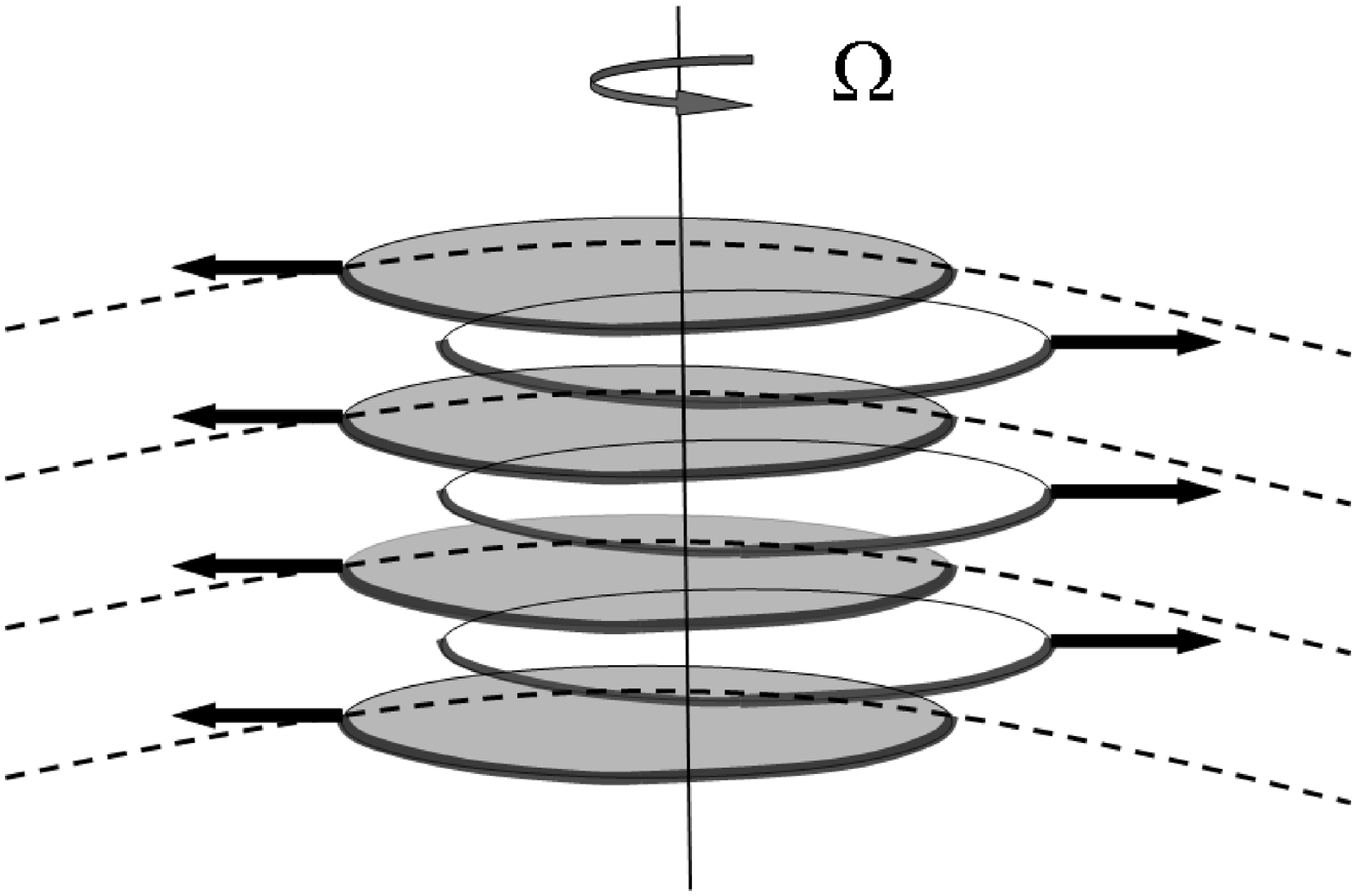}
\caption{Unstable displacements with $m=1$ of the  azimuthal magnetic field in the polar region.
The large arcs (thin broken lines) indicate the horizontal stellar  surfaces. Adapted from \citet{Spruit1999}.}
\label{pneus}
\end{figure}

A purely toroidal    field $B_{\varphi}(r, \vartheta)$, even very weak, in a stable stratified  star is unstable on an Alfv\'en timescale $1/\omega_{\mathrm{A}}$ \citep{Tayler1973}. This is the first magnetic instability to appear. It is non--axisymmetric 
of  type  $m=1$ \citep{Spruit1999,Spruit2002},
it occurs under a wide range of conditions and is characterized by a low threshold and a  short growth time.  
For an azimuthal field consisting of concentric loops around
the rotation axis, the instability appears as  low--azimuthal order displacements of  rings  (Fig. \ref{pneus}). Due to the magnetic pressure, the magnetic loops move apart from the rotation axis, like a disordered heap of tires, as described by Spruit. The pressure is released by sideways motions. 
  In a rotating star, the instability is also present \citep{Pitts1985},  however the  growth rate $\sigma_{\mathrm{B}}$ of the instability is, if $\omega_{\mathrm{A}} \ll \Omega$,
\begin{eqnarray}
\sigma_{\mathrm{B}} \, = \, \frac{\omega_{\mathrm{A}}^2}{\Omega}  \; ,
\label{sigcoriolis}
\end{eqnarray}

\noindent
instead of the Alfv\'en frequency $\omega_{\mathrm{A}}$, because the growth rate of the instability  is reduced by the Coriolis force \citep{Spruit2002}. One usually has the following ordering of the different   frequencies, 
$
N \, \gg  \, \Omega \, \gg \, \omega_{\mathrm{A}} .$
In the Sun, one has $N \approx  10^{-3}$ s$^{-1}$, $\Omega = 3 \times 10^{-6}$ s$^{-1}$ and 
 a field of 1 kG would give  an Alfv\'en frequency as low as $\omega_{\mathrm{A}} = 4 \times 10^{-9}$
 s$^{-1}$.

 The frequency $\sigma_{\mathrm{B}}$ being known, the system of Eqs. (\ref{1etaN}) and  (\ref{2sigmaN}) is defined and
  the 2  unknown 
 $\eta$ and $\omega_{\mathrm{A}}$ can be obtained \cite{magnIII2005}.
When $N_{\mu}$ dominates, which occurs around the stellar core, one has
\begin{eqnarray}
\frac{\omega_{\mathrm{A}}}{\Omega} = q \; \frac{\Omega}{N_{\mu}}   \quad \mathrm{and} \quad
\eta = r^2 \Omega  \, q^4 \; \left(\frac{\Omega}{N_{\mu}}\right)^6  \; .
\label{etaz0}
\end{eqnarray}
\noindent
This shows that the mixing of the elements decreases
 strongly for  large $\mu$ gradients and grows fast for large differential rotation. 
 The ratio $\omega_{\mathrm{A}}/\Omega$  has to be equal or larger than 
 the minimum value defined by (\ref{premier}). This  leads to a condition on the minimum differential rotation  for the
 dynamo to work \citep{Spruit2002},
 \begin{eqnarray}
q \;  > \; \left({N \over \Omega}\right)^{7/4} \left({\eta \over r^2 N}\right)^{1/4}\; ,
\label{cond1}
\end{eqnarray}
\noindent
 When $N^2$ is
large, as for example when there is a significant $\mu$ gradient, the differential rotation necessary for the dynamo to operate must also be large.

\begin{figure}[!t]
\centering
\includegraphics[height=6.0cm,width=7.2cm]{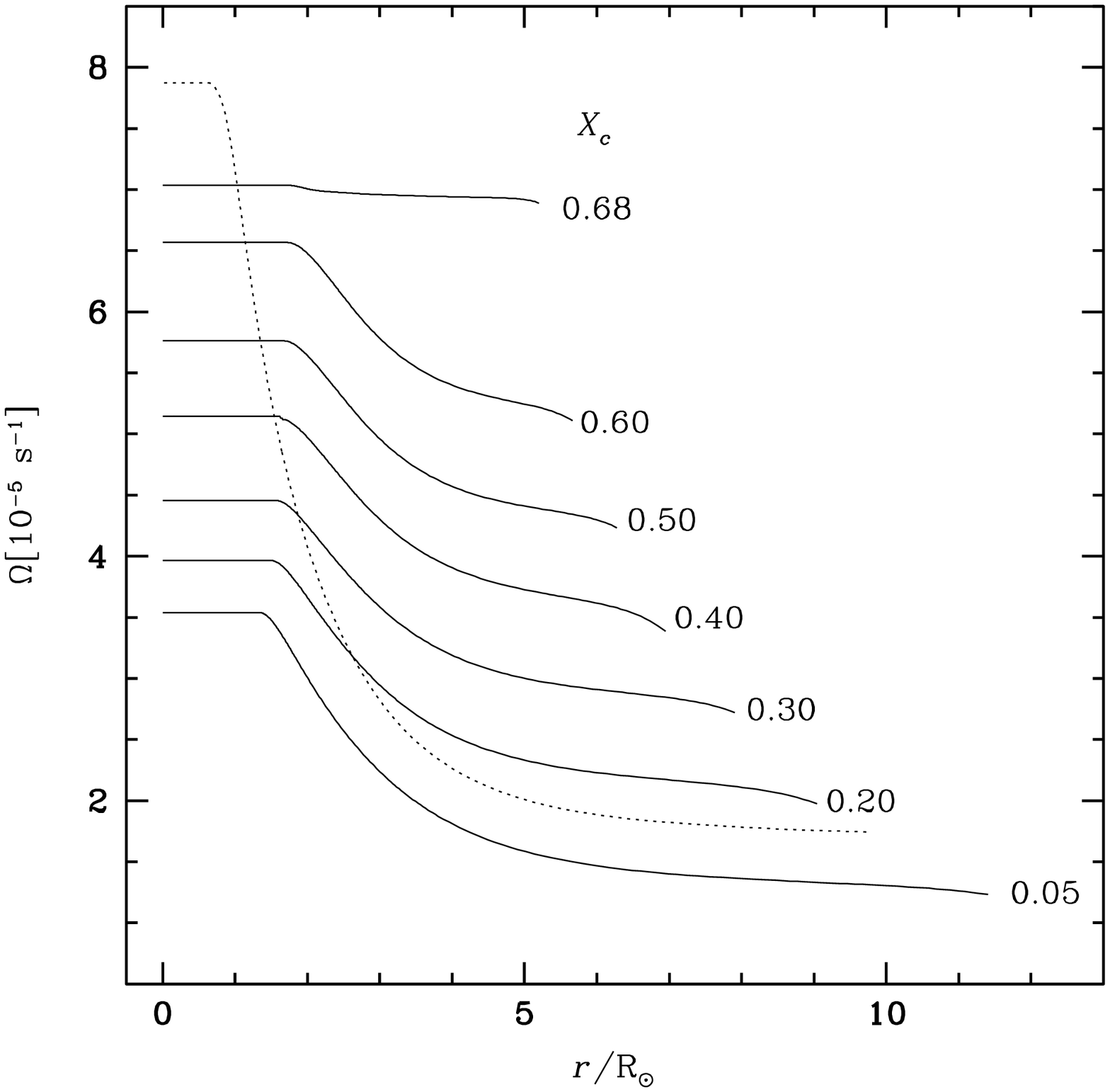}
\includegraphics[height=6.0cm,width=7.2cm]{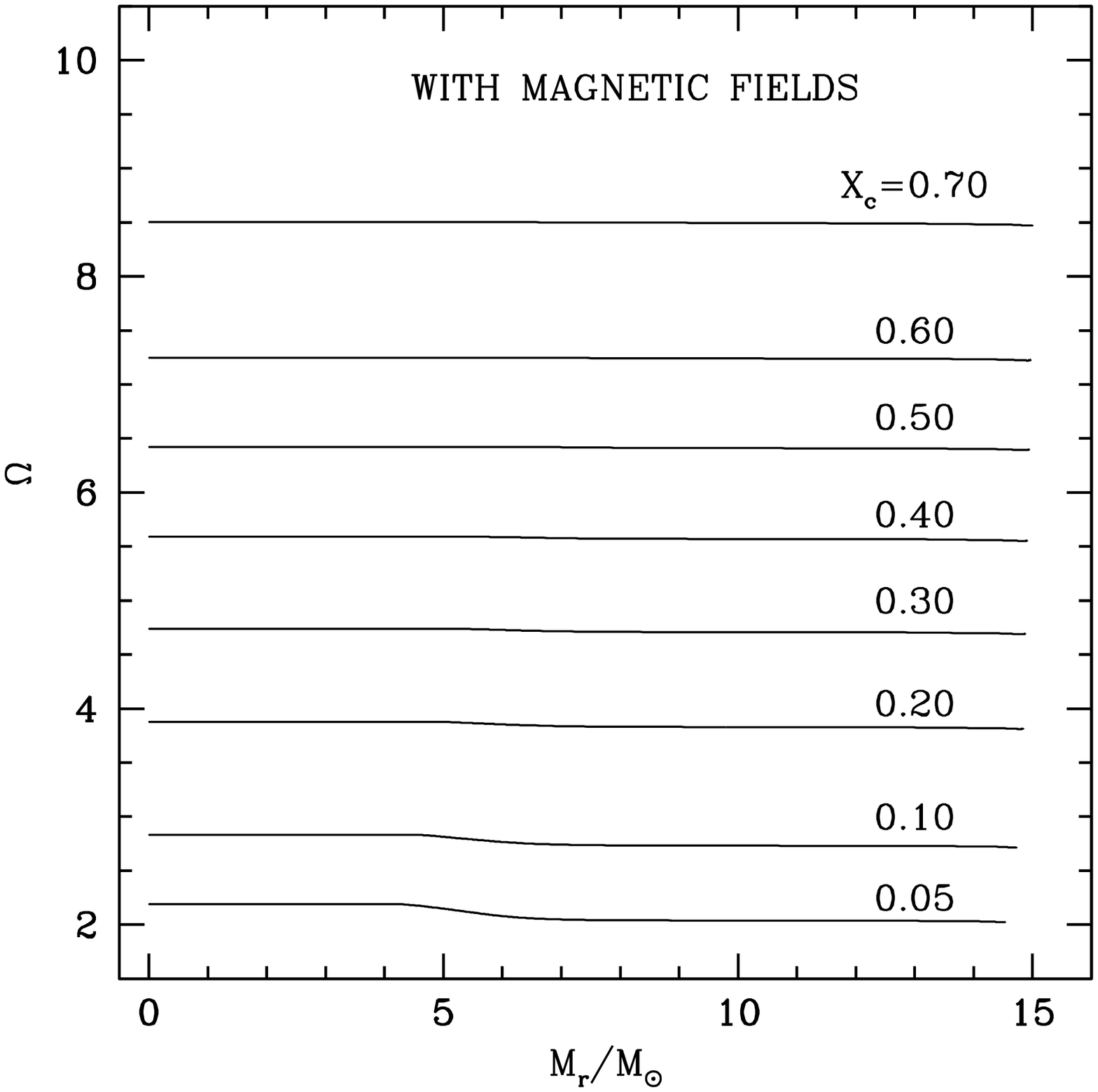}
\caption{Left: evolution of the angular velocity $\Omega$ 
 as a function of the distance to the center
in a 20 M$_\odot$ star with $v_{\rm ini}$ = 300 km s$^{-1}$. 
$X_c$ is the hydrogen mass fraction at the center.
The dotted line shows the profile when the He--core contracts at the end
of the H--burning phase \citep{paperV2000}. Right: rotation profiles
at various stages of evolution  (labeled by the central H content $X_{\mathrm{c}}$) of a 15 M$_{\odot}$ 
model with $X=0.705, Z=0.02$, an initial velocity of 300 km s$^{-1}$ and magnetic field from the Tayler--Spruit dynamo
\citep{magnIII2005}.}
\label{Omprofile}
\end{figure} 

Fig. \ref{Omprofile} shows the differences of the internal $\Omega$--profiles
during the evolution of a 20 M$_{\odot}$ star with and without magnetic field created by the Tayler--Spruit dynamo.  Without magnetic field, the star has a significant differential rotation, while  $\Omega$ is almost constant when a  magnetic field created by the  dynamo is present. It is not perfectly constant,
otherwise there would be no dynamo. In fact, the rotation rapidly adjusts itself
to the minimum differential rotation necessary to sustain the dynamo.

\section{RECENT GRIDS OF MODELS OF MASSIVE ROTATING STARS}
\label{models} 
 In the recent years, many grids of massive rotating stellar models have been computed. A non-exhaustive list is given in Table~\ref{t1}.
These models have been computed accounting for the effects of rotation in the frame of the theory exposed in the previous sections.
We have indicated those models which were computed with the the dynamo theory of \citet{Spruit1999, Spruit2002}.
Although the physics of rotation (and magnetic field when accounted for) is the same in all those models, the numerical
implementations differ.
Let us just mention here a few relevant differences.
Some codes treat the transport of the angular momentum through a diffusion equation, not accounting for the advective nature of this equation. As explained above this
 may produce error not only on the amplitude of the effect but also on its sign! However, this statement should be tempered by the fact
 that for most of the MS phase, the region where the transport of angular momentum is transported efficiently, {\it i.e.} in the Gratton--\"{O}pik cell, angular momentum is transported outwards as would do a diffusion mechanism. 2) Rotating models contain uncertain parameters, at least
 two. One is in the expression of $D_{\rm shear}$ and account for the uncertainties pertaining our knowledge of the critical Richardson number (see Section IV.D).
The other is  in the expression of $D_{\rm h}$, whose amplitude
remains difficult to obtain from first principles. This coefficient 
has to be much greater than the one of $D_{\rm shear}$ in order to allow ``shellular rotation'' to set in. 
Some models listed in Table \ref{t1} contain additional parameters, as a $f_{\mu}$, a factor multiplying the mean molecular weight gradient. A small value of this parameter decreases the effects of the $\mu-$barrier and facilitates mixing.  To constrain the values of these uncertain parameters, 
some guidelines deduced from laboratory experiments or from multi-D dimensional simulations are used. Some authors also calibrate these uncertain parameters
in order to achieve mean observed surface enrichments during the MS phase of B-type stars with models having an average rotational velocity.

\begin{table}
\caption{Grids of rotating massive star models (starting from 2000).}
\begin{center}\scriptsize
\begin{tabular}{lcccl}
 \hline
 Masses                               &  $Z$   & Initial           & Magnetic    & Ref. \\ 
           M$_\odot$              &            & rotation       & field              \\
                                             &            &   km/s          & (interior)       \\
\hline                                             
9- 200                                   & 0          & 0-800          & no            & \citet{Ekstrom2008a} \\
3-60                                      & 0           & 39-1423     & no            &  \citet{Ekstrom2008b} \\
 \hline
 9-85                                      &0.00000001 & 800    & no             & \citet{Hirschi2007}\\
\hline  
2-60                                      & 0.00001& 0-400     &   no             & \citet{paperVIII2002} \\
20-60                                   & 0.00001& 230-605 &  yes             & \citet{Yoon2005} \\
12-60                                   & 0.00001& 0 - 936    & yes              & \citet{Yoon2006} \\
3-60                                     & 0.0001 & 39-1017 &  no                & \citet{Ekstrom2008b} \\
 \hline
 20-200                                & 0.0005 & 0-800     & no                 & \citet{Decressin2007a}\\
 \hline
 20-60                                     & 0.001   & 230-605 & yes               & \citet{Yoon2005}\\
 12-60                                    & 0.001   & 0-747      & yes               & \citet{Yoon2006}\\
  \hline
 3-60                                     & 0.002 & 32-879     &  no                & \citet{Ekstrom2008b}\\ 
 12-60                                   & 0.002 & 0-653       &  yes               & \citet{Yoon2006}\\
   \hline
   5-60                                         & 0.0021 & 0-600      & yes                & \citet{Brott2011} \\    
  \hline
  9-60                                     & 0.004 & 0-300       & no                  & \citet{paperVII2001}\\
  30-120                                & 0.004 & 300           & no                  & \citet{paperXI2005}\\
  12-60                                   & 0.004 & 0-507        & yes                &  \citet{Yoon2006}\\
   \hline
   5-60                                         & 0.0047 & 0-600      & yes                & \citet{Brott2011} \\   
\hline
30-120                                     & 0.008 & 300            & no                  & \citet{paperXI2005}\\
   \hline
   5-60                                         & 0.0088 & 0-600      & yes                & \citet{Brott2011} \\   
   \hline
8-25                                         & 0.020  & 0-474          & no            & \citet{Heger00} \\
8-25                                        & 0.020  & 200             & no              & \citet{Heger2000} \\
9-120                                      & 0.020  & 0-300          & no              & \citet{paperV2000} \\
9-120                                      & 0.020                        & 0-300-500  & no             & \citet{paperX2003} \\
12-60                                     & 0.020  & 0-300           & no             & \citet{Hirschi2004a} \\
12-35                                      & 0.020  & 200               & yes \& no & \citet{Heger2005} \\
16-40                                    & 0.020  &210-556        & yes           & \citet{Yoon2006} \\
3-60                                        & 0.020  & 28-732         & no            & \citet{Ekstrom2008b} \\
   \hline
20,25,40,60,85,120                 & 0.040  & 0-300           & no            &   \citet{paperXI2005}\\
 \hline
\end{tabular}
\end{center}
\label{t1}
\end{table}

A few authors have built 2D models for rotating stars \citep{Deupree1990, Roxburgh2006, Espinosa2007}.  For instance, ZAMS models with arbitrary rotation laws have been
computed by \citet{Deupree01}. The variation with the latitude of the effective temperature for rotationally deformed stars
from a 2D stellar structure is given by \citet{Lovekin2006}. 
Non radial oscillations are computed for 2D rotating
massive models by \citet{Lovekin2008}. Effects of rotation on the pulsation frequencies has been investigated recently by
\citet{Lovekin2009, Deupree2010}.

\section{THE ELEMENT ABUNDANCES IN MASSIVE STARS}
\label{chimie}

Both mass loss and rotational mixing affect the evolution of the chemical abundances at the stellar surface of massive stars. At solar abundances, mass loss dominates in stars initially more massive than about 30 M$_{\odot}$, while rotation dominates below this mass limit. Let us first examine the effect of mass loss in the most massive stars.
\subsection{Model effects of mass loss on abundances}
\begin{figure}[!t]
\centering
\includegraphics[height=8cm,width=11cm]{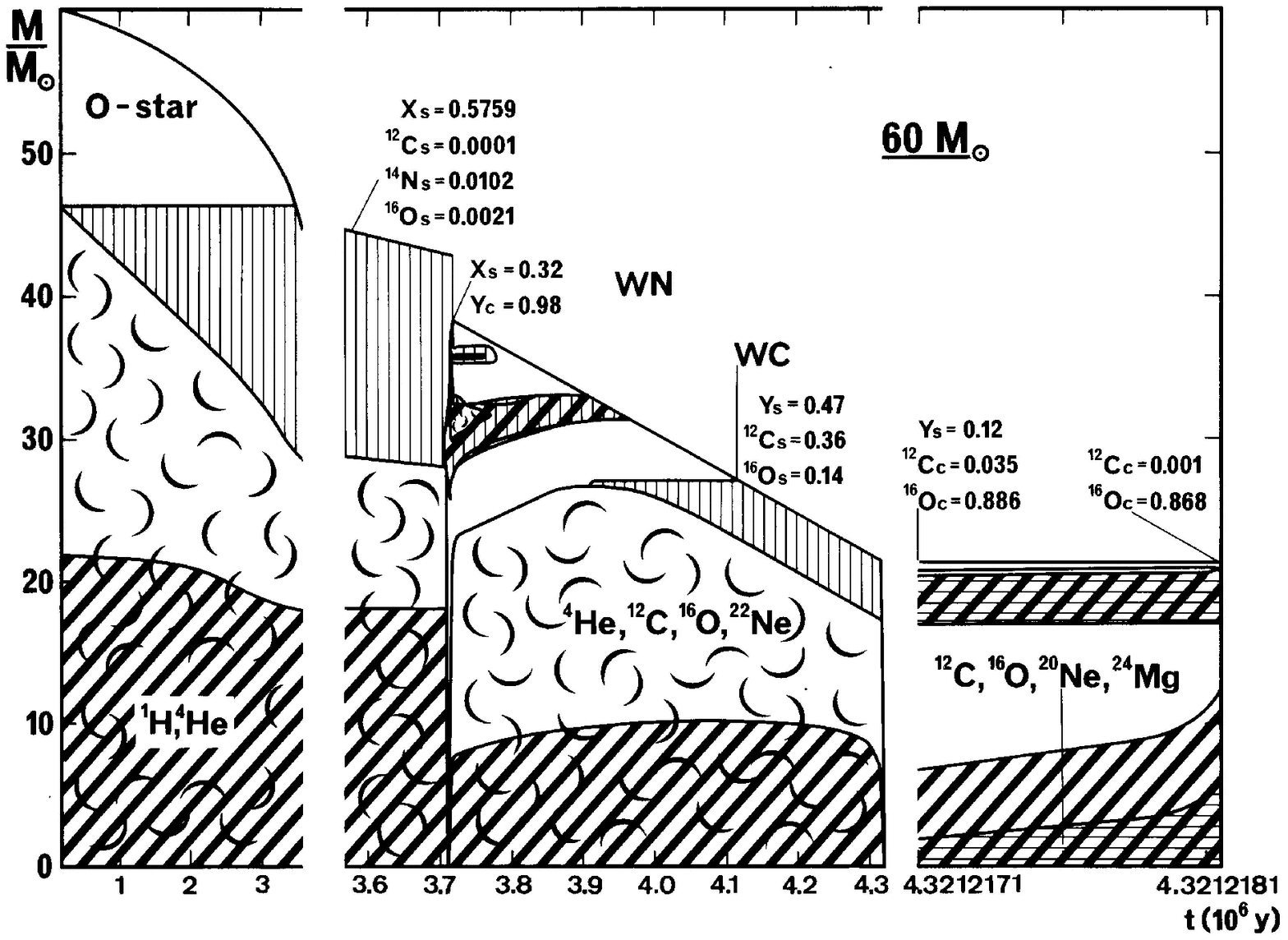}
\caption{Evolution with mass loss (only) of the total mass and internal structure as a function of time for a star with an initial mass of 60 M$_{\odot}$ \citep{Maeder1987a}. The cloudy regions indicate convection, the heavy diagonal hatching the regions of high nuclear production, the light vertical hatching regions of variable composition. }
\label{mdot60}
\end{figure}
Fig. \ref{mdot60} shows the internal evolution of a star with an initial 60 M$_{\odot}$. The removal of the outer layers
 reveals the internal composition.
There are  
 5 typical domains of composition, which result from a progression in the exposition of nuclear products.

\noindent
-- 1. \emph{The initial abundances}. Mass loss does in general not  change the   surface
composition  in OB stars during the MS phase,  except for stars above 60 M$_{\odot}$.

\noindent
-- 2. \emph{Intermediate abundances}. This stage is characterized by  partial CNO processing with possible dilution effects. There are N enrichments,  enhancements
of the $^{13}$C/$^{12}$C ratio, C depletions and  modest O depletions. This is typical of blue and red supergiants.
 
\noindent
-- 3. \emph{CNO equilibrium with H present}. CNO equilibrium is reached  before H exhaustion. The 
C/N and O/N ratios are reduced by two orders of a magnitude with respect to cosmic abundances.
This is the case of   WR stars (bare cores) of types WNL.

\noindent
-- 4. \emph{CNO equilibrium with H absent}. The  He mass fraction is  98\%,  the CNO ratios are  the same as before. 
This stage corresponds to WR stars of type WNE.
The abundances are model independent, being determined mainly  by nuclear cross--sections.

-- 5. \emph{Partial He burning}. The products  of He burning, i.e. 
$^{12}$C, $^{16}$O, $^{22}$Ne are  visible.  The changes  are abrupt (rotation make them smoother). The abundances depend strongly 
on the models (mass loss, mixing, etc.). This stage corresponds to  WC stars and to WO stars
for  O/C ratios  $>1$.

 Of course, not all stars go through this whole sequence: the smaller the initial mass and/or metallicity, the shorter the path. Rotation may accelerate and make smoother the evolution of surface 
 composition, due to the contribution of internal mixing.

\subsection{The effects of rotation on the surface abundances}

\begin{figure}[!t]
\centering
\includegraphics[height=5cm,width=5cm]{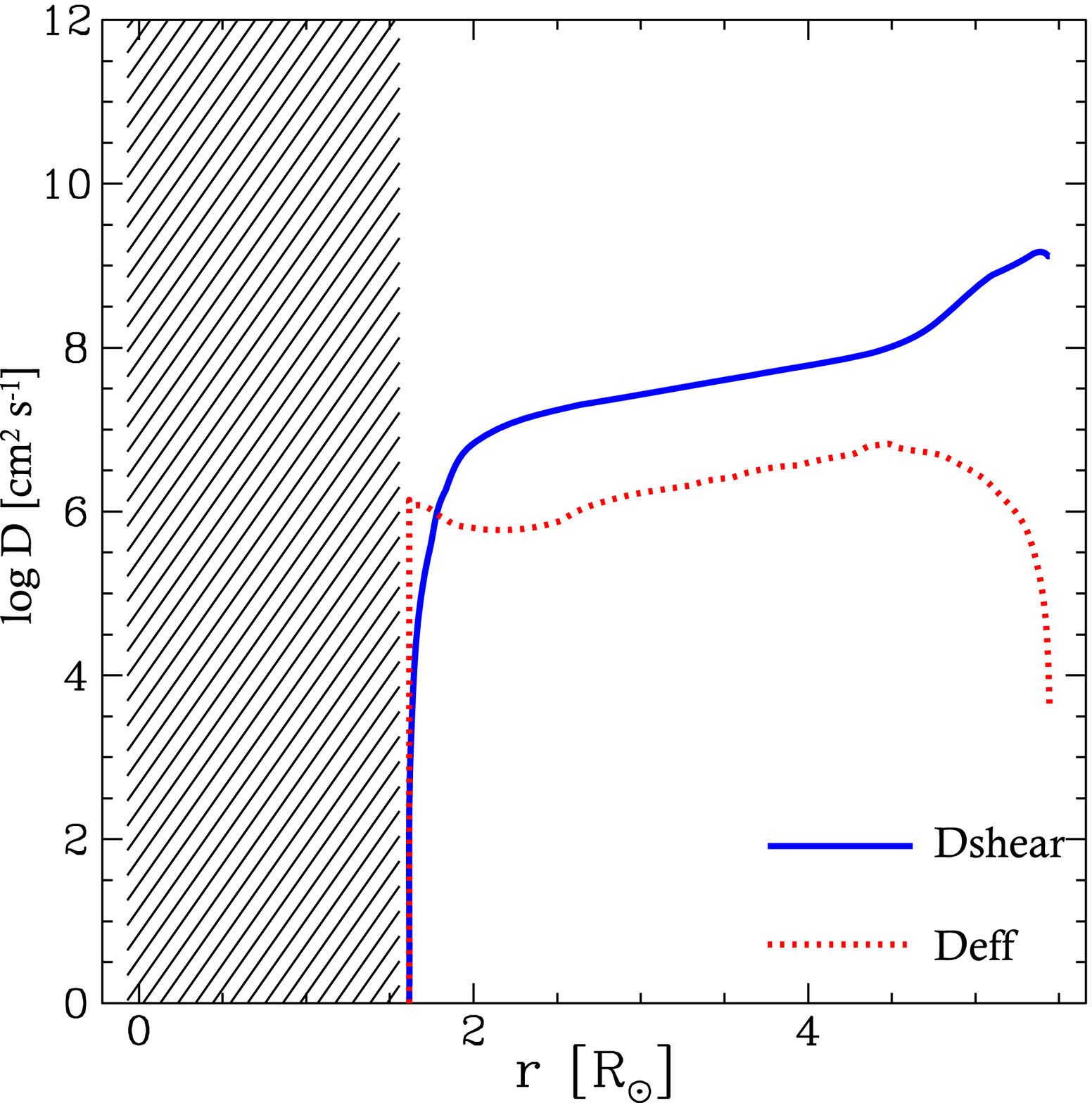}
\includegraphics[height=5cm,width=5cm]{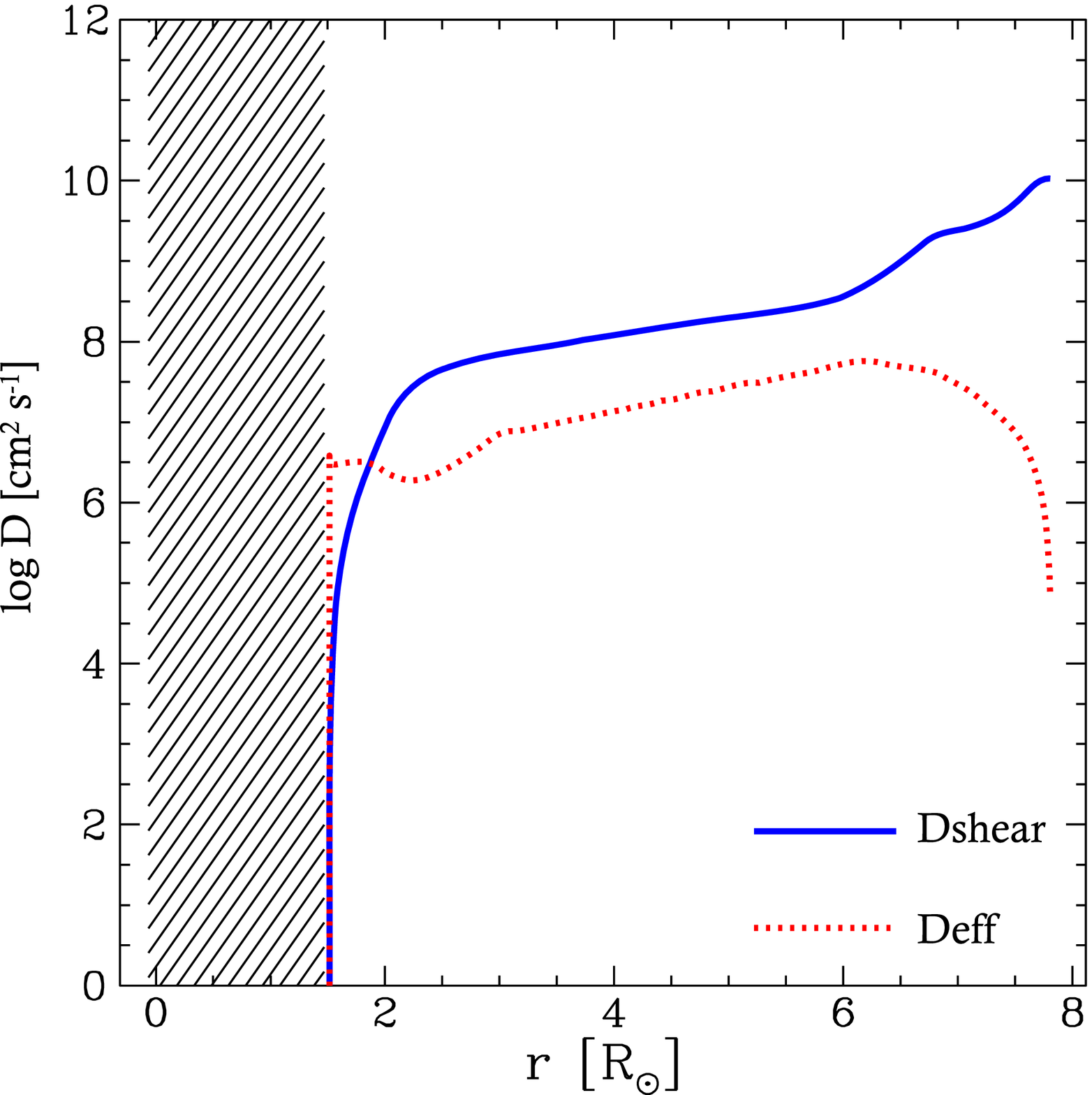}
\includegraphics[height=5cm,width=5cm]{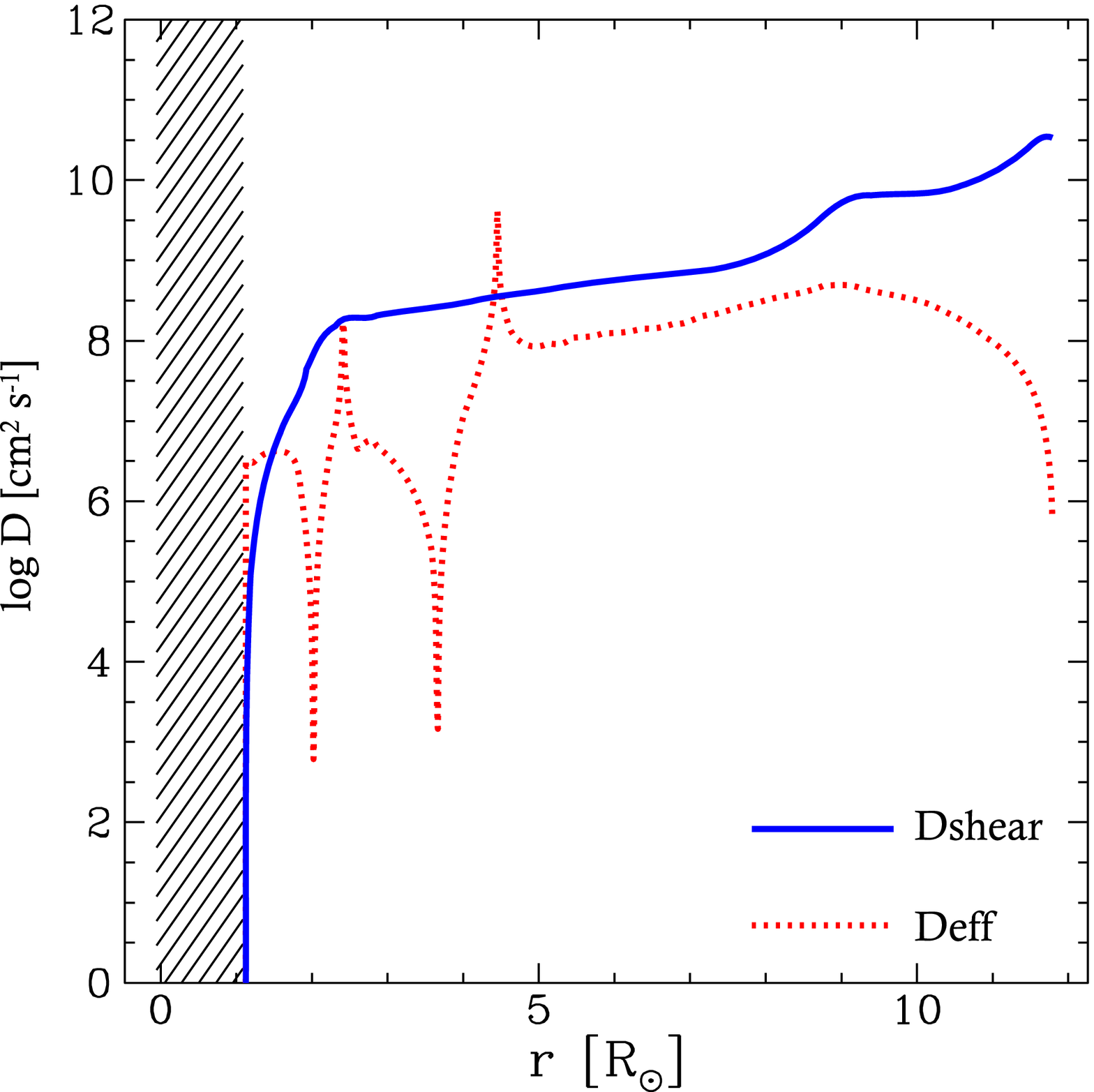}
\caption{Variation of the diffusion coefficients as a function of the radius inside a 20 M$_\odot$ model with an initial metallicity Z=0.014 and
an initial velocity on the ZAMS equal to 40\% the critical velocity. From left to right, is shown the situation corresponding to central mass fraction of hydrogen equal to
0.60, 0.30 and 0.01 respectively. Courtesy of S. Ekstrom.}
\label{coefficients}
\end{figure}

In  the mass range below about 30 M$_{\odot}$, mass loss has little effect during the MS phase. The changes of
abundances, if any, are  due to  mixing.
Fig.~\ref{coefficients} shows the run of the diffusion by shear turbulence, $D_{\rm shear}$, and by the net effect of both meridional currents and horizontal shear turbulence, $D_{\rm eff}$, at three different stages during the MS phase of a rotating 20 M$_\odot$  model.
The main mixing effect  is the diffusion by shear turbulence,  
which  results from  the internal $\Omega$ gradients built during evolution.
To a smaller extent, meridional circulation  makes some transport, however mainly of angular momentum.

Mixing brings to  the  surface the products of  CNO burning: mainly 
$^{14}$N  and $^{13}$C enrichments, $^{12}$C is depleted with limited $^4$He enrichment and $^{16}$O depletion.
Fig.~\ref{NHmass} shows the predicted  variations  of  $\log({N/H})$ during MS evolution as a function of the initial masses
\cite{paperV2000}. $(N/H)$ is here the abundance ratio of N and H in numbers. 
 Without rotational mixing, there would be no enrichment until the red supergiant stage, but  rotation produces an  increase (depending on velocity $v$)  of   N/H 
during  the MS phase. The N excesses also depend on the ages $t$. The increase is modest during the first third of the MS phase, because the elements need some time to reach the surface, then it is more rapid.  The  N enrichments are larger 
for larger  masses $M$. 

\begin{figure}[!h]
\centering
\includegraphics[height=7cm,width=10cm]{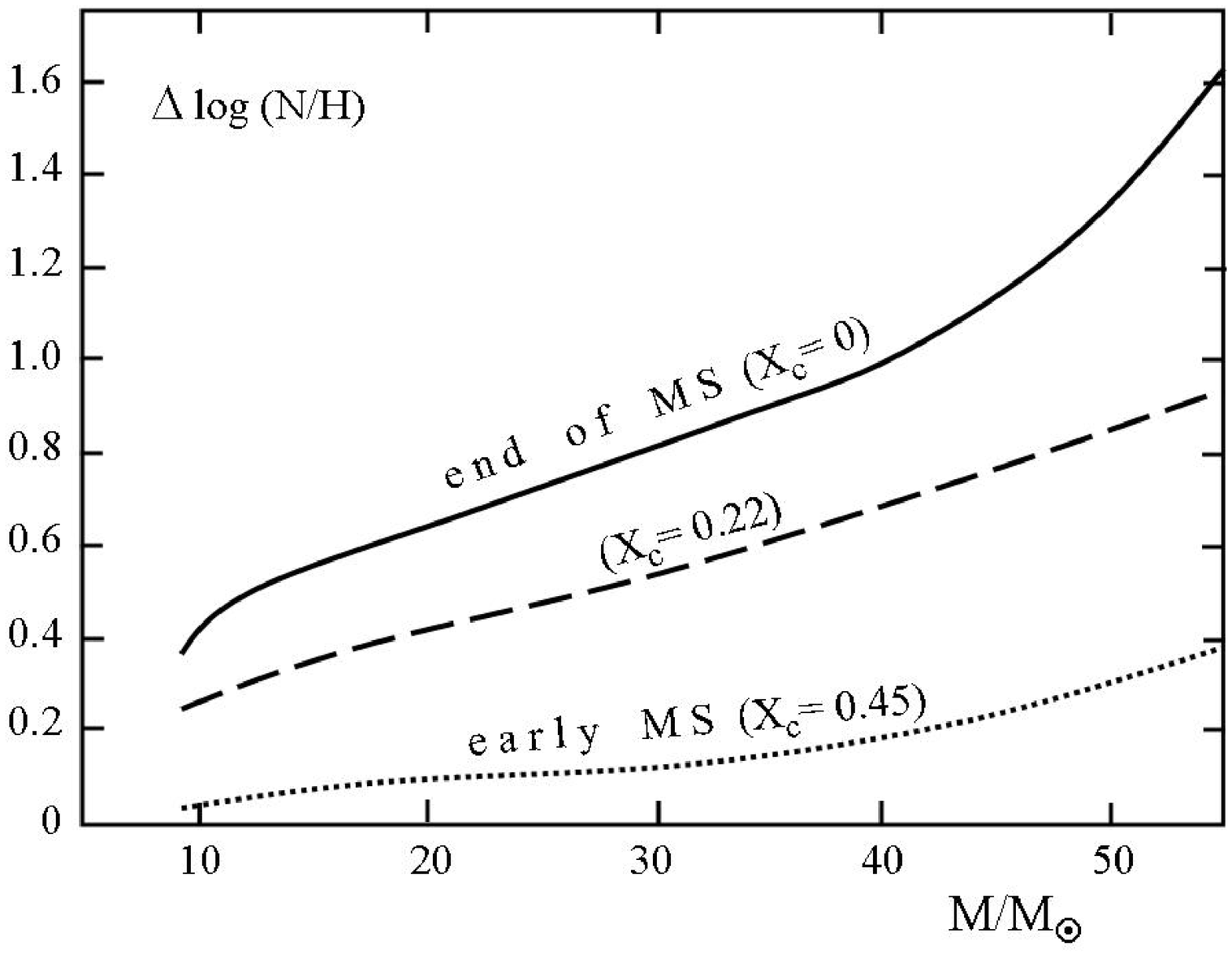}
\caption{The differences in log(N/H) as a function of the initial masses at  3 stages during the MS phase for models at $Z=0.02$ with the average rotation velocities (i.e.  
217, 209, 197, 183, 172, 168 km s$^{-1}$ for respectively 12, 15, 20, 25 40 and 60
M$_{\odot}$). The  3 stages are indicated by the value of the central H--content $X_{\mathrm{c}}$ \citep{Maeder2008c}.}
\label{NHmass}
\end{figure}

\begin{figure}[!t]
\centering
\includegraphics[height=7cm,width=10cm]{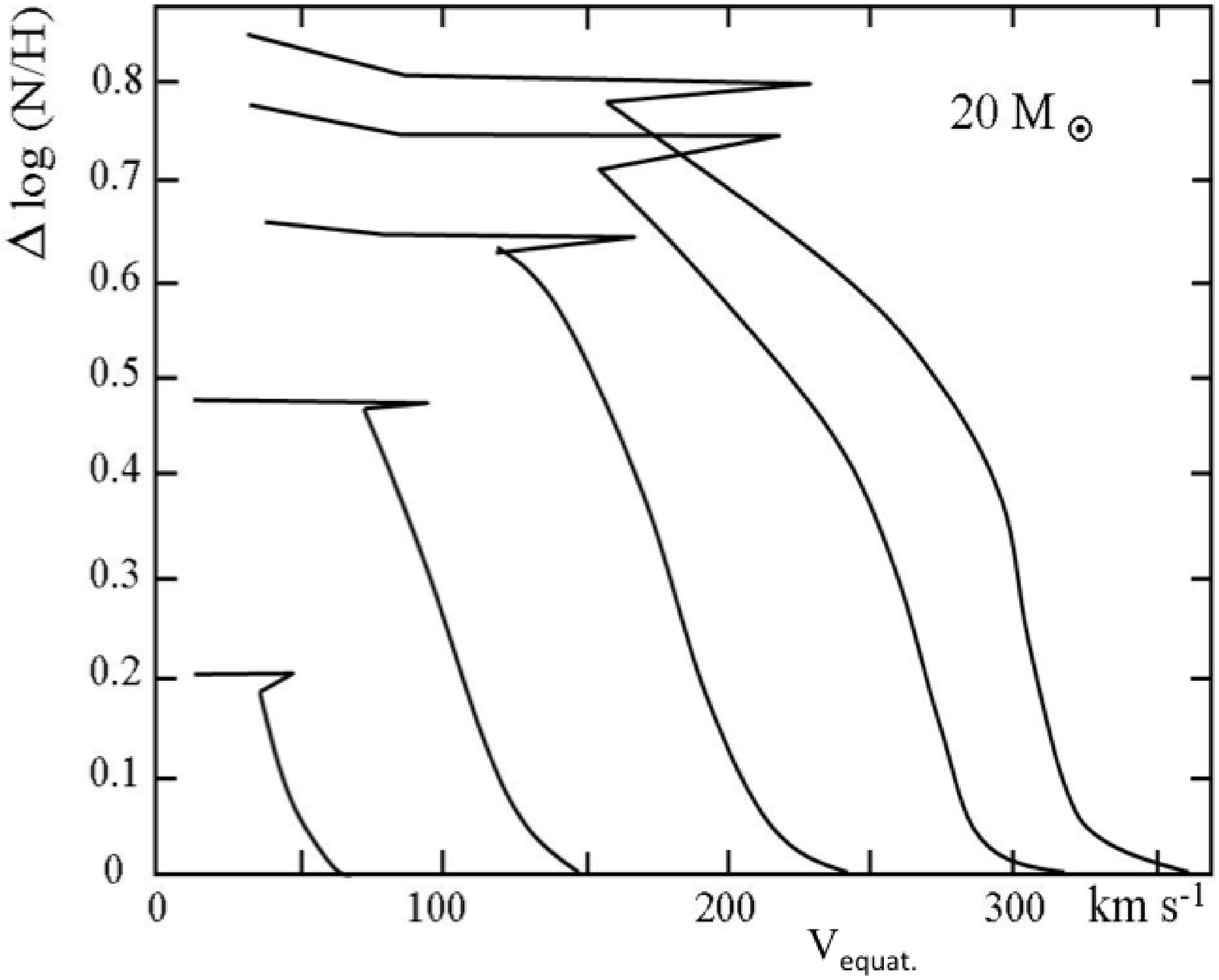}
\caption{The evolution of the differences in log(N/H) during the 
MS phase as a function of the actual rotation velocities ($\sin i$ being equal to 1 here) for models of 20 M$_{\odot}$ with $Z=0.02$ and  different initial velocities \citep{Maeder2008c}.}
\label{NHvsini}
\end{figure}

Fig.~\ref{NHvsini} shows, for  given $M$ at $Z=0.02$, the evolution of the N excesses  with age for different initial velocities. It shows that  there is no single relation between $v \sin i$ and the N/H excesses for a mixture of ages. The same is true for a mixture of different masses. 
Binarity may also affect the N and He enrichments due to tidal
mixing and mass transfer. 
A binary star with  low rotation may have a high N/H due to tidal mixing or due to the transfer of the enriched envelope of a  red giant. At the opposite, a binary star
may also have a high $v \sin i$ and no N/H excess, in  the case of the accretion of an unevolved envelope bringing a lot of angular momentum. On the whole,  the N excess is  a multivariate function 
\begin{eqnarray}
\Delta \log(N/H) =  f(M, \, t,\, v \sin i,\, multiplicity, \,Z) \;.
\label{f1}
\end{eqnarray} 
\noindent
Models with lower initial metallicities $Z$ have higher 
N enrichments for given $M$ and $v$ \citep{Maeder2001,Meynet2006}. The reason is the higher $\Omega$--gradients, resulting from the absence of the  Gratton--\"{O}pik circulation cell (Sect. \ref{mcirc}). The excesses become  very strong at metallicities $Z$ as low as $10^{-8}$. 

\subsection{The observed N/H excesses}

The amplitudes of the N enrichments  at the end of the MS phase in massive stars form a reference point
telling us the importance of  mixing. The data at different $Z$ (mean and largest values) are summarized in Table 
\ref{tblabindo} from a number of  sources (see references below the Table).   
 In the lowest mass range
considered (6.6--8.2 M$_{\odot}$), small excesses of He/H are still present  \cite{Lyubimkov2004}, they are larger  
in the Small Magellanic Cloud (SMC, $Z\approx 0.004$). \noindent
We see the following facts:

\begin{table}[!h]  
 \caption{Values of the average and largest 
 [N/H] excesses observed  for different types of stars 
 in the Galaxy, LMC and SMC.  The number of stars used to obtain the quoted values are indicated as well the sources for the observations.} \label{tblabindo}
\begin{center}\scriptsize
\begin{tabular}{lcccccccccccc}
Types of stars                                            &\multispan4  [N/H] in Galaxy $\quad$                    &\multispan4 $\quad$ [N/H] in LMC $\quad$ &\multispan4 $\quad$ [N/H] in SMC                         \\
                                                                     & mean           & max.          & N$_*$ & Ref.            & mean          & max.            & N$_*$ & Ref.    & mean             & max.              & N$_*$  & Ref.    \\
                                                                     &                      &                    &                &                    &                      &                      &                &            &                        &                        &                 &             \\
\hline 
O stars                                                        &   0.3-0.5     &   0.4-0.6     &        3      &   [1]              &                      &                      &                &            &  1.0-1.2        &   1.3-1.5         &     14       &    [8]    \\
B dwarfs $M<20$ M$_{\odot}$              &   0.2-0.4     &   0.5-0.7     &       22     &   [2,3,4,5]    &   0.3-0.5      &   0.9-1.1      &         35   &   [5]    &  0.5-0.8        &   1.3-1.5         &     34       &    [5]    \\
B giants, super.  $M<20$ M$_{\odot}$ &   0.4-0.6     &   0.7-0.9     &      10     &   [5]               &   0.6-0.8      &   1.1-1.3      &        21    &   [5]    &                       &                         &                &             \\
B giants, super.  $M>20$ M$_{\odot}$ &   0.4-0.6     &   0.8-1.0     &      38     &   [6,7]            &                     &                      &                 &           &                       &                         &                &             \\
\hline
                                                                     &                      &                    &                &                    &                      &                      &                &            &                        &                        &                 &             \\
\multispan9 [1] \citet{Villa05, Villa02} \hfill\hfill&  &  &  &  \\
\multispan9 [2] \citet{Gies92} \hfill\hfill&  &  &  &  \\
\multispan9 [3] \citet{Kilian92} \hfill\hfill&  &  &  &  \\
\multispan9 [4] \citet{Morel08} \hfill\hfill&  &  &  &  \\
\multispan9 [5] \citet{Hunter2009} \hfill\hfill&  &  &  &  \\
\multispan9 [6,7] \citet{Crowther06, Searle08} \hfill\hfill&  &  &  &  \\
\multispan9 [8] \citet{Heap2006} \hfill\hfill&  &  &  &  \\
\end{tabular}
\end{center}
\vspace*{-5mm}
\end{table}

\begin{itemize}
\item On the average, the N enrichments are larger for larger masses.
\item The N enrichments are larger at lower $Z$.
\item Away from the ZAMS, but still in the Main Sequence, 
 the  He and N enrichments are larger  and they are even larger in the supergiant stages. These various features are quite consistent with
 the predicted properties of rotational mixing as can be seen in Fig. \ref{ncp}.
\item Correlations between N or He excesses and the observed $v \sin i$ have been found  in the upper part of the MS band  \citep{Lyubimkov2004}, in agreement with model predictions.
\end{itemize}
 
 \begin{figure*}
 \includegraphics[height=.600\textheight]{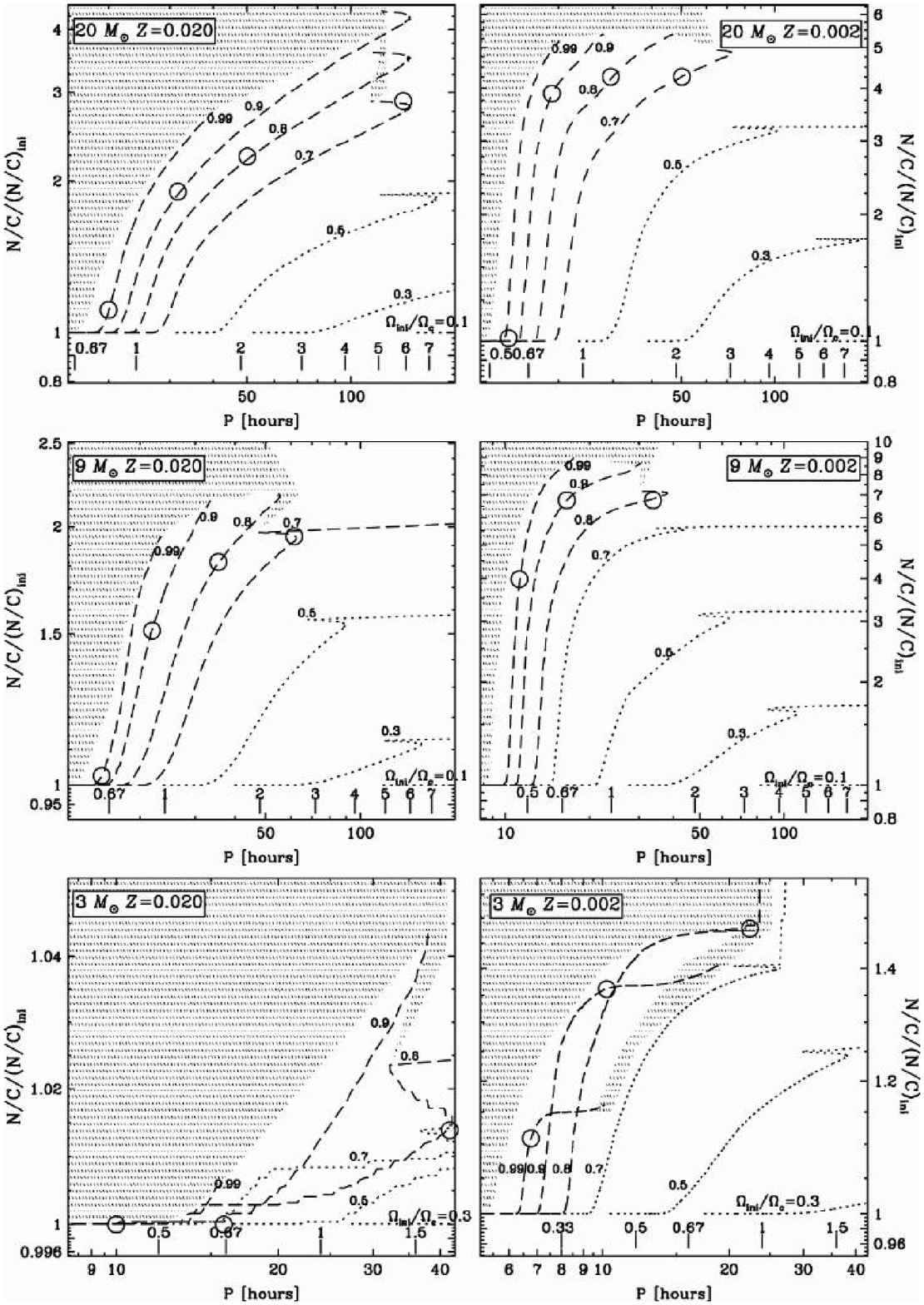}
      \caption{Evolutionary tracks in the plane surface N/C ratio, normalized to its initial value, versus the rotational period in hours for different initial mass stars, various initial velocities and for the metallicities $Z$=0.02 and 0.002. Positions of some periods in days are indicated at the bottom of the figure. The dotted tracks never reach the critical limit during the MS phase. The short dashed tracks reach the critical limit during the MS phase. The dividing line between the shaded and non-shaded areas corresponds to the entrance into the stage when the star is at the critical limit during the MS evolution. Big circles along some tracks indicate the stage when $\upsilon/\upsilon_\mathrm{crit}$ becomes superior to 0.7. If Be stars are rotating at velocities superior to 70\% of the critical velocity, present models would predict that they would lie in the region comprised between the big circles and the dividing line. Beware the different vertical scales used when comparing similar masses at different metallicities. Figure taken from \citet{Ekstrom2008b}.}
       \label{ncp}
  \end{figure*}


\begin{figure}[!t]
\centering
\includegraphics[height=6.2cm,width=10cm]{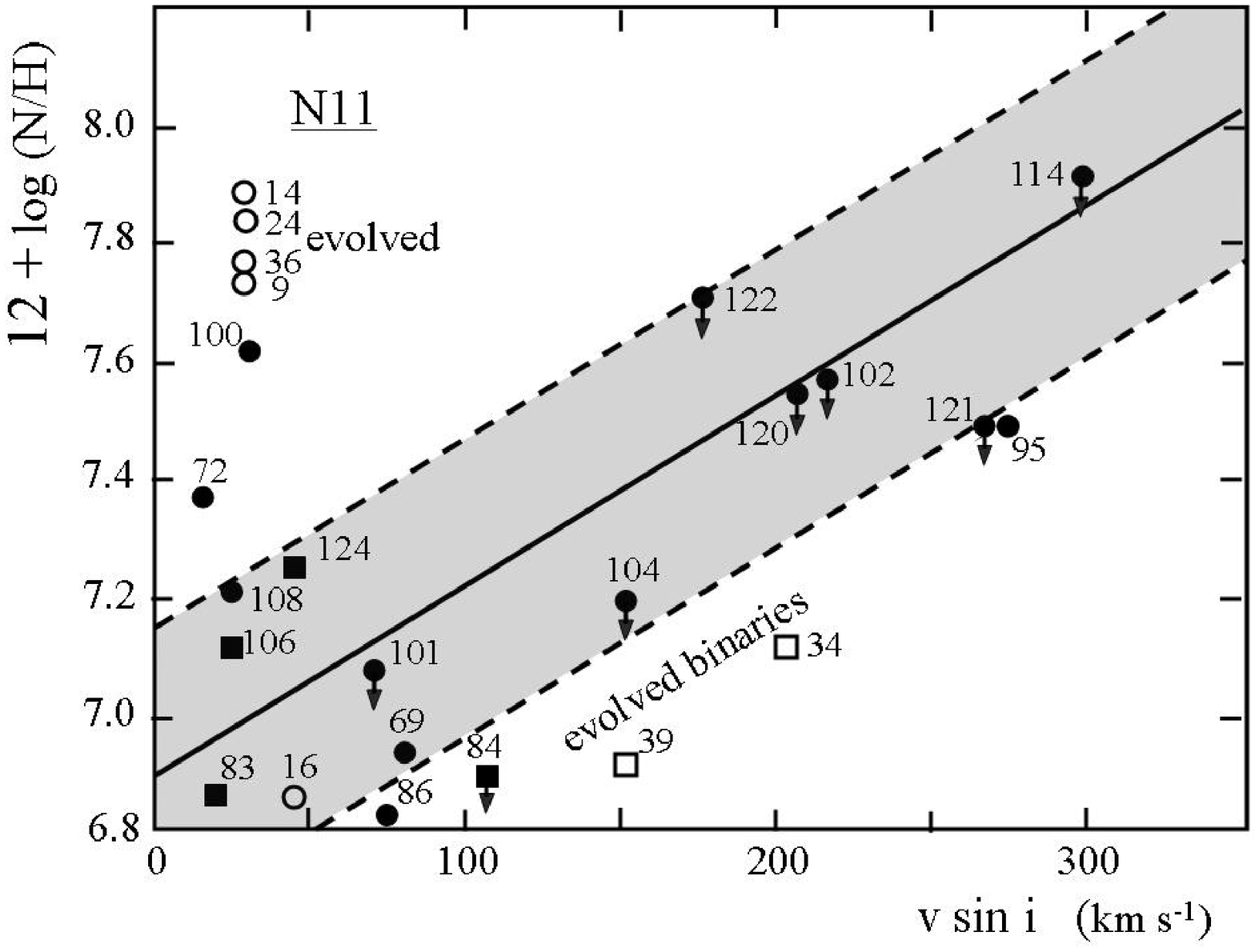}
\caption{The N abundance (in a scale where $\log H=12.0$) 
as a function of $v \sin i$ for the MS stars (black dots) in N11 with masses between
14 and 20 M$_{\odot}$ according to \citet{Hunter2008a}. The downward arrows indicate abundance upper limits.
The binaries are shown by a square. The evolved stars
in a band of 0.1 dex in $\log T_{\mathrm{eff}}$ beyond the end of the MS are shown with open symbols.
The grey band indicates uncertainties of $\pm 0.25$ dex \citep{Maeder2008c}.}
\label{NHN11}
\end{figure}

\noindent

To find a correlation for a multivariate function like N/H with some parameter like $v \sin i$, it is necessary to limit as much as possible the range of the other involved parameters. Otherwise, the conclusions may
be erroneous. From Fig.~\ref{NHN11}, we note that data samples limited in masses and ages support a N enrichment depending on rotational velocities. Stars beyond the end of the MS phase
do not obey to  such a relation, because their velocities 
converge toward low values \cite{paperV2000}. 
A fraction, which we estimate to be  less than $\sim20$ \%
of the stars, may escape to the relation as a result of binary evolution or due to some
other mechanisms as for instance magnetic braking \citep{Meynet2011}. This latter mechanism might explain
some magnetic stars showing large N--excesses
and low $v \sin i$ \citep{Henrichs2003a,Henrichs2003b}. 

To obtain more tight constraints on the models, very careful and accurate abundance analysis are required. Very interestingly
\citet{Przy2010} showed that observed ratios of N/O and N/C follows a well defined relation in the N/O versus N/C plane, which does not depend on the models but just on the CNO nuclear reactions properties. In the future, this characteristic can be used to test the surface abundances obtained from spectral analysis before these abundances are used
to constrain the stellar models.


According to \citet{Hunter2007} main-sequence binary objects have close to baseline nitrogen surface abundances. These systems thus do not present apparent signs of 
extra-mixing. In contrast several evolved binary objects have high nitrogen enhancements. These
abundances are similar to those observed in apparently single stars. Thus it appears difficult to
discriminate among the possible causes of the enrichments in binaries, {\it i.e.} between extra-mixing operating in single stars and mass-transfer events in close binary systems.
The same result has been obtained by \citet{Trundle2007}.

\subsection{The observed He excesses}

Surface enrichments in helium have been observed with the following
main trends:
\begin{itemize}
\item He in O-type stars in the SMC: In the SMC, for 31 O-type stars, \citet{Mokiem2006}
find values of $y=n_{\rm He}/(n_{\rm H}+n_{\rm He})$ between 0.09 and 0.24, where $n_{\rm He}$ is the density number of helium and $n_{\rm H}$ of hydrogen . Note that $n_{\rm He}/(n_{\rm H}+n_{\rm He})=Y/(Y+4X)$, where $Y$ and $X$ are respectively the mass fraction of helium and of hydrogen. 
Setting $Y\sim 1-X$, one obtains values for $Y=4y/(1+3y)$ between 0.28 and 0.56
(here 0.28 would correspond to y=0.09 {\it i.e.} to the initial helium mass fraction).
These authors conclude that while rotation can qualitatively account for such enrichments, the observed enrichments are in many cases much stronger than those predicted by the models.
\item He in O-type stars in the Large Magellanic Cloud (LMC): In the LMC, for 28 O-type stars, $y$ values between about 0.09 and 0.28 
are obtained by \citet{Mokiem2007}, {\it i.e.} helium mass fractions between
0.28 and 0.61.
\item He in OB-type stars in the Milky Way (MW): 
\citet{Huang2006b} determine He abundances for OB galactic stars.
In their high mass range ($8.5$ M$_\odot < M < 16$ M$_\odot$), the He enrichment progresses through the main sequence and is greater among the faster rotators\footnote{These authors also found many
helium peculiar stars (He-weak and He-strong). These stars were not used to study the process of He-enrichment.}. On average He abundance increases of 23\% $\pm$13\% between ZAMS and TAMS.
These authors also obtain that He enrichments are higher for higher $v \sin i$ values.
\citet{Lyubimkov2004} find an increase during the MS phase in He abundance of 26\% for stars in the mass range 4-11 M$_\odot$ and 67\% for more massive stars in the range
12-19 M$_\odot$.
\end{itemize}

From a theoretical point of view, one expects that the N-enrichment occurs very rapidly at the surface well before any surface helium enrichment. This is of course due to the fact that very rapidly the nitrogen abundance increases at the centre creating thus a strong chemical gradient between the core and the envelope.
Since diffusive velocity is greater when the gradient of abundance is greater, the presence of such a strong gradient favors a rapid mixing. 
The gradient of helium in stellar interiors is built up on much longer timescales than the gradient in nitrogen. As a consequence, the diffusion of this element is expected to be much less rapid than that of nitrogen.
Thus any strong He-enhancement should be accompanied by a very strong nitrogen enhancement. At the moment, the observations
of both He- and N-enrichments in the same stars are too scarce to check this trend.

\section{EVOLUTION OF ROTATION AT THE SURFACE AND IN THE INTERIOR}
\label{sec:surfvelo}

\subsection{Initial conditions for rotating models}

Ideally one would like to know the shape of the distribution of rotational velocities on the ZAMS for the whole mass and metallicity ranges spanned by the
stars during the whole cosmic history. This is of course not an easy task and many difficulties have to be overcomed. Let us just mention the main ones here:
\begin{itemize}
\item First, most observations concern non-ZAMS stars, whose surface velocity is not  representative of the initial stellar
velocity. In order to obtain ZAMS velocities some theoretical guidelines have to be used. For instance, using stellar models, it is possible to
deduce which are the velocities needed on the ZAMS in order to achieve a given mean velocity during the MS phase. 
\item What is measured through Doppler broadening is $\upsilon \sin i$ and thus a sufficiently great number of stars need to be measured in order to correct
in a statistical way for the effect of the inclination angle.
\item There are theoretical reasons for believing that beyond a velocity limit the line broadening due to rotation saturates, making
this method inappropriate for measuring velocities near the critical limit. This is because when the velocity approaches the critical one, the equator, which has the higher linear velocities,
darkens and thus contributes less to the line formation than the regions of the star near to the pole which have the smaller linear velocities.
\end{itemize}

It is also worthwhile to underline the fact that even if we knew very well the distributions of the  surface velocities, this would not help
in determining  the initial distribution of $\Omega$ inside the star.
Hopefully asteroseismology will allow us to obtain such information, but presently the indications are still scarce
 \cite[see however results presented in][]{Aerts2008}\footnote{See Table 1 in this reference. Two stars present a high contrast between the angular velocity of the core and that of the envelope,
between  a factor 4 and 5. One star does not show any contrast. All the three stars are slow rotators and are B2-B3 stars with class luminosity between V and III. }.  Also interferometry, which allows  us to measure the shape of fast rotating stars, can give some indications on how the interior rotate. For instance a ratio of the equatorial radius to the polar radius equal to 1.5 when
the star rotates at the critical limit implies that the interior is not much deformed, in agreement with the Roche approximation. 

The knowledge of  the interior distribution of $\Omega$ would be important for estimating the total angular momentum of the star on the ZAMS.
At present time, stellar models make the simplifying hypothesis that the evolution starts with a flat profile of $\Omega$.
The exact shape of the initial profile of $\Omega$ is not so important for the further evolution of $\Omega$ inside the star.
It has been found that the initial distribution of the angular momentum inside the star is rapidly forgotten and erased by a rapid convergence of the profile of $\Omega$ towards an equilibrium profile \citep{Denissenkov1999}. This results from two
counteractive effects: on one hand meridional currents build $\Omega$ gradient (which tends to increase $\Omega$ in the inner layers)  and on the other hand shear turbulence
tends to decrease the $\Omega$ gradient. 

The mean initial angular momentum of a model star on the ZAMS, $\mathcal{L}_{ini}$,  is estimated as the value 
needed for the star to rotate at the observed average rotational velocity during the MS phase.
The inertia momentum of a 10 M$_\odot$ star on the ZAMS is of the order of 10$^{56}$ g cm$^2$.  Considering a radius $R \sim 4 R_\odot$ and a surface equatorial velocity, $\upsilon_{eq}$,
of 200 km s$^{-1}$ (see Fig. \ref{distrv}), one obtains for $\Omega=\upsilon_{eq}/R=0.00007$ s$^{-1}$, and a total angular momentum content of the order of 10$^{52}$ g cm$^2$ s$^{-1}$ assuming the star rotates as a solid body on the ZAMS. 
These are the typical initial angular momentum which are considered in stellar models.

\subsection{Evolution of the surface velocities}

Once the initial conditions have been chosen, models predict the evolution  of the surface
velocity as a function of time. The evolution of $v$ in the HR diagram is shown in
Fig.~\ref{v_HR} ($\upsilon_{\rm ini}= 300$ km s$^{-1}$ on the
zero--age main sequence). Lines of constant $v$ are indicated.
On the MS, we notice 
the decrease of the surface velocity as a function of time.
This trend is confirmed by the observations of
 \citet{Huang2006b} who show that all OB stars of their sample experience 
 a spin-down during the MS phase. A few relatively fast rotators are found 
 near the end of the MS phase. According to these authors, these stars may be 
 spun up by a short contraction phase or by mass transfer in a close binary.
In Fig.~\ref{v_HR}, we see also  that the decrease of the surface velocity is
much faster for the most massive stars than for stars with
M $\leq$ 15 M$_{\odot}$, as a consequence of stronger mass losses in the
higher mass range. This difference remains also present in the domain of B--supergiants.
Interestingly, according to 
\citet{Hunter2008a} there is some evidences that the most massive objects 
rotate slower than their less massive counterparts.
\citet{Dufton2006} confirm that the mean rotational 
velocity of stars which have strong winds is lower than that of the lower mass stars.

A few other points can be noted:
the average 
$\overline v$ is lower for O--type stars than for the early
B--type stars \citep{Slettebak1970}. Again this may be the consequence of the higher 
losses of mass and angular momentum in the most massive stars.
Also, we remark that the increase of $\overline v$ from O--stars to
B--stars is larger for the stars of luminosity class IV than for class V 
\citep{Fukuda1982}. This is consistent with the models,
which show (cf. Fig.~\ref{v_HR}) that
 the differences  of $\overline v$ beween O-- and 
B--type stars are much larger at the end
of the MS phase.  Another fact in the observed data is the
 strong decrease
of $\overline v$ for the massive supergiants of OB--types. Also
during the crossing of the HR diagram, the rotational velocities decrease
fast, to become very small, i.e. of the order of a few km s$^{-1}$, in the
red supergiant phase. 
This is predicted by all stellar models \citep[cf. also][]{Langer1998}
due to the growth of the stellar radii. 
This is also confirmed by the recent observations of
 \citet{Mokiem2006}  and \citet{Hunter2008a}. These last authors find  from a sample of 400 O- and early B-type stars in the Magellanic Clouds  that supergiants are the slowest rotators in the sample, typically having rotational velocities less than 80 km s$^{-1}$. 
 
 Note that the models shown in Fig.~\ref{v_HR} are models without magnetic fields.
When magnetic fields are accounted for as in the Tayler-Spruit dynamo, the evolution of the surface velocity
is different. For instance,  a 15 M$_\odot$ model with $\upsilon_{\rm ini}$=300 km s$^{-1}$ on the ZAMS and
computed with magnetic fields,  keeps a nearly constant surface
velocity around 300 km s$^{-1}$ all along the MS phase . Without
magnetic fields, the velocity decreases during the MS phase. The average velocity is around 240 km s$^{-1}$ \citep[see Fig. 3 in][]{magnII2004}.

\begin{figure*}[tb]
\includegraphics[height=15cm,angle=-90]{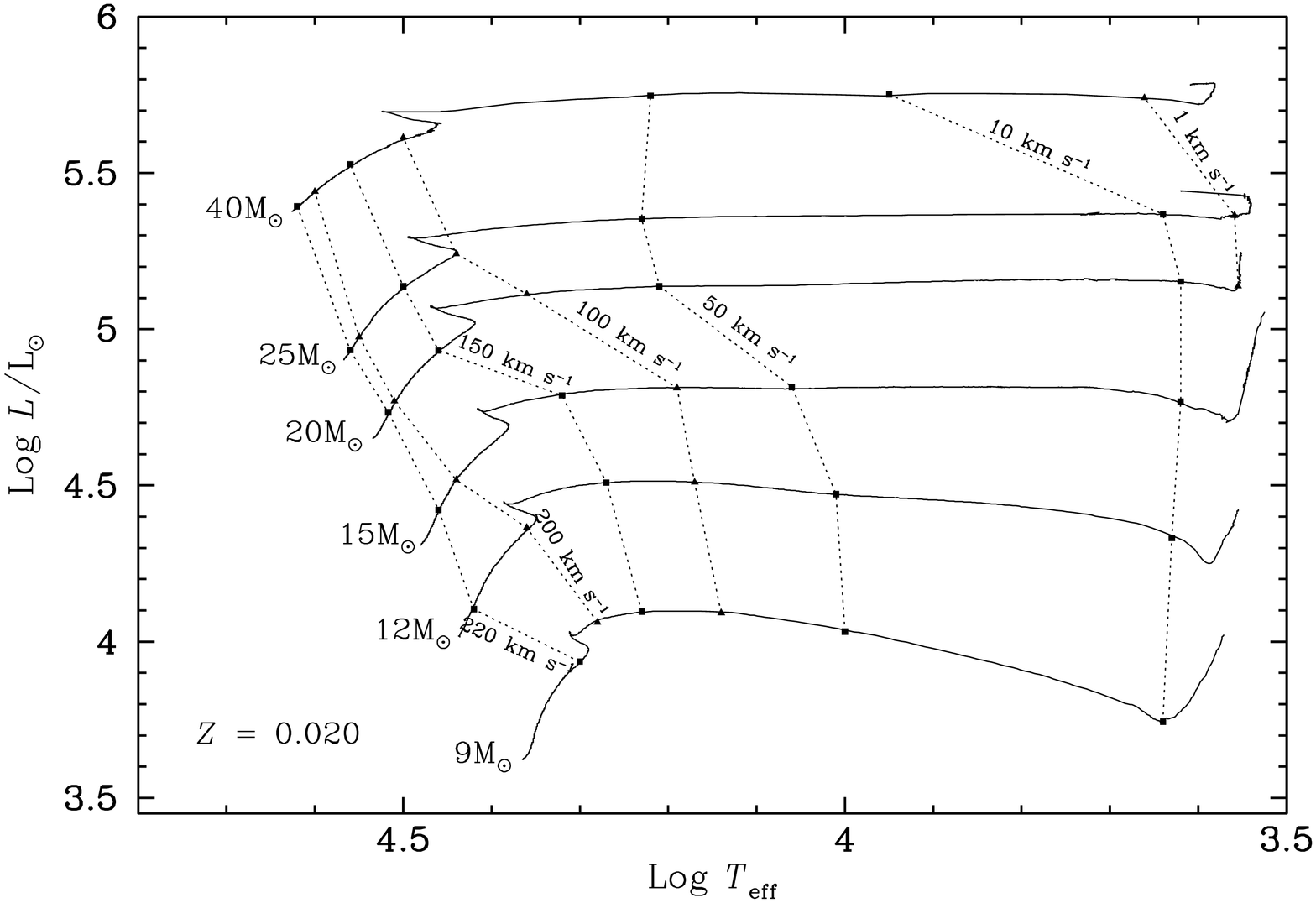}
  \caption{Evolution of the equatorial surface velocities along the evolutionary
tracks in the HR diagram starting from $v_{\rm ini}$ = 300 km s$^{-1}$.
For purpose of clarity, only the first part of the 40 M$_\odot$ track
is shown. Figure taken from \citet{paperV2000}.
}
  \label{v_HR}
\end{figure*}

Fig.~ \ref{v60} clearly illustrates 
the very different evolution of the rotational velocities
of a 60 M$_{\odot}$ at various metallicities.
At low $Z$ like in the  models at $Z = 10^{-5}$ ,
 the growth of $\frac{\Omega}{\Omega_{\rm c}}$
is possible because of the very small mass loss by stellar winds.
In view of these results, it is likely  that at very low $Z$  a large fraction
of the massive stars reach their break--up velocities (see right panel of Fig. \ref{fig2}).  
This question is of high
importance, because if the massive stars reach their break--up velocity,
most of their evolutionary and structural  properties will be affected.
For example, they could also
 lose a lot of mass and produce some Wolf--Rayet
(WR) stars. They would have a relatively small remaining mass
at the time of the supernova explosion, like their counterparts at
solar composition.
Interestingly, the trend illustrated above is supported by 
\citet{Hunter2008a} who obtain that SMC metallicity stars rotate on average faster than galactic ones (mainly field objects)
and by \citet{Martayan2007a}, who find that, for B and Be stars, the lower the metallicity, the higher the rotational velocities.

\begin{figure}[tb]
 \includegraphics[height=10cm]{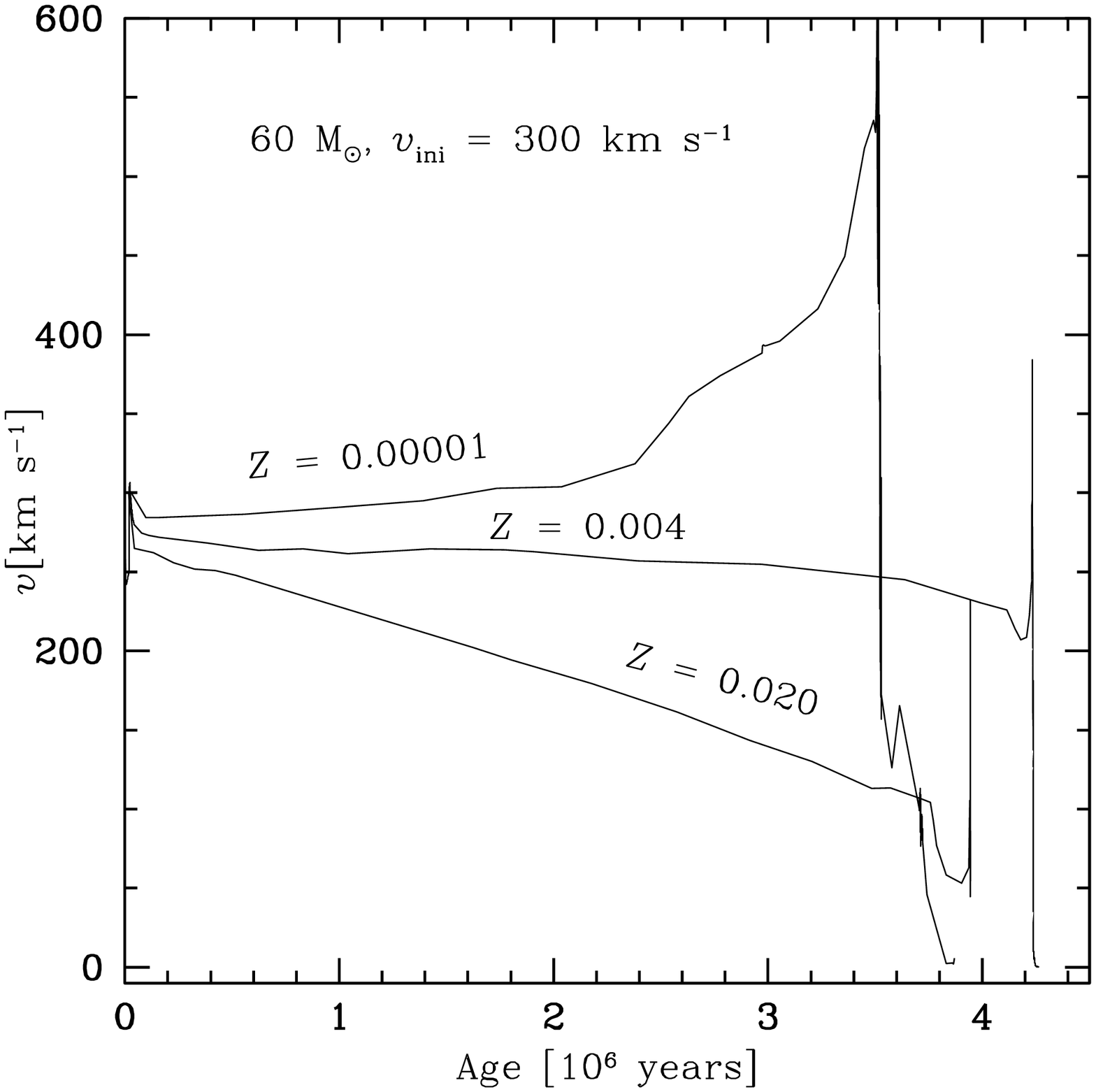}
  \caption{Evolution of the surface equatorial velocity as 
a function of time for 60 M$_\odot$ stars with $v_{\rm ini}$ = 300
km s$^{-1}$ at different initial metallicities. Figure taken from \citet{paperVIII2002}.
}
  \label{v60}
\end{figure}

\subsection{The Be stars and the evolution towards the critical velocity}

Be stars are emission line stars. Emission originates in a circumstellar outflowing disk.
How do these disks form~? How long are their lifetimes~? Are they intermittent~? 
Are they Keplerian~? Many of these questions are still subject of a lively debate. A point however which seems well accepted is the fact that
the origin of a disk might be connected to the fast
rotation of the star. This view is supported by the fact that objects with Be phenomena are the fastest rotators in the sample studied by \citet{Hunter2008a} (400 OB stars in the MCs). This trend has also been found by \citet{Martayan2006b, Martayan2007a} who obtain that Be stars rotate faster than B stars whatever the metallicity. 

\citet{Maederetal1999} and \citet{Wisniewski2006}
find that the fraction of Be stars with respect to the total number of B and Be stars
in clusters with ages (in years)  between 7.0 and 7.4 (in logarithm) increases when the metallicity decreases. This fraction passes from about 10\% at solar metallicity to about 35\% at the SMC metallicity. These significant proportions imply that Be stars may drastically affect the mean $v \sin i$ obtained for a given population of B-type stars. 
Very interestingly there appears to be a correlation between the frequency of Be stars and that of red supergiants.

These observations indicate that metallicity plays a role in the Be phenomenon and
provides hints on the way surface velocity may evolve differently for stars of different
initial metallicities. The correlation of Be star populations with those of red supergiants
can be seen as an indication that fast rotation not only favors the formation of Be stars
but also that of red supergiants (see Sect. \ref{BR}).

The evolution of surface velocities during the Main Sequence lifetime results from an interplay between meridional circulation (bringing angular momentum to the surface) and mass loss by stellar winds (removing it). The dependence on metallicity of these two mechanisms plays a key role in determining for each metallicity, a limiting range of initial masses (spectral types) for stars able to reach or at least approach the critical limit. 
\citet{Ekstrom2008b} computed models (with no account for the Tayler-Spruit dynamo) for initial masses between 3 and 60 M$_\odot$ and considered, for each initial mass, models with $\Omega/\Omega_{\rm crit}$ equal to 0.1, 0.3, 0.5, 0.7, 0.8, 0.9 and 0.99. The following conclusions have been obtained based on these models: 

\begin{enumerate}
\item Meridional currents bring angular momentum inwards only at the very beginning of the core H-burning phase. During most of the MS phase, meridional currents transport angular momentum from the inner regions to the outer layers.
As long as the mass loss rates are not too important, this brings the $\upsilon/\upsilon_\mathrm{crit}$ ratio close to 1 during the MS phase for stars
having on the ZAMS $\Omega/\Omega_{\rm crit} \ge \sim 0.8$.
\item Since the velocity of the meridional currents in the outer layers scales with the inverse of the density, the process becomes more efficient for stars of higher initial mass and/or higher initial metallicity.
\item At high metallicity however, mass loss becomes more and more important and can prevent the stars from reaching the break--up limit.
\item In stellar clusters, supposing an initial distribution of the rotational velocities as given by \citet{Huang2006a}, one expects that the fraction of stars having $\upsilon/\upsilon_\mathrm{crit} \ge 0.8$ becomes maximum for ages between 20-32 Myr at $Z=0.020$. This range of ages shifts to older ages at higher metallicities and to younger ages at lower metallicities. Such a shift results from the dependence on  metallicity and  mass of the intensity of the mass loss rates and of the velocities of the meridional currents. If we  look at stars with $\upsilon/\upsilon_\mathrm{crit} \ge 0.7$, the maximum number will be found in clusters of ages around 10 Myr and below, and this at all non-zero metallicities.
\item Be stars might be the natural outcome of stars with an initial rotational velocity in the upper tail of the initial velocity distribution. 
\item Depending on when the critical velocity is reached, one expects more or less high surface enrichments (see Fig. \ref{ncp}). If
the limit is reached very early during the MS phase, no enrichment is expected, while if this limit is reached at the end of the MS phase, high N/C and N/O ratios are expected. 
However, the enrichment also depends on the initial mass. For stars originating from small initial mass stars (typically from 3 M$_{\odot}$ at standard metallicity) one expects no or small surface enrichments, at whatever
time the critical velocity is reached.
\item To reproduce the higher fractions of Be stars at low metallicity, we must assume that in metal poor regions, a higher number of stars are born with high values of $\Omega/\Omega_\mathrm{crit}$.
\end{enumerate}

\subsection{Evolution of the interior angular momentum content}
\label{pulsar}

Figure \ref{omevo} shows the evolution of $\Omega$ inside the
25 M$_\odot$ model from the ZAMS until the end of the core Si--burning phase \citep{Hirschi2004a}.
The evolution of $\Omega$ results from many different processes:
convection enforces solid body rotation,
contraction and expansion respectively increases and 
decreases $\Omega$ as a consequence of local conservation of the
angular momentum,
shear (dynamical and secular) erodes
$\Omega$--gradients while meridional circulation may erode or build 
them up and finally
mass loss may remove angular momentum from the surface.
If during the core H--burning phase, all these processes may be important,
from the end of the MS phase onwards, the evolution
of $\Omega$ is mainly determined by convection, the local conservation of the angular momentum
and, for the most massive stars by mass loss.

During the MS phase, $\Omega$ decreases in the whole star. 
When the star becomes a red supergiant (RSG), $\Omega$ at the surface
decreases significantly due to the expansion of the outer layers. Note that the envelope
is gradually lost by winds in the 25 M$_{\odot}$ model. 
In the centre, $\Omega$ significantly increases when the core contracts
and then the $\Omega$ profile flattens due to convection. $\Omega$ reaches values of the 
order of $1\,s^{-1}$ at the end of Si--burning. It never
reaches the local break--up angular velocity limit, $\Omega_c$, although, when
local conservation holds, $\Omega/\Omega_c \propto r^{-1/2}$.

 Figure \ref{jevo} shows the evolution of the specific angular momentum,
$j=2/3\, \Omega  r^2$, in the central region of a 25 M$_\odot$ stellar model. 
The specific angular momentum remains constant under the effect
of pure contraction or expansion, but varies when transport mechanisms are active.
One sees that the transport processes remove angular momentum from the
central regions. Most of the removal occurs during
the core H--burning phase.
Still some decrease occurs during the core He--burning phase, then the evolution is mostly
governed by convection, which 
transports the angular momentum from the inner part of 
a convective zone to
the outer part. 
This produces the teeth seen in Fig.~\ref{jevo}.
The 
angular momentum of the central regions of the star at the end of Si--burning is 
essentially the same as at the end of He--burning (by end of
He--burning, we mean the time when the central helium mass fraction
becomes less than $10^{-5}$). As a numerical example,
we calculated, for the 25 M$_{\odot}$ model, the angular
momentum of its 3 M$_{\odot}$ remnant. 
We obtained 
${\cal L}_{\rm rem}=2.15\,10^{50}$g\,cm$^2$\,s$^{-1}$
at the end of He--burning and 
${\cal L}_{\rm rem}=1.63\,10^{50}$g\,cm$^2$\,s$^{-1}$
at the end of Si--burning. This corresponds to a loss of only 24\%. 
In comparison, the angular momentum
is decreased by a factor $\sim$5 between the ZAMS and the end of
He--burning. This shows the importance of correctly treating the 
transport of angular momentum during the Main Sequence phase.
It means also that, in the frame of the present models,
we can obtain a first order of magnitude estimate of the pre--supernova angular momentum of the core at the end of He--burning.

\begin{figure}[!tbp]
\centering
   \includegraphics[width=8.8cm]{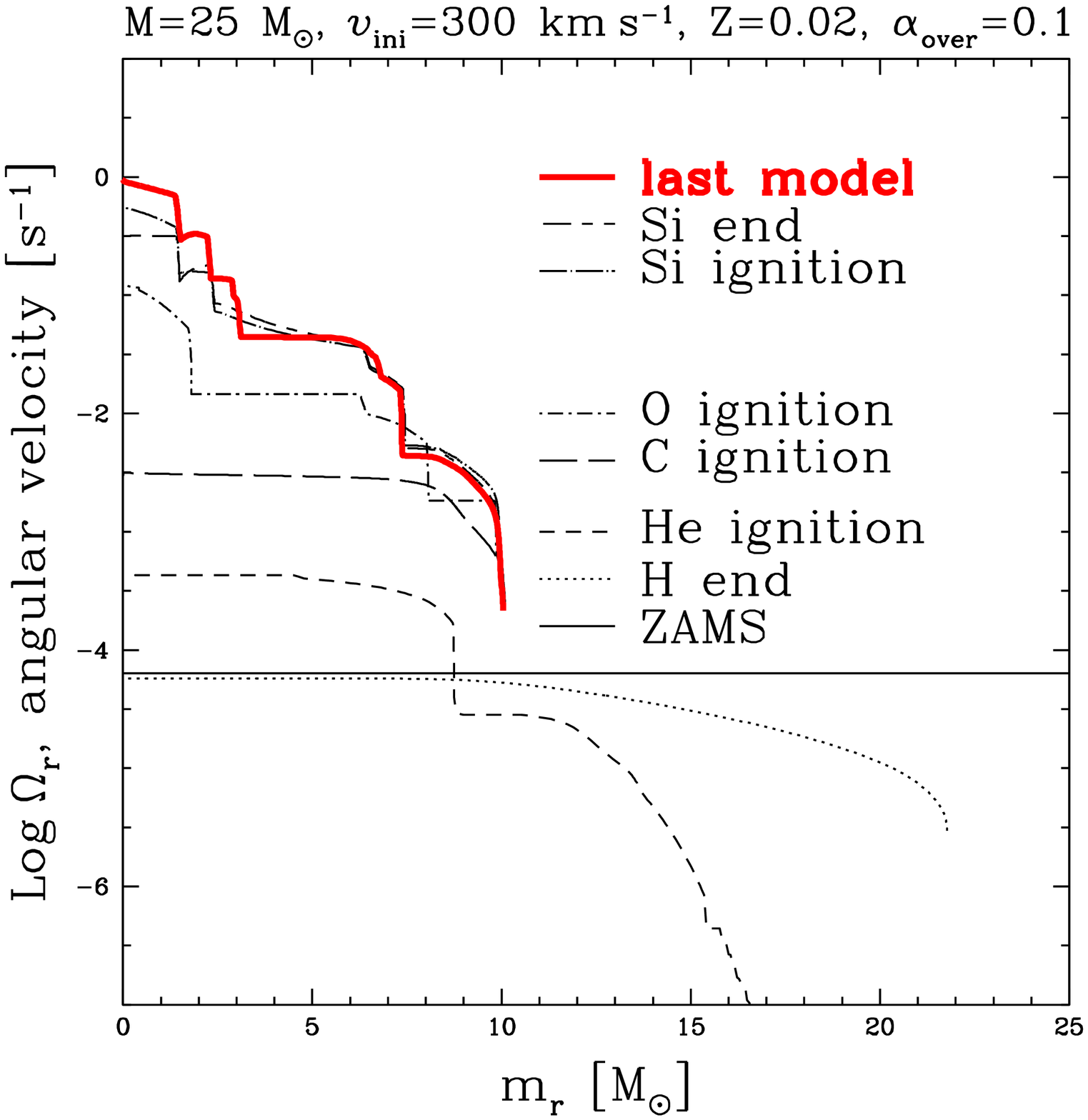}
\caption{Angular velocity as a function of the lagrangian mass
coordinate, $m_r$ inside the 25 $M_{\odot}$ model ($v_{\rm{ini}}$=
300 km\,s$^{-1}$) at various evolutionary stages \citep{Hirschi2004a}.
}
\label{omevo}
\end{figure}
\begin{figure}[!tbp]
\centering
   \includegraphics[width=8.8cm]{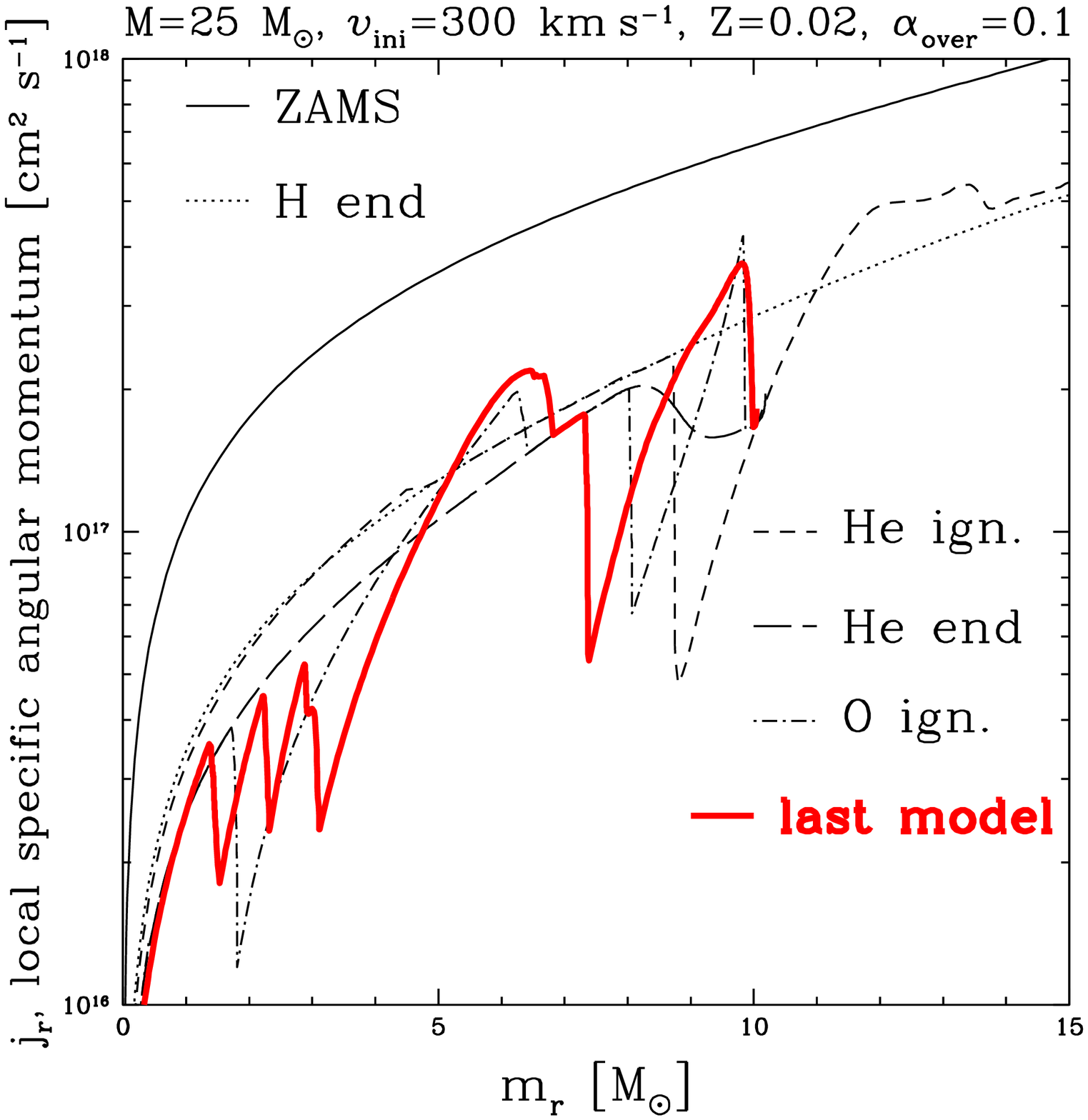}
\caption{Local specific angular momentum profiles for the 25 M$_{\odot}$ model ($v_{\rm{ini}}$=
300 km\,s$^{-1}$) at different evolutionary stages \citep{Hirschi2004a}.
}
\label{jevo}
\end{figure}

The present models have difficulties in accounting for
  the observed rotation rates of young pulsars and white dwarfs \citep[see
  e.g.][]{Kawaler1988,Heger2005,Suijs2008}. More precisely,
  they predict too fast rotation of the stellar cores in the advanced
  phases. This may be due to different causes that could lead to additional
  angular momentum loss from the central regions at different evolutionary
  phases:
\begin{itemize}
\item already during the nuclear lifetime;
\item at the time of the supernova explosion in the case of neutron stars,
  or during the Thermal Pulse Asymptotic Giant Branch (TP-AGB) phase at the time of the superwind episode in the
  case of white dwarfs;
\item during the early evolution of the new born neutron star or white dwarf.
\end{itemize} 
During the MS phase, angular momentum can be efficiently extracted from the core in case the star
 rotates as a solid body. Such a  situation is realized in numerical simulations 
where the Tayler-Spruit dynamo is accounted for as can be seen in Fig. \ref{Omprofile} \citep[see e.g.][]{magnIII2005}. 
\citet{Heger2005} find that magnetic torques decrease the final rotation rate of the collapsing iron core by about a factor 30-50 when compared with the nonmagnetic counterparts, allowing to obtain rotation periods of young pulsars which better fit the observations.

\section{EVOLUTIONARY TRACKS AND LIFETIMES}

\subsection{The Main-Sequence phase}

\begin{figure*}[tb]
 \includegraphics[height=10cm,angle=-90]{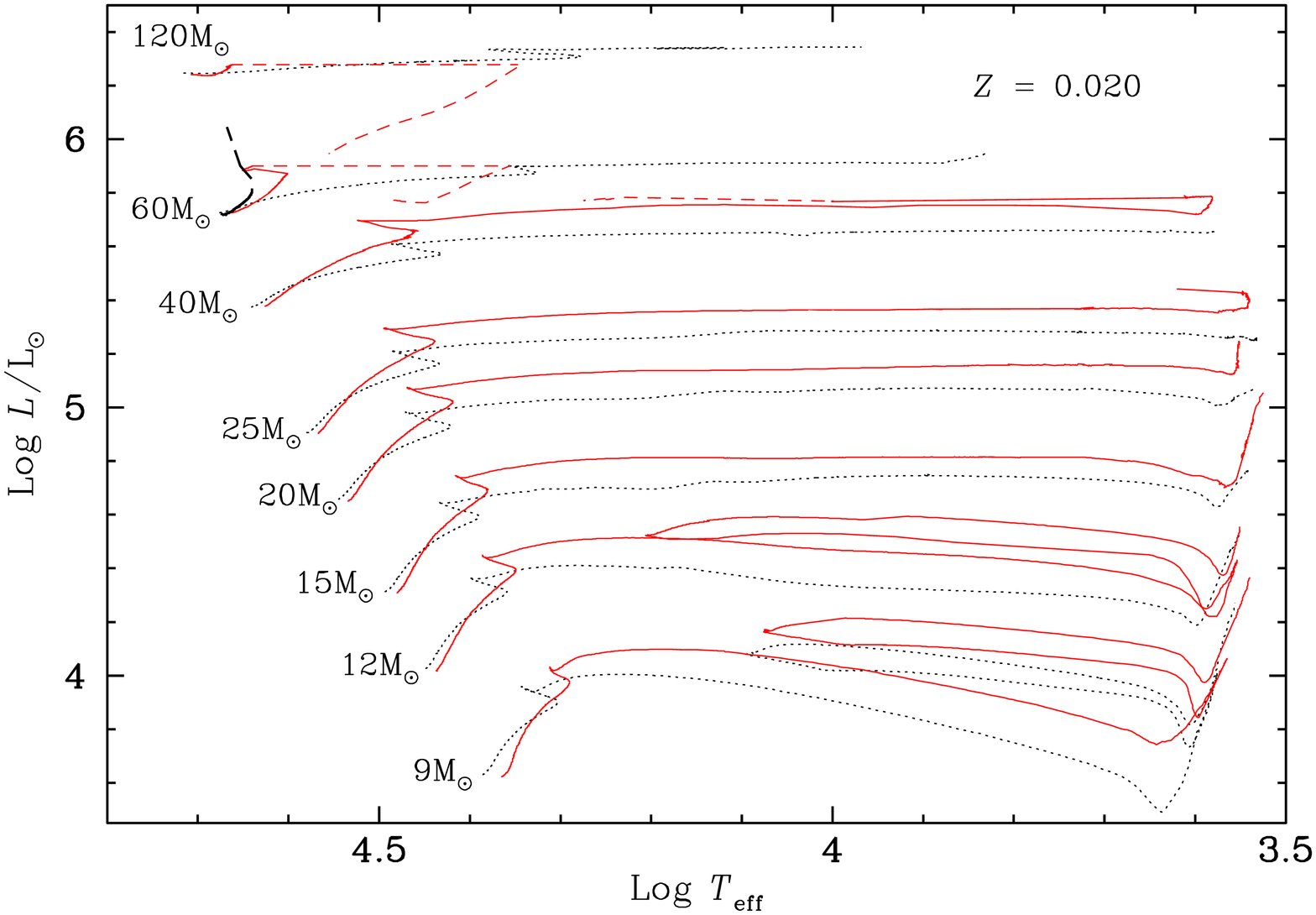}
  \caption{Evolutionary tracks for non--rotating 
(dotted lines) and
rotating (continuous lines) models with solar metallicity. 
The rotating models
have an initial velocity $v_{\rm ini}$ of 300 km s$^{-1}$.
For purpose of clarity, only the first part of the tracks for the most massive stars (M $\ge 40$ M$_\odot$) is shown.
Portions of the evolution during 
the WR phase for the rotating massive stars are indicated by
short--dashed lines. 
The long--dashed track for the 60 M$_\odot$ model corresponds to a very fast rotating star
($v_{\rm ini}$ $\sim 400$ km s$^{-1}$), which
follows a nearly homogeneous evolution. Only the beginning of its evolution is 
shown. Figure taken from \citet{paperV2000}.}
  \label{HRge}
\end{figure*}

Figure~\ref{HRge} shows the evolutionary tracks of non--rotating and rotating
stellar models for initial masses between 9 and 120 M$_\odot$ at Z=0.020. For the rotating stellar models, 
the initial velocity is $v_{\rm ini}$ = 300 km s$^{-1}$.
There is little difference between tracks with 
$v_{\rm ini}$ = 200 or 300 km s$^{-1}$.
If the effects behaved  like $v_{\rm ini}^2$, there 
would be  larger differences.
The present saturation effect occurs because outward transport of angular momentum by shears are larger when rotation is larger, also
a larger rotation produces 
more mass loss, which makes a larger reduction of rotation during the evolution. Let us note however that for some surface abundance ratios as
N/C or N/O, the increase from $v_{\rm ini}$ = 200 to 300 km s$^{-1}$
produces significant changes. Thus, the similarity of the evolutionary 
tracks does
not necessarily imply the similarity of the surface abundances for these
elements. 

On and near the ZAMS, rotation produces a small shift of the tracks
towards lower luminosities and $T_{\rm eff}$. This effect is due to both
atmospheric distorsions and to the lowering of the effective gravity 
\citep[see e.g.][]{Kippenhahn1970, MaederP1970, Collins1977}. At this stage the star
is still nearly homogeneous.
When evolution proceeds, the tracks with rotation become more luminous than for
no rotation.
This results from essentially two effects. On one side, rotational mixing
brings fresh H--fuel into the convective core, slowing down its decrease in mass
during the MS. For a given value of the central H--content,
the mass of the convective core in the rotating model
is therefore larger than in the non--rotating one and thus the stellar luminosity is higher
\citep{Maeder1987b, Talon1997, Heger2000}.

Rotation reduces the MS width in the high mass range (M $\sim 40$ M$_\odot$). 
Let us recall that
when the mass increases, the ratio of the diffusion timescale for the
chemical elements to the MS lifetime decreases. 
This can be shown in the following way: let us call $\tau_{\rm diff}$ the diffusion timescale which is proportional to $R^2/D$ where $R$ is the radius and $D$ the diffusion coefficient. For shear mixing, $D$ is proportional to $K=4acT^3/(3\kappa\rho^2 c_P)$, the thermal diffusivity.
Using homology relations, it can be shown that $\tau_{\rm diff}$ varies as $1/M^{1.8}$. Now the MS lifetime, $\tau_{\rm MS}$, for stars with initial masses greater than 15 M$_\odot$ varies as $1/M^{0.8}$, therefore the ratio $\tau_{\rm diff}/\tau_{\rm MS}$ varies as $M^{-1.1}$.}
As a consequence, starting
with the same $v_{\rm ini}$ on the ZAMS,
massive stars will be more mixed than low mass stars at an identical stage
of their evolution. This reduces the MS width since greater
chemical homogeneity makes the star bluer. 
Moreover, due to both rotational mixing and mass loss, their surface will be
rapidly enriched in H--burning products. These stars will therefore enter the 
Wolf--Rayet phase while they are still burning their hydrogen in their core. This again reduces the MS width.
For initial masses between 9 and 25 M$_\odot$, the MS shape can be widened by rotation mimicking the effect
of overshooting \citep{TZMM1997}.

\subsection{The post- Main-Sequence evolution}

The post--MS evolution of the most massive stars (M $\ge 40$ M$_\odot$) which become WR stars
will be discussed in Sect. \ref{WRS} below. We mention here some points of general interest. For low or moderate rotation, the
convective core shrinks  as usual during MS evolution,
while for high masses ($M \sim 40$ M$_{\odot}$) and  large initial rotations 
($\frac{\Omega}{\Omega_{\rm crit}}
\geq 0.5 $), the convective core grows in mass  during 
evolution. This latter situation occurs in the fast rota\-ting 60 M$_\odot$ model
shown on Fig.~\ref{HRge}.
These behaviours, i.e.\ reduction or growth of the core,
determine  whether the star will follow 
respectively the usual redwards MS tracks in the HR diagram, 
or whether it will bifurcate to the blue \citep{Maeder1987b}
towards the classical tracks of homogeneous evolution \citep{Schwarzschild1958}.

The stars with initial masses between 15 and 25 M$_\odot$
become red supergiants (RSG). Rotation does not change qualitatively this behaviour but
accelerates the redwards evolution, especially for the 15 and 20 M$_\odot$ models. As a numerical example, for an initial
$v_{\rm ini}$ = 300 km s$^{-1}$, the model stars burn all their helium as red supergiants at $T_{\rm eff}$
below 4000 K, while
the non--rotating models spend a significant part of the He--burning phase in the blue part
of the HR diagram: for the non--rotating 15 and 20 M$_\odot$ models, respectively 25 and 20\% of the total He--burning
lifetime is spent at $\log T_{\rm eff} \ge 4.0$.  Note that rotation is not the only factor able to affect the blue-red supergiant lifetimes. Mass loss, convection, overshooting, semiconvection, are, among others, all important factors in that respect.
The above behaviour of the rotating models results 
mainly from the enhancement 
of the mass loss rates. 
This effect prevents the formation of a big intermediate convective zone
and therefore favours a rapid evolution toward the RSG phase 
\citep{Stothers1979, Maeder1981}. 
Let us note that the dispersion of the initial rotational
velocities produces a mixing of the above behaviours.
As we shall see below (see Sect. \ref{BR}), a redwards evolution is also favored at low metallicity.
But in that case, this is mainly due to rotational mixing and not to a mass loss effect.


The problem of the mass discrepancy \citep{Herrero1992}, 
{\it i.e.} of the difference obtained between spectroscopic and evolutionary masses, has been
significantly reduced thanks to improvements brought to stellar atmosphere models. However
still some discrepancies are reported in general linked with strong helium surface enrichments:
\begin{itemize}
\item  In SMC: \citet{Mokiem2006} find a mild mass discrepancy for stars with 
spectroscopic masses inferior to about 20 M$_\odot$, which correlates with the surface helium
abundance. These authors find that the discrepancies are consistent with the predictions of chemically homogeneous evolution. Most of the stars observed by \citet{Heap2006} exhibit the mass discrepancy problem although no surface He enrichment.
\item In LMC: \citet{Mokiem2007} from the analysis of O-type stars in the LMC find that bright giants and supergiants do not show any mass discrepancy, regardless of the surface helium abundance. In contrast they find that the spectroscopically determined masses of the dwarfs and giants are systematically smaller than those derived from non-rotating evolutionary tracks. All dwarfs and giants having $y > 0.11$ ($Y >0.33$) show this mass discrepancy.
\end{itemize}
These mass discrepancies may arise 
from the use of stellar models with too small convective cores. Convective overshooting and rotation
produce larger convective cores and thus can help in explaining this mass discrepancy.


\subsection{Lifetimes}

Generally we can say that
 the MS lifetime duration is affected by rotation at least through
three effects:
\begin{enumerate}
\item Rotation increases the quantity of hydrogen burnt in the core. 
This increases the MS lifetime.
\item The hydrostatic effects of rotation make a star of a given initial mass
to behave
as a non--rotating star of a smaller initial mass. This tends to increase the
MS lifetime.
\item  Rotation increases the helium abundance in the outer radiative
envelope. This tends
to make the star overluminous with respect to its non--rotating counterpart and
thus
to reduce the MS lifetime.
\end{enumerate}
Depending on the initial mass and metallicity, rotation can either increase or decrease
the MS lifetimes with respect to the durations obtained from non-rotating stellar models.
For $Z=0.020$, the effect number 1 dominates and rotation increases the lifetimes.
For instance,
from the models by \citet{paperV2000},
one can deduce a nearly
linear relation between the relative enhancement of the MS lifetime,
$\Delta t_{\rm H}$ and $\overline{v}$, where 
$$\Delta t_{\rm H}(\overline{v})=[t_{\rm H} (\overline{v})-t_{\rm H}(0)]/t_{\rm H}(0).$$
One obtains,
$${\Delta t_{\rm H} (\overline{v}) \over t_{\rm H}(0)}=0.0013\cdot \overline{v},$$
where $\overline{v}$ is in km s$^{-1}$. 
When the metallicity decreases, the effect number 3 tends to
become the most important one. Typically
at $Z=10^{-5}$, the MS lifetimes are 
decreased by about 4--14\% for the mass range between 
3 and 60 M$_\odot$ when   $v_{\rm ini}$
increases from 0 to 300 km s$^{-1}$ \citep{paperVIII2002}.
The rotating 2 M$_\odot$ model of the same grid has  a longer MS phase than its non--rotating counterparts,
in contrast with what happens for higher initial mass stars.
This is because,
when the initial mass decreases, the hydrostatic effects (effect number 2)
become more and more important.

The He--burning lifetimes are less affected by rotation than the MS lifetimes. 
The changes are less than 10\%. The ratios $t_{He}/t_{H}$ of the
He to H--burning lifetimes are only slightly decreased by rotation and
remain around 10--15\%.

\section{MASSIVE STAR POPULATIONS AND THE CASE OF FIRST STARS}

\subsection{The ratio of blue to red supergiants}
\label{BR}

As noted by \citet{Langer1995}, the current models 
(without rotation)
with Schwarzschild's criterion predict no red supergiants 
in the SMC. This is
also seen in Fig.~\ref{vvv20/age} which illustrates for models
of 20 M$_{\odot}$ at $Z$ = 0.004  the variations of the
$T_{\mathrm{eff}}$ as a function of the fractional lifetime in the 
He--burning phase for different  rotation velocities \citep{paperVII2001}.
For zero rotation,  we see that the star
only moves to the red supergiants
 at the very end of the He--burning phase,
so that the blue to red supergiant ratio $B/R$
is  $ \simeq 47$.  This is 50-100 times greater than
the observed ratio 
 in the SMC cluster NGC 330 which
lies between 0.5 and 0.8, according to the various sources
discussed in \citet{Langer1995} and \citet{Eggenberger2002}. 

This disagreement implies that no reliable predictions can be made concerning
the nature of the supernova progenitors in different environments, or the populations of supergiants in galaxies. 
The B/R ratio also constitutes an important and sensitive test for stellar evolution models, because it is very sensitive to mass
loss, convection and mixing processes \citep{Langer1995}. 
Thus, the problem of the blue to red supergiant ratio ($B/R$ ratio)
remains one of the most severe problems in stellar evolution.

Looking at Fig.~\ref{vvv20/age}, we see that
for  average rotational velocities 
$\overline{v}$ during the MS,  $\overline{v}$ = 152, 229 and 311
km s$^{-1}$, one has respectively  $B/R =$ 1.11, 0.43 and 0.28. 
Thus, \emph{ the $B/R$ ratios
are much smaller for higher initial rotation velocities}, as
rotation favours the formation of
red supergiants and reduce the lifetime in the blue.
We notice in particular
 that for 
 $\overline{v} =   200$ km s$^{-1}$, we have a B/R ratio  of about 0.6
well corresponding to the range of the observed values.

\begin{figure}[tb]
 \includegraphics[height=10cm]{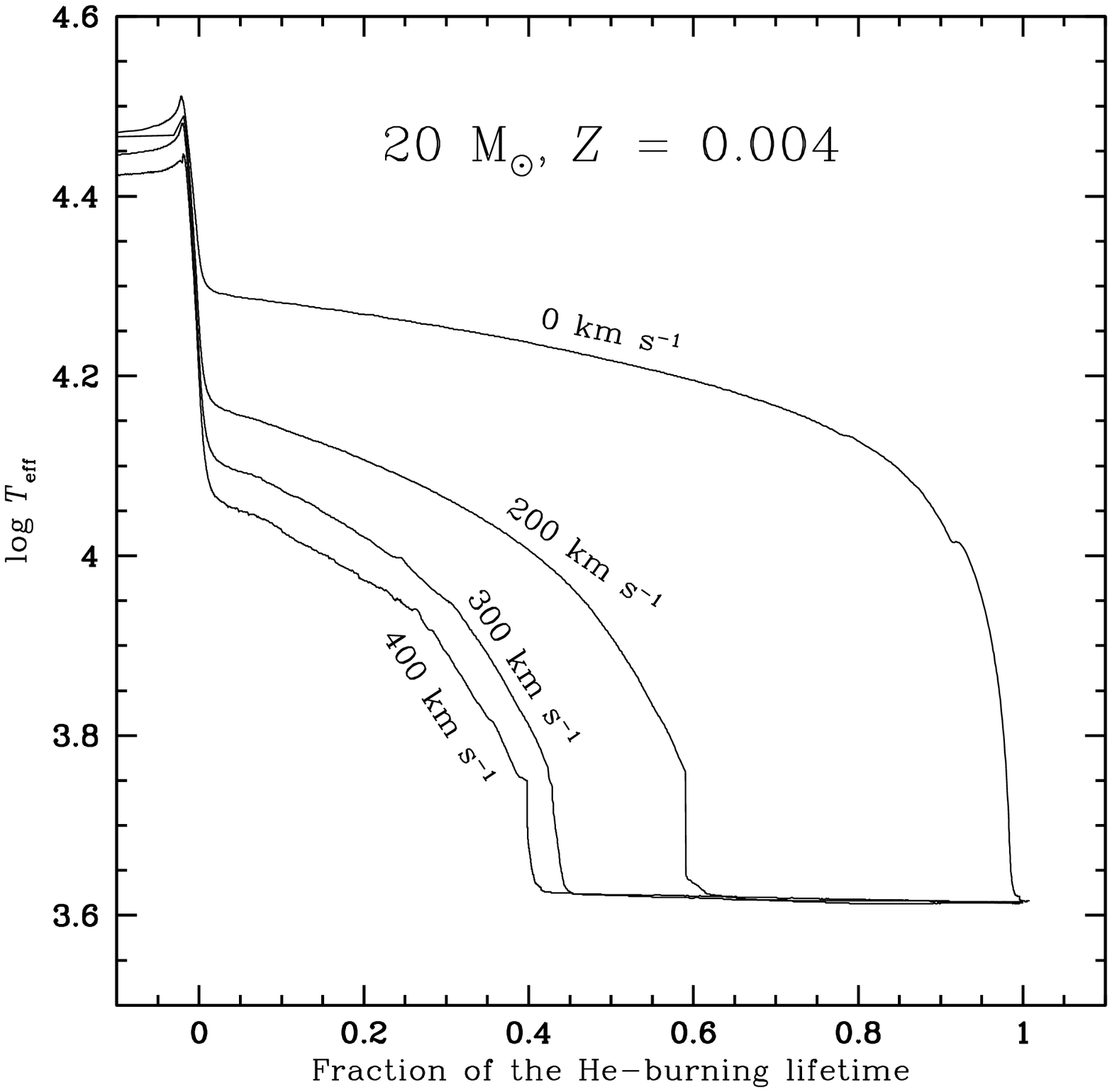}
  \caption{Evolution of the $T_{\mathrm{eff}}$
as a function of the fraction of the lifetime spent
in the He--burning phase for 20 M$_\odot$ stars with different
initial velocities. Figure taken from \citet{paperVII2001}.
}
  \label{vvv20/age}
\end{figure}

\begin{figure}[tb]
 \includegraphics[height=10cm]{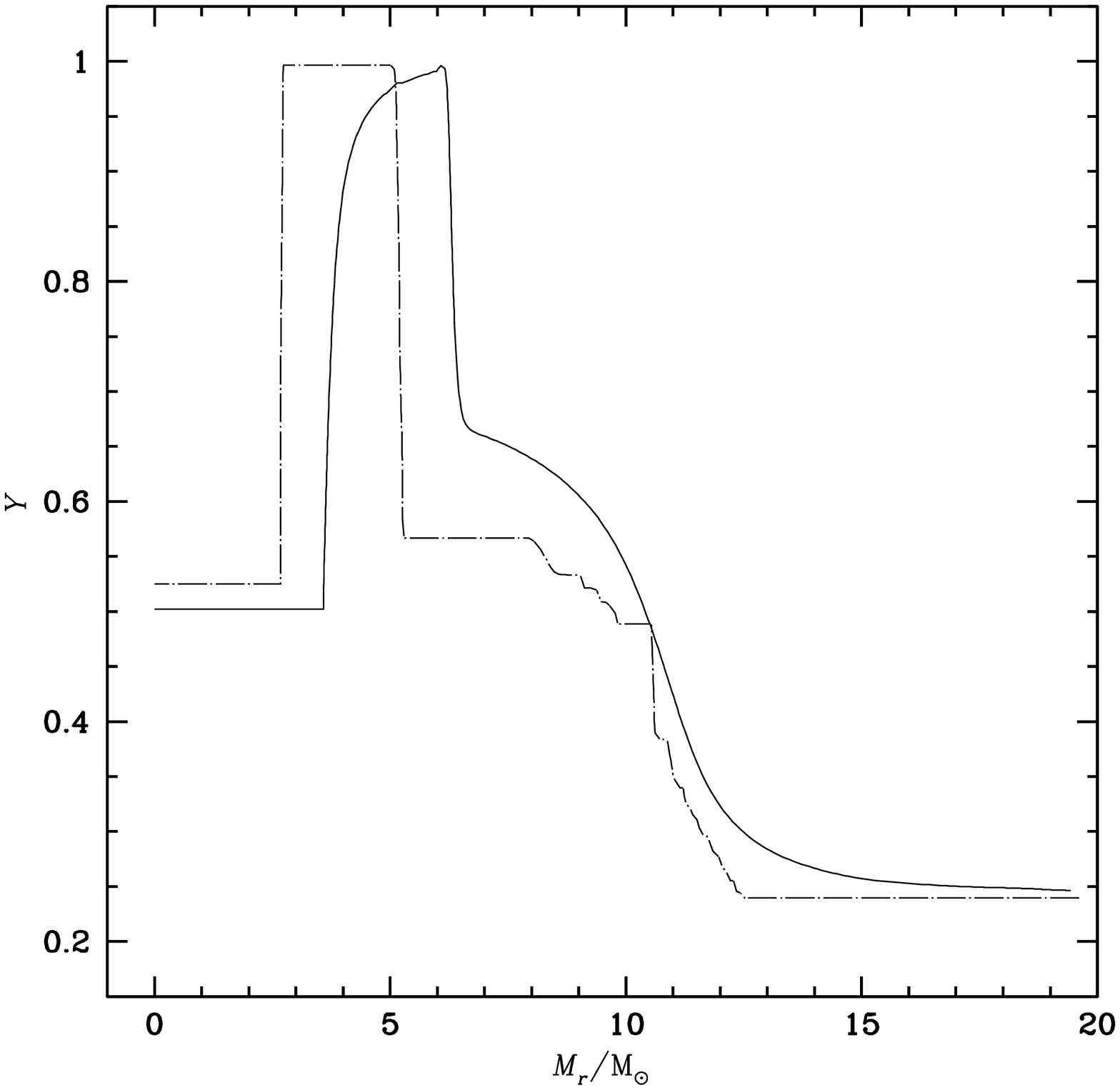}
  \caption{ Comparison of the internal distribution of helium
in two models of 20 M$_\odot$ at the middle of the
He--burning phase. The dashed--dot line concerns the models
with zero rotation and the continuous line represents
the case with $v_{\mathrm{ini}}$ = 300 km s$^{-1}$.  Figure taken from \citet{paperVII2001}.
}
  \label{Hedistr}
\end{figure}

The reason for this variation of the B/R ratio when the rotation velocity changes
is discussed in details in \citet{paperVII2001}. Let us just mention here
that when an extended intermediate convective zone is associated to the
H-burning shell, the star tends to stay in the blue part of the HR diagram 
Rotational mixing tends
to reduce the extension of this intermediate convective zone and
thus to accelerate the redwards evolution in the HR diagram.
How rotation limits the convective zone associated to the H-burning shell
can be understood looking at Fig.~\ref{Hedistr}.
The larger He--core in the rotating 
models means that a larger fraction of the total luminosity
is made in the core (0.42 instead of 0.31 for the models
of Fig.~\ref{Hedistr}). This means that the H--burning shell
in the rotating model produces a smaller fraction of the total
luminosity and this  contributes to reduce the importance
of the convective zone  above the H--burning shell. Simultaneously,
the  higher He--content near the shell in the
rotating case leads to a
decrease of its H-content and opacity,  and this also contributes to
reduce
the importance of the  convective zone 
associated with  the H--burning shell.


Thus models with rotation well
account for the long standing problem of the large numbers
of red supergiants observed in low $Z$ galaxies, while current
 models with mass loss were predicting no red supergiants.

\subsection{Current problems with Wolf-Rayet stars}
\label{WRS}

Wolf-Rayet stars are the bare cores of initially massive stars, whose H-rich envelope has been removed by strong stellar winds or through Roche lobe overflow in a close binary system \citep[see e.g. the review by][]{Crowther2007}.
Wolf--Rayet stars play a very important role in Astrophysics, as signatures
of star formation in galaxies and starbursts, as injectors of chemical elements, 
as  sources of kinetic energy into the interstellar medium and 
as progenitors of supernovae and, likely, as progenitors of long soft $\gamma$--ray bursts.

Evolutionary scenarios, at solar metallicity, for Wolf-rayet stars follow the sequences shown below

\eject
\vskip 5mm
\noindent
{\bf \underline{{$M >90  M_{\odot}$}}}:  O –- Of –- WNL –- (WNE) -– WCL –- WCE -– 
SN (Hypernova low Z )\\
\noindent
{\bf \underline{{$60-90 \; M_{\odot}$}}}: O –- Of/WNL$\Leftrightarrow$LBV -– WNL(H poor)-– WCL-E -– SN(SNIIn?)\\
\noindent
{\bf \underline{{$40-60 \; M_{\odot}$}}}: O –- BSG –-  LBV $\Leftrightarrow$ WNL -–(WNE) -- WCL-E –- SN(SNIb) \\
\hspace*{5.9cm}  - WCL-E - WO – SN (SNIc) \\
\noindent
{\bf \underline{{$30-40 \; M_{\odot}$}}}:  O –- BSG –- RSG  -- {LBV ?} -- WNE –- WCE -– SN(SNIb)\\
\noindent
{\bf \underline{{$25-30 \; M_{\odot}$}}}: O -–(BSG)–-  RSG  -- BSG (blue loop) -- RSG  -- SN(SNIIb, SNIIL)\\
\noindent
{\bf \underline{{$10-25 \; M_{\odot}$}}}: O –-  RSG -– (Cepheid loop, $M < 15 \; M_{\odot}$) – RSG -- 
SN (SNIIL, SNIIp)\\ 

Of-stars are O-type stars showing emission lines. WN stars present emission lines from nitrogen and helium. 
Depending on the line ratios between N III-IV and He I-II, they are classified as WNE (WN2-WN5) or WNL (WN7-WN9),
with WN6 being early (E) or late (L) type. WC spectral type depend on the line ratios of C III and C IV (WC4-WC6 are early,
WC7-WC9 are late). Oxygen-rich WO 
stars extend the  WCE sequence, exhibiting strong O VI
emission lines \citep[see the review by][and references therein]{Crowther2007}. LBV stands for Luminous Blue Variables 
\citep[see the review by][and references therein]{Vink2009}. BSG and RSG for blue and red supergiants respectively. 
The sign $\Leftrightarrow$ means back and forth motions between the two  stages. The various types of supernovae are tentatively indicated \citep[see the review by][for the core-collapse supernova classification]{Smartt2009}. The limits 
between the various scenarios  depend on metallicity $Z$ and rotation. The limiting masses indicated
corresponds to the standard metallicity case, {\it i.e.} $Z=0.020$.

From the above evolutionary scenarios, we see that at the standard metallicity, the minimum initial mass for a
star to enter the WR phase is between 25 and 30 M$_\odot$. This inferior limit is however dependent
on the stellar models considered and in particular on the mass loss rates used and the mixing processes 
(size of the convective zone  and other mixing processes in radiative zones as for instance those induced by rotation).
These processes, mass loss and mixing,  also affect the durations of the WR phases.

Let us recall
some difficulties faced by standard stellar models concerning the WR stars.
  A good agreement between 
the predictions of the stellar models for the WR/O number ratios and the observed 
ones at different metallicities in regions of constant star formation was achieved 
provided the mass loss rates were enhanced by about a factor of two during the MS
and WNL phases \citep{Maeder1994}. This solution, which at that time appeared 
reasonable in view of the uncertainties pertaining the mass loss rates, is no longer
applicable at present, at least if we take for granted the fact that the mass loss rates during the WR phase are
reduced by a factor 2 to 3, when account is given to the clumping effects
in the wind \citep{Nugis2000}.
Also, the mass loss rates for O--type stars have been substantially revised 
(and in general reduced) by the results of 
\citet{Vink2001}.
With these revised mass loss rates
the predicted numbers of WR stars by the standard non--rotating models of  \citet{paperX2003} are much
too low with respect to the observations.

A second difficulty of the standard models with mass loss concerns
the observed number of transition WN/WC stars. These stars show simultaneously some 
nitrogen characteristic of WN stars and some carbon of the further WC stage.
The observed frequency of WN/WC stars among WR stars turns around 4.4 \% \citep{VanDerHucht2001a}, while
the frequency predicted by the standard models without extra--mixing processes 
are lower by 1--2 orders of magnitude \citep{Maeder1994}. This feature cannot be accounted for by a change
of the mass loss rates and indicates that mixing is not adequately incorporated in the standard stellar models.
A third difficulty of the standard models as far as WR stars were concerned was that
there were relatively too many WC stars with respect to WN stars predicted at low metallicity 
\citep[see the review
by][]{Massey2003}. These difficulties are the signs that some process is missing
in standard models.

\subsection{Solutions of WR problems in terms of rotation?}
\label{WR2}

What could be missing? Binarity, changes of the mass loss history, mixing processes can be invoked
and there is little doubt that playing with these different features, a given observational feature can be reproduced.
We think however that good stellar models have to reproduce all the main observed features exhibited by massive stars
simultaneously  (as the MS width, the changes of the surface abundances during the MS phase, the 
variation with the metallicity of the of blue to red supergiant ratio, or of the WR/O number ratio  to just cite a few of them) without having to change the physics included in order to adapt it to resolve a particular question. This makes the problem
much more constrained and discard some solutions.
As discussed previously, there are
some evidences that some extra mixing process is still missing in stellar models and that this extra-mixing is probably 
driven by rotation. Thus in the following we shall focus on the consequences for the WR population of rotational mixing.
We first indicate some general trends induced by rotation and then illustrate the consequences obtained in some
specific stellar models.

Rotational  mixing favours the entry into the
WR phase in two ways, firstly
by allowing chemical species produced in the core to diffuse in the radiative envelope and, secondly, by
making the mass of the convective core larger. 
In the non--rotating model,
mass loss by stellar winds is the key physical ingredient which
allows internal chemical products to appear at the surface and thus the  formation of a WR star. The star becomes a WR star
only when sufficiently deep layers are uncovered. In rotating models, the characteristic surface abundances
of WR stars can be obtained through  the effects of mass loss by stellar winds and of
rotational mixing. The action of rotation allow WR stars to appear through the single star scenario even when
the mass loss rates are reduced. To realize that, imagine a star rotating so fast that it would follow a homogenous
evolution. Such a star can become a WR star, i.e. being a star with a log T$_{\rm eff}$ greater than about 4
and a mass fraction of hydrogen at the surface below $\sim$ 0.4 without losing mass!

When models with a time-averaged rotation velocity during the MS phase in the range of the observed
values are considered, then a reasonable number of WR stars can be produced through the single
star scenario even using low mass loss rates. This is illustrated in Fig. \ref{WRO}
Also the rotational models well
reproduce the WN/WC ratio at low metallicity, the observed fraction of WR stars in the transition stage WN/WC and the variation with the metallicity of the number ratios of type Ibc to type II supernovae 
\citep[see more detailed discussion in][and in Sect. \ref{IBC}]{paperX2003,paperXI2005}. 

\begin{figure}[t]
 \includegraphics[height=10cm, angle=-90]{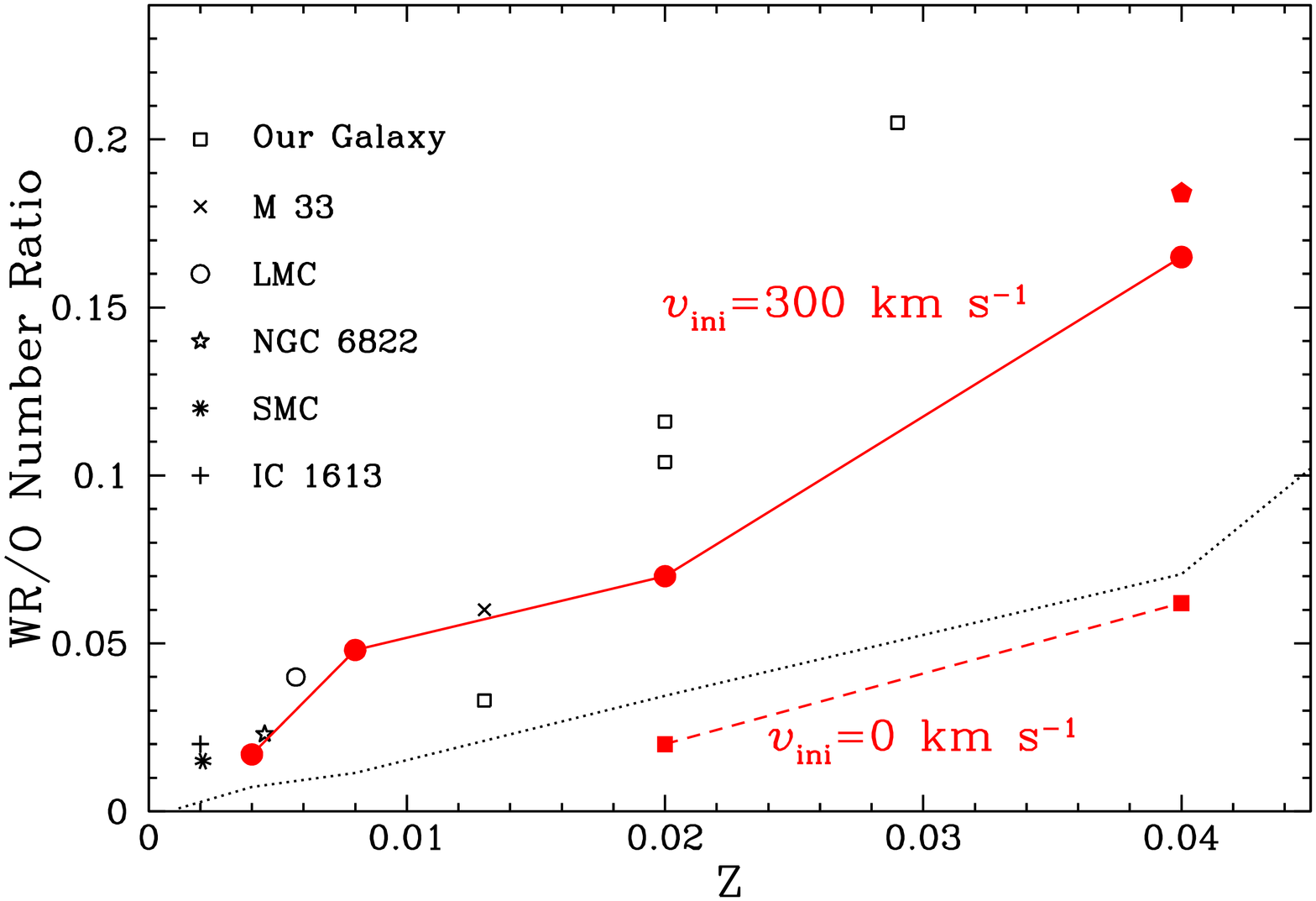}
  \caption{Variation of the number ratios of Wolf--Rayet stars to O--type stars as a function of the metallicity.
The observed points are taken as in \citet{Maeder1994}.
The dotted line shows the predictions of the models
of \citet{Meynet1994} with normal  mass loss rates.
The continuous and the dashed lines show the predictions of the rotating and 
non--rotating stellar models
respectively of \citet{paperX2003, paperXI2005}. The black pentagon shows the ratio predicted by Z=0.040 models computed
with the metallicity dependence of the mass loss rates during the WR phase proposed
by \citet{Crowther2002b}.
}\label{WRO}
\end{figure}

\begin{figure}
\includegraphics[height=10cm,angle=-90]{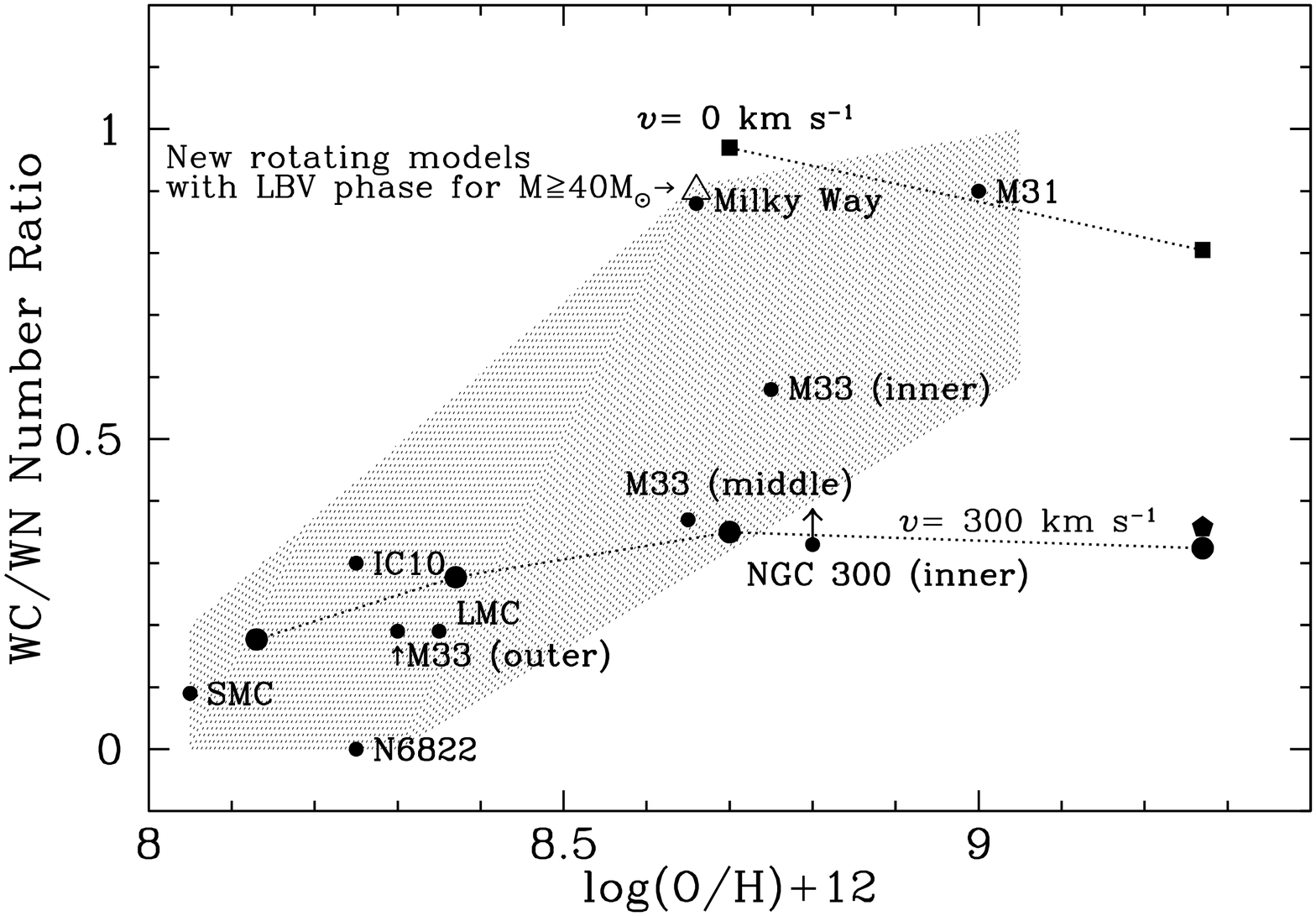}
\caption{Variation of the number ratios of WN to WC stars as a function of metallicity. The grey area
encompasses the observed ratios. Individual measures are indicated by
black circles labeled with the name of the Galaxy \citep[see references in][]{paperXI2005}. Solar (O/H) value
is taken from \citet{Asplund2005}, the (O/H) values for the SMC and LMC are taken from \cite{Hunter2007}.
The dotted lines show the predictions of the rotating and non--rotating stellar models
of \citet{paperX2003, paperXI2005}. The black pentagon shows the ratio predicted by Z=0.040 models computed
with the metallicity dependence of the mass loss rates during the WR phase. The open triangle shows the WC/WN ratio obtained from the rotating models allowing an LBV phase during the WR phase \citep{Meynet2008b}.
}\label{WNWC}
\end{figure}

The number of WC to WN stars  increases with the metallicity (see Fig.~\ref{WNWC} \footnote{We consider here regions having reached a stationary situation, {\it i.e.} regions where the star formation rate can be considered to have remained constant for the last twenty million years.}). 
Many attempts have been performed to reproduce the observed trend: for instance the enhanced mass loss rate models of \citet{Meynet1994} provided a good agreement for solar and higher than solar metallicity but produced too few WN stars in metal-poor regions. The inclusion of rotation together with reduced mass loss rates accounting for the effects of clumping improved the situation in the metal poor region, but produced too many WN stars at solar and higher metallicities
\citep{paperXI2005}.
\citet{Eldridge2006} show that  models that include the mass-loss metallicity scaling during the WR phase closely reproduce the observed decrease of the relative population of WC over WN stars at low metallicities. However such models severely underestimate the fraction of WR to O-type stars. In that case, to improve the situation, a high proportion of
Wolf-Rayet stars  originating from mass transfer in close binaries would have to be assumed at all metallicities. For instance at solar metallicity about 75\% of the WR stars should be produced in close binary systems \citep[see Fig.~5 in][]{Eldridge2008}. This is clearly in contradiction with the results of \citet{Foellmi2003a,Foellmi2003b} that locally in the Milky Way only 24\% of the WR stars are binaries.
Thus, as shown below, another explanation has to be found.

The WN/WC number ratio depends also on other factors, in particular on the evolutionary scenario. 
In \citet{paperX2003, paperXI2005}, the most massive rotating stars enter into the WR regime already during the MS phase. This feature has good and bad effects. On one hand, it allows these models to well reproduce the variation
of the number fraction of WR to O-type stars since it significantly increases the WR lifetimes. On the other hand, it produces very long WN phases since the star enters into the WR phase having still a huge H-rich envelope. As a consequence,
too low values for the WC/WN ratio are obtained at solar and higher metallicities. 

In \citet{paperX2003,paperXI2005}, the hypothesis has been made that 
when a star enters into the WR stage during the MS phase,  it
avoids the Luminous Blue Variable phase. Actually, stars may behave differently. It may well be 
that a star which becomes a WR star during the MS phase, enters a LBV phase after
the core H-burning phase, before evolving back into the WR regime. 
When this evolutionary scenario is followed, reasonable values for both the WR/O and the WC/WN ratios are obtained.
Indeed the ratios of WR/O and of WC/WN given by these models at the solar metallicity are
0.06 and 0.9 which compare reasonably well with the observed values of 0.1 and 0.9 respectively.
Both ratios are not reproduced by the non-rotating models to which a similar scenario is applied.
This discussion illustrates the possible key role that the LBV phase may play in shaping the
WC/WN ratio. 

\subsection{Pop III star models with rotation}

Understanding the evolution of massive stars at low and very low metallicity is a requirement to address questions such as the nature of the sources of the reionization in the early Universe, 
the evolution of the interstellar abundances during the early phases of the evolution of galaxies, for finding possible signatures of primordial stellar populations in the integrated light of
very distant galaxies and for discovering which objects are the progenitors of the long soft Gamma Ray Bursts. At present, the most ``iron'' poor objects known in the Universe are not very far from us since they are galactic halo field stars. Provided these stars are trustworthy very metal poor stars \citep[a view recently challenged by][]{Venn2008}, these objects offer a unique opportunity to study the yields of the first generations of stars.

In this section, we focus the discussion on {Pop III stars, i.e. on}
 stars which were made up of material having been processed only by 
primordial nucleosynthesis.
In an environment with primordial composition, one expects the following differences with respect
to the more classical evolution at higher metallicity:
\begin{itemize}
\item At strictly $Z=0$, the cooling processes, so important for allowing the
evacuation of the energy produced when the molecular clouds collapse and thus
its fragmentation, are
not so efficient than when metals are present. This favors the formation of more
massive stars. The initial mass function is probably different depending on the mass
of the ``minihaloes'' \citep[see][]{Greif2006}. In minihaloes with masses between 10$^6$ and
10$^8$ M$_\odot$, virial temperature is between 10$^3$ and 10$^4$ K and the cooling is due mainly to
the molecules H$_2$. This allows the formation of stars with characteristic masses $\ge 100$ M$_\odot$. In minihaloes where ionization occurs prior to the late stages of the protostellar accretion process, namely those with a virial temperature superior to 10$^4$ K and thus with masses
above 10$^8$ M$_\odot$\footnote{This may also happen in relic HII regions left by the first stars, \citep[see][]{Johnson2006}.}, the hydrogen deuteride (HD) molecule provides an additional cooling channel.
In those minihaloes, metal-free gaz can cool more efficiently. This leads to the formation of stars
with masses superior to 10 M$_\odot$. Thus at the very beginning, we would have first, during a quite
short period, only very massive Pop III stars and, when gas assembles in more massive haloes (or is
reionized by the first stars), a second Pop III star generation appears with smaller characteristic masses (about 10 M$_\odot$ and above). When the
metallicity becomes higher than 10$^{-3.5}$ Z$_\odot$, {\it i.e.} for $Z$ above about 5 10$^{-6}$
a normal IMF governs the mass distribution of newly born stars\footnote{Some authors argue that
star formation switches to more classical mode already when $Z=10^{-6}$ due to dust production in the early Universe, 
\citep[see][]{Schneider2006}.}. According to \citet{Greif2006} the very massive Pop III stars only contribute marginally to feed the reservoir of ionizing photons and to the chemical enrichment of the interstellar medium. Much more important are on this respect the less massive Pop III stars born in more massive haloes or in relic HII regions.
\item The (nearly) absence of heavy elements implies that massive stars cannot begin burning hydrogen through the CNO cycle but through the pp chains. However the energy output extracted from the pp chains is not sufficient to compensate for the high luminosity of these stars. The stars must compensate the deficit  of nuclear energy by extracting energy from the gravitational reservoir, {\it i.e.} they contract. At a given point however, due to this contraction, the central temperature becomes high enough for activating triple $\alpha$ reactions. Some carbon is then produced. When the mass fraction of carbon is of the order of 10$^{-12}$, the CN cycle becomes the main source of energy and the evolution then proceeds as in a more metal rich massive stars. For stars above about 20 M$_\odot$, activation of the CN cycle intervenes very early during the core H-burning \citep[typically before five percents of the hydrogen at the centre is consumed, see][]{Marigo2003}.
\item Pop III stars are more compact due to the contraction they undergo at the beginning of the core H-burning phase (see above) and to the fact that the opacity of primordial material is smaller than that of metal rich one. Typically a Pop III 20 M$_\odot$ star on the ZAMS has a radius reduced by a factor 3.5 with respect to the radius of the corresponding star with $Z=0.020$. Even passing
from the very low metallicity $Z=10^{-5}$ to 0 already produces a decrease of the radius by a factor 2! Smaller radius favors more efficient mixing in two ways: first, for a given value of the diffusion coefficient, $D$, the mixing timescale decreases with the square of the radius as $\sim R^2/D$, second as can be seen from Fig.~\ref{UdiffZ}, in pop III models, the meridional currents have smaller amplitudes
and present a very different configuration than at higher metallicities. Typically at high metallicity (look at the curve for $Z=0.020$),
there is an extended outer cell which tends to reduce the $\Omega$-gradient. This is no longer the case in metal poor models.
\end{itemize}

\begin{figure}[t]
\includegraphics[width=2.6in]{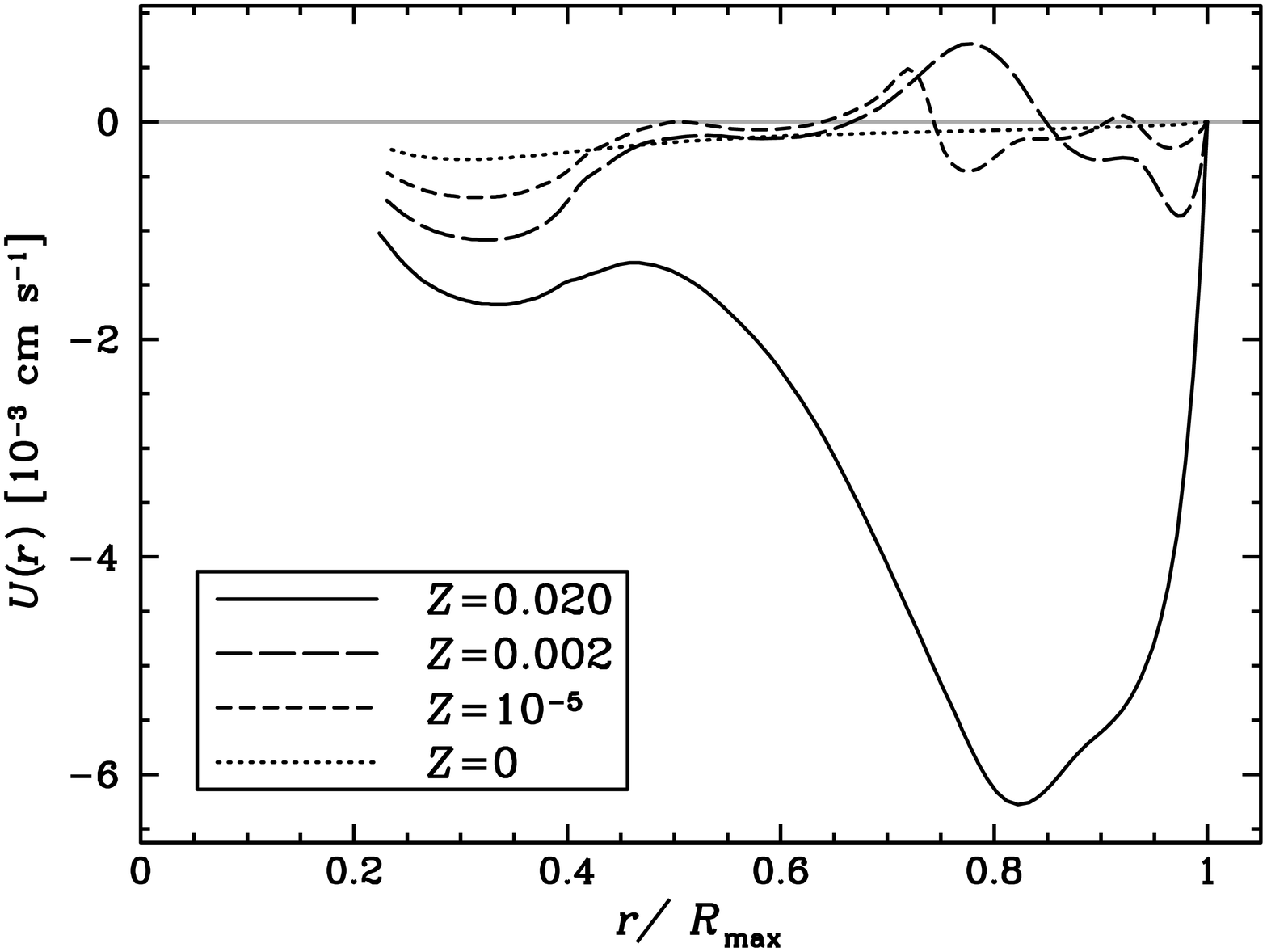}
\caption{Internal profile of $U(r)$ in models of 20 $M_\odot$ at various metallicities, with $\Omega_\mathrm{ini}/\Omega_\mathrm{crit}= 0.5$, where $u(r,\theta)$ the vertical component of the velocity of the meridional circulation is $u(r,\theta)= U(r)\ P_2(\cos \theta)$ with $P_2$ the second Legendre polynomial. The radius is normalized to the outer one. All the models are at the same evolutionary stage, when the central H mass fraction is about 0.40.
Figures taken from 
\citet{Ekstrom2008a}.
}
\label{UdiffZ}
\end{figure}

\begin{figure}[t]
\includegraphics[width=2.6in]{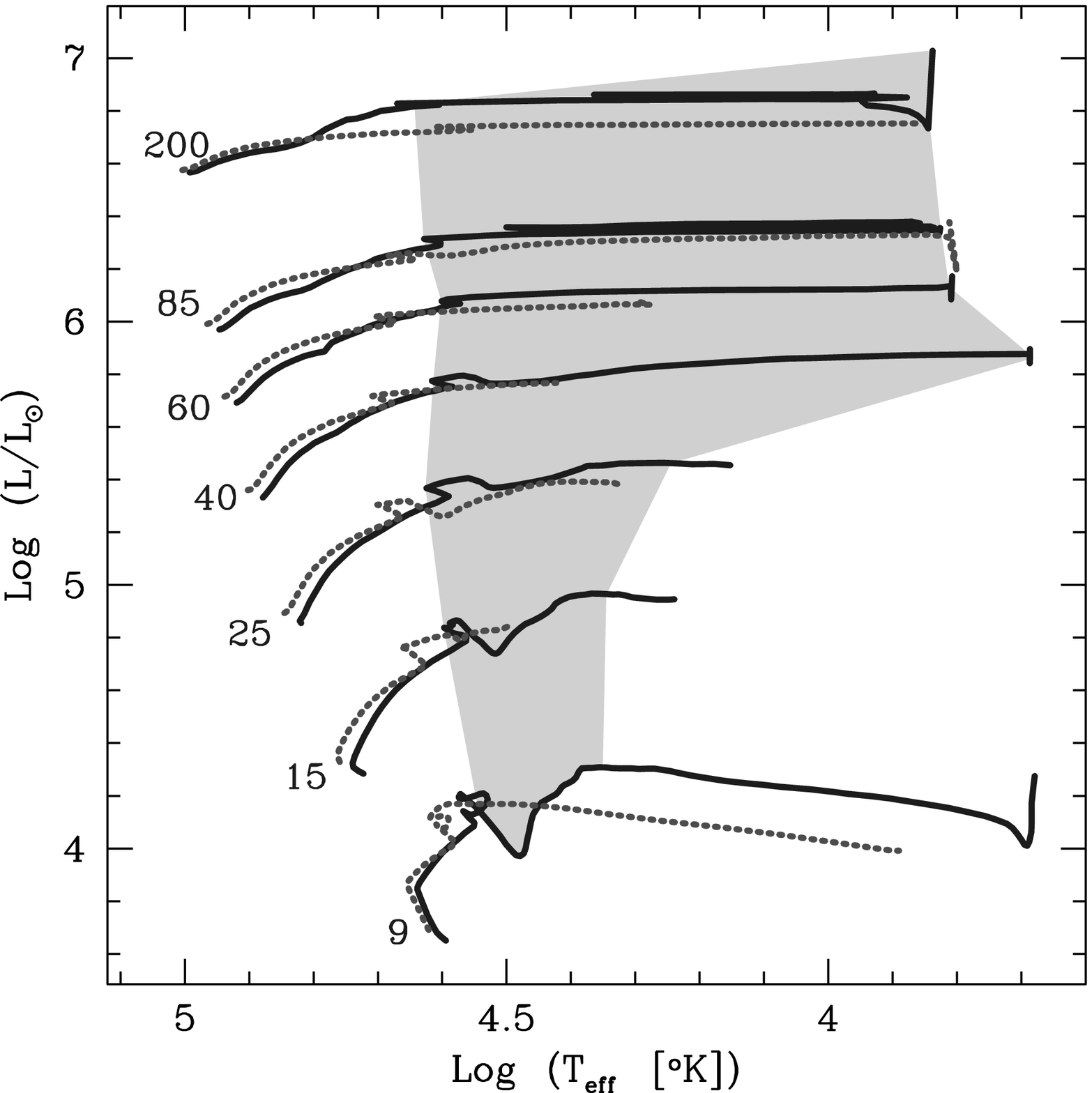}
\includegraphics[width=2.6in]{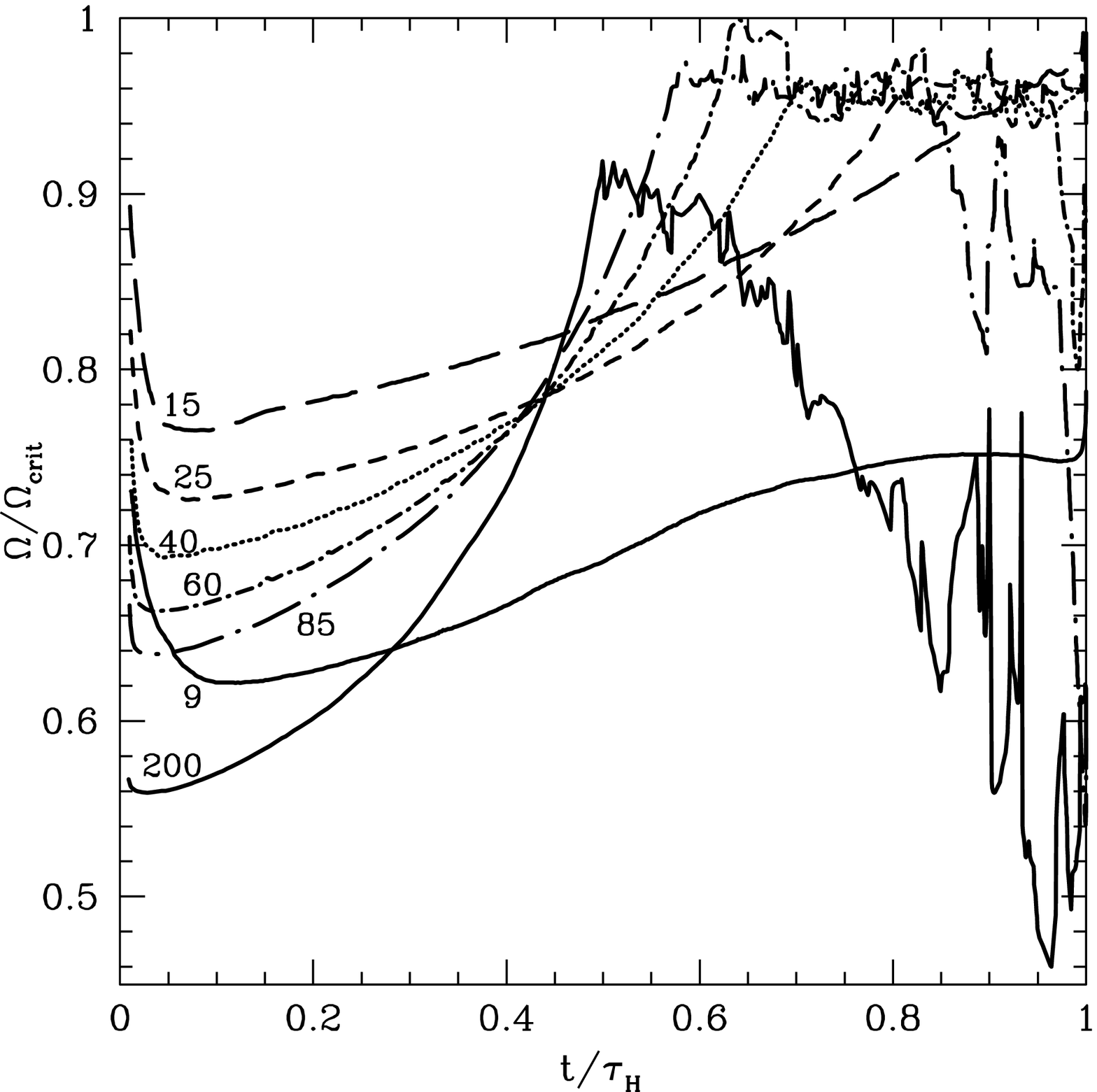}
\caption{
{\it Left panels}: Evolution of $Z=0$ models (with rotation: continuous lines; without rotation: dotted lines) in the Hertzsprung Russell diagram. The grey area shows the zone of the diagram where He burns in the core of the rotating models.
{\it Right panel}: Evolution of the $\Omega/\Omega_\mathrm{crit}$ ratio during the MS. All the models start the MS with $\upsilon_\mathrm{eq}=800$ km s$^{-1}$, except the 9  M$_\odot$ which starts the MS with 500 km s$^{-1}$. Figures taken from 
\citet{Ekstrom2008a}.
}
\label{fig2}
\end{figure}

Recently \citet{Ekstrom2008a} presented a grid of Pop III stellar models including the effects of rotation. 
Evolutionary tracks of non-rotating and rotating Pop III stellar models are shown in Fig.~\ref{fig2}
(left panel). The models are computed until the end of the core Si-burning, except the 9 M$_\odot$ that has developed a degenerate core before carbon ignition and has thus been stopped then, and the 15 M$_\odot$ model that has been stopped at the end of O-burning also because of a too degenerate core at that time. An initial velocity of 800 km s$^{-1}$ on the ZAMS was chosen.
For the 60 M$_\odot$ model, this corresponds to a value of $\upsilon/\upsilon_{\rm crit}=0.52$ on the ZAMS, which is slightly superior to 0.4, the value required at solar metallicity to obtain averaged velocities during the MS phase corresponding to observed values. 

We notice that the ZAMS is shifted toward lower effective temperature and luminosity with respect to the non-rotating case as was the case at higher metallicities\footnote{We recall that this shift is due to the sustaining effect of the rotation: the gravity is counter-balanced both by the gas pressure and the centrifugal force in such a way that the star behaves like a lower mass one.}. Then, when the evolution proceeds, the tracks become more luminous, and the main-sequence turn-off is shifted to cooler temperature: the core of the rotating models is refueled by fresh H brought by the mixing. It thus grows, leading to an enhancement of the luminosity.

The onset of the CNO cycle described above can be seen in the HRD:  the tracks evolve toward the blue side of the diagram, until the energy provided by the CNO cycle stops the contraction and bends the tracks back in the usual MS feature. In the rotating 9 M$_\odot$ model, this happens at an age of 12.2 Myr (when the central H mass fraction is $X_\mathrm{c}=0.439$) while in the non-rotating one it happens a little earlier, at an age of 10.9 Myr (but at a similar burning stage: $X_\mathrm{c}=0.439$). In the case of the non-rotating 15 M$_\odot$ model, it happens after merely 1.5 Myr ($X_\mathrm{c}=0.695$), while it takes 2 Myr ($X_\mathrm{c}=0.677$) in the case of the rotating one. 

After central H exhaustion, the core He-burning phase (CHeB) starts right away: the core was already hot enough to burn a little He during the MS and does not need to contract much further.
This prevents the models to start a redward evolution, so they remain in the blue part of the HRD at the beginning of CHeB. Then, something particular happens to the rotating models: because of rotational mixing, some carbon produced in the core is diffused toward the H-burning shell, allowing a sudden ignition of the CNO-cycle in the shell. This boost of the shell leads to a retraction of the convective core and a decrease of the luminosity. At the same time, it transforms the quiet radiative H-burning shell into an active convective one. Some primary nitrogen is produced
(see more on that point in the next section). 

All the models, except the 9 M$_\odot$, reach the critical velocity during the MS phase (see the right panel of Fig.~\ref{fig2}). Once at critical limit, all the models remain at the critical limit until the end of the Main-Sequence phase. On the right panel of Fig.~\ref{fig2}, the 85 and especially the 200 M$_\odot$ seem to depart from $\Omega/\Omega_\mathrm{crit}=1$, but this is due to the limit shown here being only the $\Omega$-limit, where the centrifugal force alone is taken into account to counterbalance the gravity. In the two above models however, the radiative acceleration is strong and the models reach the so-called $\Omega\Gamma$-limit, that is the second root of the equation giving the critical velocity: $\vec{g_\mathrm{eff}}\,[1-\Gamma]=\vec{0}$ \citep{Maeder2000}. The true critical velocity is lowered by the radiative acceleration, and though the $\Omega/\Omega_\mathrm{crit}$ ratio plotted becomes lower than 1, these models are actually at the $\Omega\Gamma$ critical limit and remain at this limit till the end of the MS phase. At central H exhaustion, the moderate inflation of the radius brings the surface back to sub-critical velocities. 
The mass which is lost by the mechanical winds amounts only to a few percents of the initial stellar mass and thus does not
much affect neither their evolution, nor their nucleosynthetic outputs. Much more mass can be lost by mechanical mass loss (see next subsection) when the effects of magnetic fields are accounted for as prescribed in the Tayler-Spruit dynamo theory \citep{Spruit2002}, or when the metallicity is non-zero.

\subsection{Strong mass loss in Pop III stars?}

\begin{figure}[t]
\begin{center}
 \includegraphics[width=1.0\textwidth]{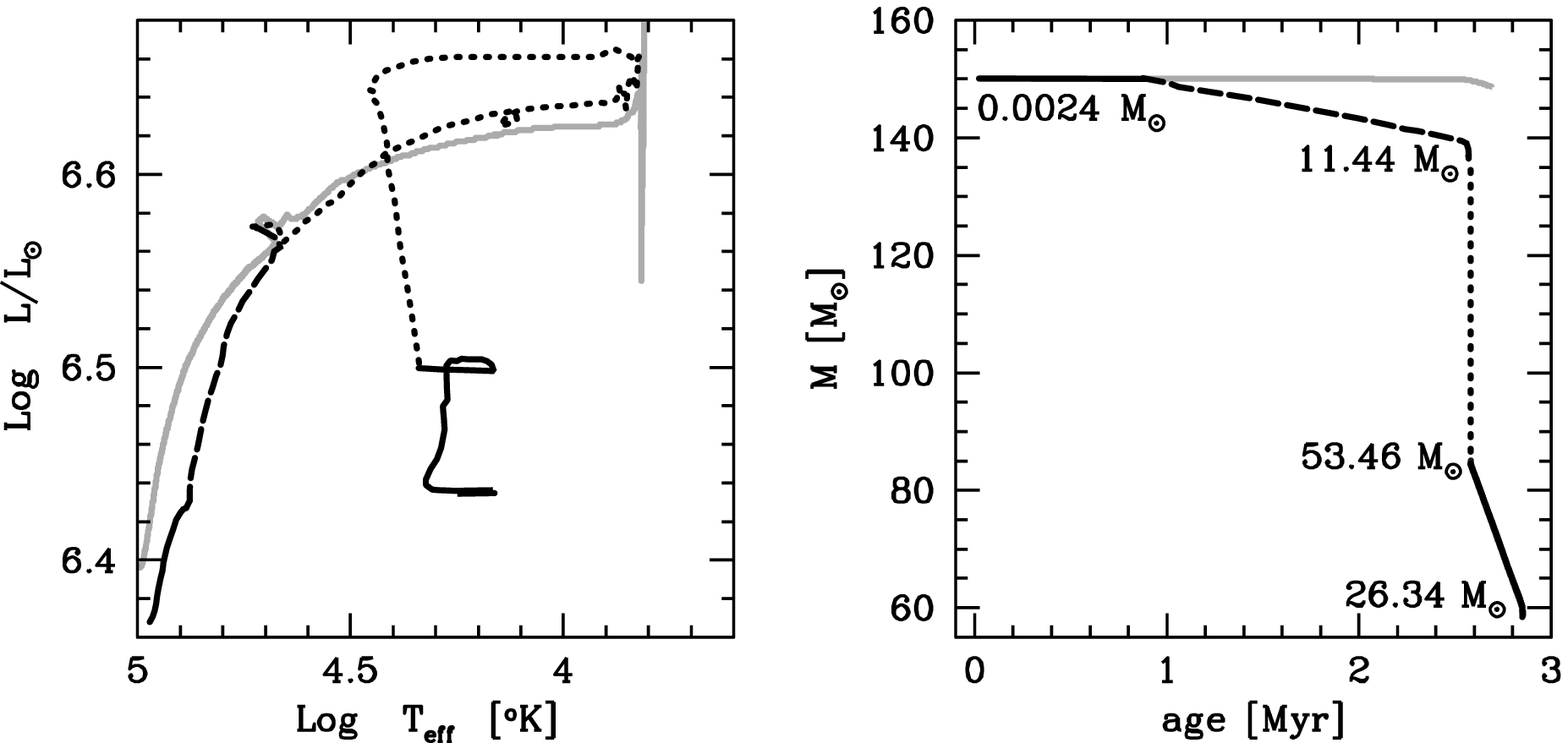}
 \caption{ \textit{Left panel}: evolution in the Hertzsprung-Russell diagram; \textit{Right panel}: evolution of the mass of the model. The mass indicated is the mass lost at each stage, not a summation. 
Black line: rotating model; \emph{continuous part}: beginning of MS ($X_{\rm c}=0.753$ down to 0.58; \emph{dashed part}: rest of the MS; \emph{dotted part}: beginning of core He-burning phase ($Y_{\rm c}=1.00$ down to 0.96); \emph{continuous part}: rest of the He-burning. Grey line: non-rotating model for comparison.Figures taken from \citet{Ekstrom2008a}.}
   \label{fevol150}
\end{center}
\end{figure}

According to \citet{Heger2003}, the fate of single stars depends on their He-core mass ($M_{\alpha}$) at the end of the evolution. They have shown that at very low metallicity, the stars having 64 M$_{\odot} < M_{\alpha} < 133$ M$_{\odot}$ will undergo pair-instability and be entirely disrupted by the subsequent supernova. This mass range in $M_{\alpha}$ has been related to the initial mass the star must have on the main sequence (MS) through standard evolution models: 140 M$_{\odot} < M_{\rm ini} < 260$ M$_{\odot}$. However the link between the masses of the He-core mass and the initial mass
can be very different depending on the physics considered.  \citet{Ekstrom2008a} showed that
a Z=0, 150 M$_{\odot}$ stellar model, having an initial ratio between the equatorial velocity and the critical one equal to $\upsilon_{\rm ini}/\upsilon_{\rm crit} = 0.56$, computed accounting
for the Tayler-Spruit dynamo mechanism \citep{Spruit2002} and the effects of wind anisotropy \citep{Maeder1999a} will lose such great amount of mass that it will reach the end of the core He-burning phase with a mass of $M_{\alpha}$ too small to go through a pair-instability process.

In Fig. \ref{fevol150}, we present the evolution in the HR diagram (left panel) and the evolution of mass with time (right panel) for the 150 M$_\odot$ stellar model of  \citet{Ekstrom2008a}. The grey line shows a non-rotating model computed with the same physics for comparison.
During its whole evolution up to the end of core He-burning, the non-rotating model loses only 1.37 M$_{\odot}$. This illustrates the weakness of radiative winds at $Z=0$.
The evolution of the rotating model (black line) can be described by four distinct stages:
\begin{enumerate}
\item \emph{(continuous part}, lower left corner) The model starts its evolution on the MS with only radiative winds, losing only a little more than 0.002 M$_{\odot}$. During this stage, the ratio
of the surface velocity to the critical one increases quickly, mainly because of the strong coupling exerted by the magnetic fields.

\item \emph{(dashed part)} When the central content of hydrogen is still about 0.58 in mass fraction, the star reaches the critical velocity and starts losing mass by mechanical mass loss. It remains at the critical limit through the whole MS, but the mechanical wind removes only the most superficial layers that have become unbound, and less than 10\% of the initial mass is lost at that stage (11.44 M$_{\odot}$). The model becomes also extremely luminous, and reaches the Eddington limit when 10\% of hydrogen remains in the core. Precisely, it is the so-called $\Omega\Gamma$-limit that is reached here. 

\item \emph{(dotted part)} The combustion of helium begins as soon as the hydrogen is exhausted in the core, then the radiative H-burning shell undergoes a CNO flash, setting the model on its redward journey. The model remains at the $\Omega\Gamma$-limit and loses a huge amount of mass. The strong magnetic coupling keeps bringing angular momentum to the surface and even the heavy mass loss is not able to let the model evolve away from the critical limit. The mass lost during that stage amounts to 53.46 M$_{\odot}$. When the model starts a blue hook in the HR diagram, its surface conditions become those of a WR star ($X_{surf} < 0.4$ and $T_{\rm eff} > 10'000$ K). The luminosity drops and takes the model away from the $\Gamma$-limit, marking the end of that stage.

\item \emph{(continuous part)} The rest of the core He-burning is spent in the WR conditions. The mass loss is strong but less than in the previous stage: another 26.34 M$_{\odot}$ are lost.
\end{enumerate}

At the end of core He-burning, the final mass of the model is only $M_{\rm fin}=$ 58 M$_{\odot}$, already below the minimum $M_{\alpha}$ needed for PISN ($M_{\alpha} \geq 64 $ M$_{\odot}$). Note that the contraction of the core after helium exhaustion brings the model back to critical velocity, so this value for $M_{\rm fin}$ must be considered as an upper limit.

This result shows that a fast rotating Pop III 150 M$_\odot$ may avoid to explode as a PISN. Of course it is by far not certain that the conditions required for such a scenario to occur are met in the first stellar generations but it underlines the fact that fast rotation may drastically change the picture. 

The yields of PISN is discussed by \citet{HW2005} and are dominated by alpha nuclei. The iron-to-oxygen ratio show a strong increase with the initial mass. The PISN yields show a strong decrease beyond the iron group,
have essentially no s-process and r-process element production.
These nucleosynthetic signatures of PISN are not observed in the most metal poor halo stars. Is this due to the above scenario? To the fact that the signature was very quickly erased by the next generations of stars?\footnote{Maybe the metal-poor stars we observe are enriched by more SNe than we actually think, and the later contributions are masking the primordial ones.} Or were such high mass stars not formed? These various hypotheses cannot be disentangled at the present time, but the observation of more and more metal-deficient stars will probably provide elements of response to these questions.

\section{SUPERNOVAE AND GAMMA RAY BURSTS}

\subsection{The type Ib and Ic supernovae}
\label{IBC}

Type Ib supernovae are core-collapse supernovae whose spectrum shows no hydrogen lines. The spectra of type Ic show no hydrogen and helium lines \citep[see e.g.][]{Wheeler1987a,Nomoto1994a}. The progenitors of these core collapse supernovae are thus believed to be stars stripped of their original H-rich envelope for type Ib's and also of their He-rich envelope for type Ic's. Progenitors are therefore naked stellar cores as e.g. Wolf-Rayet stars. The observed frequency at solar metallicity of type Ibc supernovae is about 20\% the frequency of type II supernovae \citep[see e.g.][]{Cappellaro1999a}, which represents a significant fraction of all core collapse supernovae.  In four cases, the typical spectrum of a type Ic supernova has been observed associated with a long soft gamma ray burst (GRB) event \citep{Woosley2006a}, indicating a privileged link between type Ic's and the most powerful supernova explosions observed in the Universe. 

\begin{figure}[t]
\centering
\includegraphics[angle = 0, width=9cm]{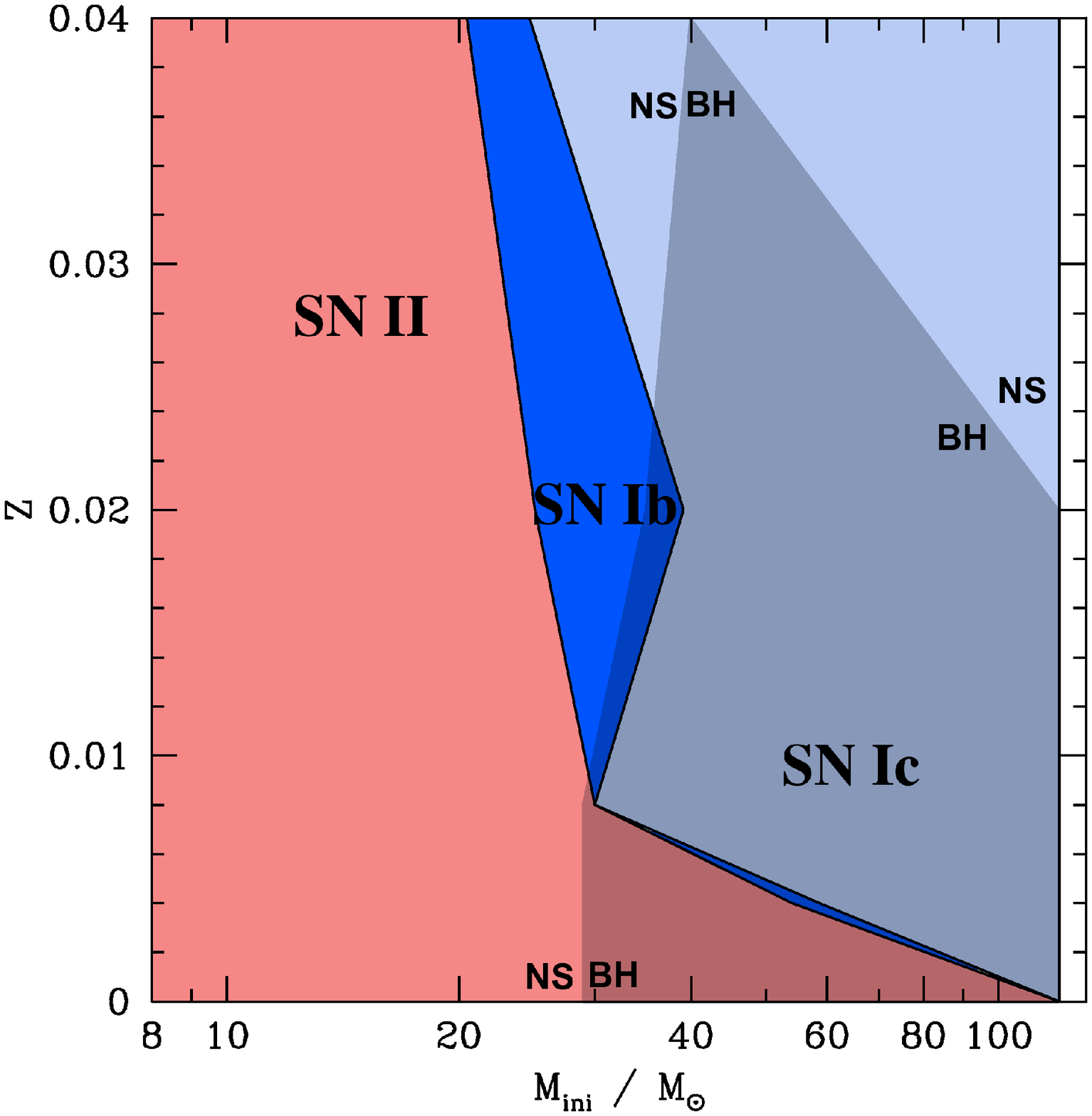}
\caption{Ranges of masses of different types of supernovae at different metallicities. The shading on the right indicates the area where formation of a black hole (BH) is expected ; elsewhere, the remnant is a neutron star (NS). Figure taken from \citet{Georgy2009}.}
\label{SNtype}
\end{figure}

Fig. ~\ref{SNtype} shows the different supernova types expected for various initial masses and metallicities as given
from rotating models by \citet{Georgy2009}. 
Type Ic SNe cover a much broader range of initial masses than type Ib. This trend reflects that most stars which, at a given stage, in their evolution have peeled off their H-rich envelope do not stop at that stage. Their evolution drives them beyond, up to the stage where most of their He-rich layers have been peeled off.

One immediately sees that the number fraction of type Ibc to type II supernovae should increase with  metallicity.
This is an effect of the stronger winds at higher metallicities, which favor the formation of H-free stars at the end 
of their evolution. Very interestingly, this expected increase of the Ibc/II ratio has been put in evidence by 
\citet{Prantzos2003a}. When one compares the theoretical ratios obtained from single stars with the observed ones,
a good fit is obtained only for the rotating models \citep{paperXI2005, Georgy2009}.

However such an observed trend can also be reproduced invoking close binary-star evolution \citep[see for example][]{Podsiadlowski1992a,Vanbeveren2007a,Eldridge2008}. In this scenario, the hydrogen-rich envelope is removed either through a Roche lobe overflow process or during a common envelope phase, producing a WR star.
Thus we face here the situation where two very different models (single stars with rotation/close binary evolution with mass transfer) are both able to give a reasonable fit to the data. Actually both scenarios probably contribute to the observed populations of type Ibc supernovae. However, it would be interesting to know their relative importance and how their relative importance changes with the metallicity. It might be that both scenarios predict different behaviors for the way the frequencies of the type Ib and type Ic SNe vary as a function of the metallicity. Predictions of single star models for the variations
with the metallicity of the seperate ratios of type Ib/II and of Ic/II are discussed in \citet{Georgy2009}. 


\subsection{The progenitors of long soft Gamma Ray Bursts}
\label{grb}

One of the most promising model for the progenitors of the long soft Gamma Ray Bursts  (GRB) is the collapsar model by \citet{Woosley1993a}. In this model, GRB are the product of the evolution of massive stars which, at the end of their
evolution, produces a fast rotating black hole. In addition,  for the GRB to be visible, the massive star should have lost
its envelope. This model received a strong support
when observations showed that    
there is a connection between so-called long-soft GRB events and type Ic supernovae \citep{Woosley2006a}. This kind of supernova can be produced by both WC or WO stars. In this way, WC and WO stars are natural candidates to be GRB progenitors. However, GRBs are primarily found in metal--poor environments \citep{Modjaz2008a}, while type Ic's SNe appear mainly at high metallicity.  Many physical reasons have been invoked to explain why GRBs seem to occur only at low metallicities. Among them are the following \citep[see the review by][]{Woosley2006a}:
\begin{itemize}
\item At low metallicity, stellar winds (even during the WR phases) are weaker, thus bringing away small quantities of angular momentum. Black holes are also more easily formed since higher final masses are obtained.
\item As already emphasized previously, in absence of any magnetic coupling, the transport of the angular momentum is
mainly due to meridional currents, while the mixing of the elements is mainly driven by shear instabilities.
At low metallicity the transport of angular momentum between the core and the envelope is less efficient than at high metallicities
(see Fig.~\ref{UdiffZ}), because of slower meridional currents in metal poor stars. This favors a higher angular momentum content
of the core when the star is at the end of its nuclear lifetime, it also favors steeper $\Omega$-gradient in the interior triggering
more efficient shear instabilities and thus chemical mixing.
\item Since the chemical mixing due to rotation is more efficient at low $Z$, homogeneous evolution is more easily obtained in metal-poor regions. Homogeneous evolution allows massive stars to produce a type Ic SN event without having to lose large amounts of mass (and thus of angular momentum). Indeed, a perfectly homogeneous evolution (actually never realized) would allow the formation of a pure CO core (and then lead to a type Ic SN event) at the end of the core He-burning phase without the need for the star to lose any mass! Note that such homogeneous evolution allows massive stars to produce
much more ionizing photons than standard one \citep[see e.g.][]{Meynet2008b}.
\item The distribution of initial velocities at low metallicity might contain more fast rotators than at high metallicities \citep[see Fig. 9 in ][]{Martayan2007a}.
\end{itemize}

It is interesting to compare the observed GRB frequency with the observed frequency of  potential candidates. First, as mentioned above, type Ic supernovae at low metallicity do appear interesting candidates. From Fig.~\ref{GRBRate}, one can see that the observed rate of type Ic supernovae from single star models is still above the estimated number ratio GRB / core collapse supernovae (CCSNe) even when one only considers low-metallicity, type Ic SNe. Here it is supposed that the formation of a BH does not prevent a SN event. In case, only rare circumstances would allow a SN to occur when a black hole is formed, then the situation might be very different.

Other interesting candidates are the WO stars that primarily occur at low metallicity. These stars will explode as a type Ic SN. The frequency of type Ic's SNe with WO star progenitors is shown in Fig.~\ref{GRBRate}. The expected rate is only marginally compatible with the observed GRB rate (assuming that the aperture angle of the bipolar jet is very small, typically around $1^\mathrm{o}$). This conclusion has been obtained by \citet{Hirschi2005a}, so even restraining the progenitors of GRB to WO stars would still not match the observed frequency of GRBs. 

From the above discussion, it can be deduced that exploding as a type Ic SN in metal-poor regions (or having a WO progenitor) is not a sufficient condition for obtaining a GRB. Physical characteristics shared by a subsample of the metal poor type Ic events exists that are needed to obtain a GRB event. Probably this physical characteristic is the high angular momentum in the core \citep{Yoon2006, Woosley2006b}. 

If GRB would only occur for initially very fast-rotating stars, rotating so fast that these stars would follow a homogeneous evolution \citep{Maeder1987a,Yoon2006,Woosley2006b,paperGRBWa2007}, can we expect to find any peculiar feature in the chemical composition of the ejecta testifying this previous homogeneous evolution? Or in other words, is there any difference in the chemical composition of the ejecta between a type Ic having ``normal WC or WO" progenitors and those arising from a model that followed a homogeneous evolution during the MS phase? The answer is probably no. It has been shown by
\citet{Georgy2009} that the differences in the masses of He, CNO elements, and $Z$ are very small. Thus there is little chance from the observations of the composition of the ejecta in CNO elements to be able to distinguish between a normal and a homogeneous evolution.

A point that would be interesting to check is the following: since the star rotates very fast during a great part of its evolution, it will produce anisotropic stellar winds (see Fig. \ref{PolWind}), typically bipolar winds and probably equatorial mass loss when the critical limit is reached \citep{Maeder1999a}. These anisotropic winds will shape the circumstellar environment of the star \citep[][and see also Fig.~\ref{PolWind}]{vanMarle2008a} and it might be that some traces of the resulting particular morphology will still be present at the time of the SN event (for instance, equatorial mass loss probably gives rise to a slow equatorial expanding disk whose traces might still be present when the star explodes). In that case the circumstellar environment of GRBs may be peculiar. This can in turn have an impact on some features in the spectrum. More generally, the circumstellar environment of stars that have been in a not too remote past very fast rotators may also be characterized by such features. 

\begin{figure}
\centering
\includegraphics[angle=0,width=9cm]{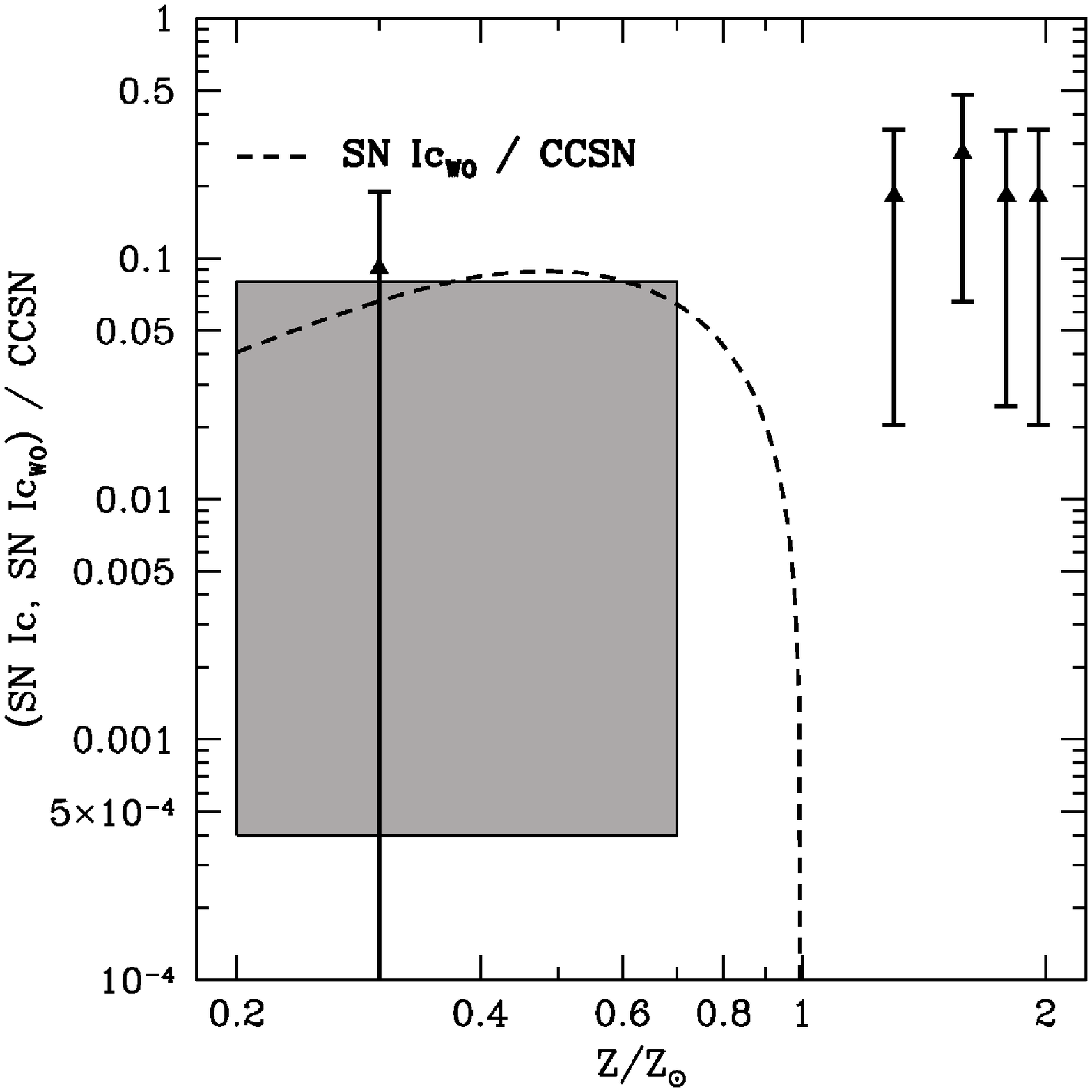}
\caption{Observed rate of type Ic SNe (triangles with error bars) and predicted rate of type Ic SNe whose progenitor is a WO star (dashed line) with respect to the total number of core collapse SNe (CCSNe). The grey rectangle represents the extension in metallicity \citep{Modjaz2008a} and rate \citep{Podsiadlowski2004a} of GRB events. Figure taken from \citet{Georgy2009}.}
\label{GRBRate}
\end{figure}

\begin{figure}
\centering
\includegraphics[angle=0,width=9cm]{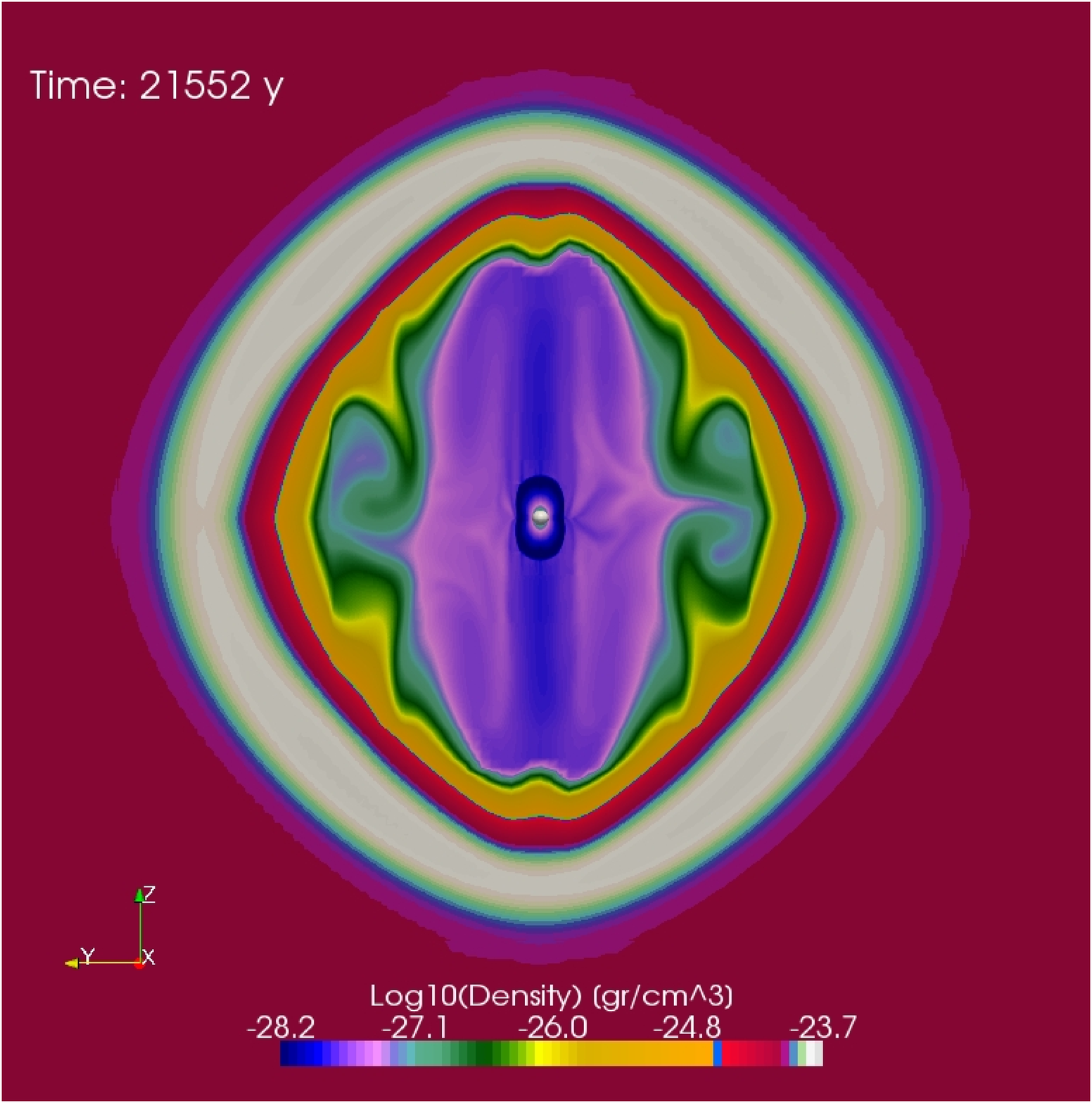}
\caption{Young bi-modal shaped nebula blown from the highly aspherical, polar dominated winds around a $20\,\mathrm{M}_\odot$ star at $Z=10^{-5}$. The star has evolved during $21552\,\mathrm{yr}$ since it has left the ZAMS. The maximal extension of the wind is $\sim1.5\cdot 10^{18}\mathrm{cm}$. This model was computed by the A-Maze code \citep{Walder2000a}. Figure taken from \citet{Georgy2009}.}
\label{PolWind}
\end{figure}

\section{STELLAR NUCLEOSYNTHESIS}

\subsection{Rotation and stellar yields at Z=0.020}

Figure \ref{yres} displays the total stellar yields (i.e. the mass of an element newly synthesized by the star and ejected into
the interstellar medium)
divided by the initial mass of the star, $p^{\rm{tot}}_{im}$, 
as a function of its initial mass, $m$, for the non--rotating (left) 
and rotating (right) models \citep{Hirschi2005b}. 
The different shaded areas correspond from top
to bottom to $p^{\rm{tot}}_{im}$ for $^{4}$He, $^{12}$C, $^{16}$O and
the rest of the heavy elements. The fraction of the stars locked in the
remnant as well as the expected explosion type are shown at the bottom. 
The dotted areas show the wind contribution for 
$^{4}$He, $^{12}$C and  $^{16}$O. 

The remnant masses have been estimated using the relation between
the mass of the carbon-oxygen core and the mass of the remnant given
by \citet{Maeder1992a} \citep[see details in][]{Hirschi2004a, Hirschi2005b}. This value is quite uncertain and may affect
the yields, especially those of the iron-group elements and heavier. The yields shown in
Fig.~\ref{yres} are those obtained before explosive nucleosynthesis occurs.

For $^4$He (and other H--burning products like $^{14}$N), the wind
contribution increases with mass and dominates for $M \gtrsim 22$
M$_{\odot}$  for rotating stars and $M \gtrsim 35$
M$_{\odot}$  for non--rotating stars, i. e. for the stars which enter
the WR stage. For very massive
stars, the SN contribution is negative and this is why
$p^{\rm{tot}}_{^4\rm{He} m}$ is smaller than 
$p^{\rm{wind}}_{^4\rm{He} m}$.
In order to eject He--burning products, a star must not only become a WR 
star but must also become a WC star. 
Therefore for $^{12}$C, the wind contributions only start to be 
significant above
the following approximative mass limits: 
30 and 45 M$_{\odot}$ for rotating and
non--rotating models respectively. 
Above these mass limits, the
contribution from the wind and the pre--SN are of similar importance.
Since at solar metallicity, no WO star is produced, for 
$^{16}$O, as for heavier elements, the wind contribution remains very small in the whole mass range. 

So we see that, below $\sim 30$ M$_{\odot}$, rotation increases the  
total metal yields, $Z$, and in particular the yields of carbon and oxygen 
by a factor of 1.5--2.5.
As a rule of thumb, the yields of a rotating 20 M$_\odot$ star are similar
to the yields of a non--rotating 30 M$_\odot$ star, at least for the light
elements shown in  Fig.~\ref{yres}. We see therefore that at $Z=0.020$, the effects of rotation
are qualitatively  similar to those obtained by models with larger convective cores as for instance those
computed with an overshoot. This is of course not surprising since rotational mixing tends also
to increase the core size at a given evolutionary phase.
For very massive stars ($\sim 60$ M$_{\odot}$), 
rotation increases the yield of helium but does not significantly 
affect the yields of heavy elements. The similarities of the results between the rotating
and non-rotating models for the  60 M$_{\odot}$ model  comes from the fact that, although the mass is not lost
at the same stages in both series of models, the total mass loss is nevertheless not very different. 
The non-rotating model goes through an LBV phase after the MS phase where a significant
fraction of the mass is lost. The rotating model enters into the WR phase already during the MS phase
and no LBV phase has been accounted.

Figure \ref{yimf} presents the stellar yields convolved with the Salpeter 
initial mass
function (IMF) ($dN/dM\propto M^{-2.35}$). This reduces the importance
of the very massive stars. Nevertheless, the differences between rotating and
non--rotating models remain significant, especially around 20 M$_\odot$.
Despite these differences, it will be difficult to discriminate between these two series of yields
from observations of the evolution of the abundances as a function of the increasing metallicity, because
in such comparisons, the yields are only one of the ingredients of the chemical evolution models. Other factors
as the star formation history, the initial mass function, the rate of infall and outflow etc... will blurr the picture.
More constraining are the early phases of the chemical evolution of the galaxies during which only massive stars
had time to contribute.

\begin{figure*}[!tbp]
\centering
\includegraphics[width=8.8cm]{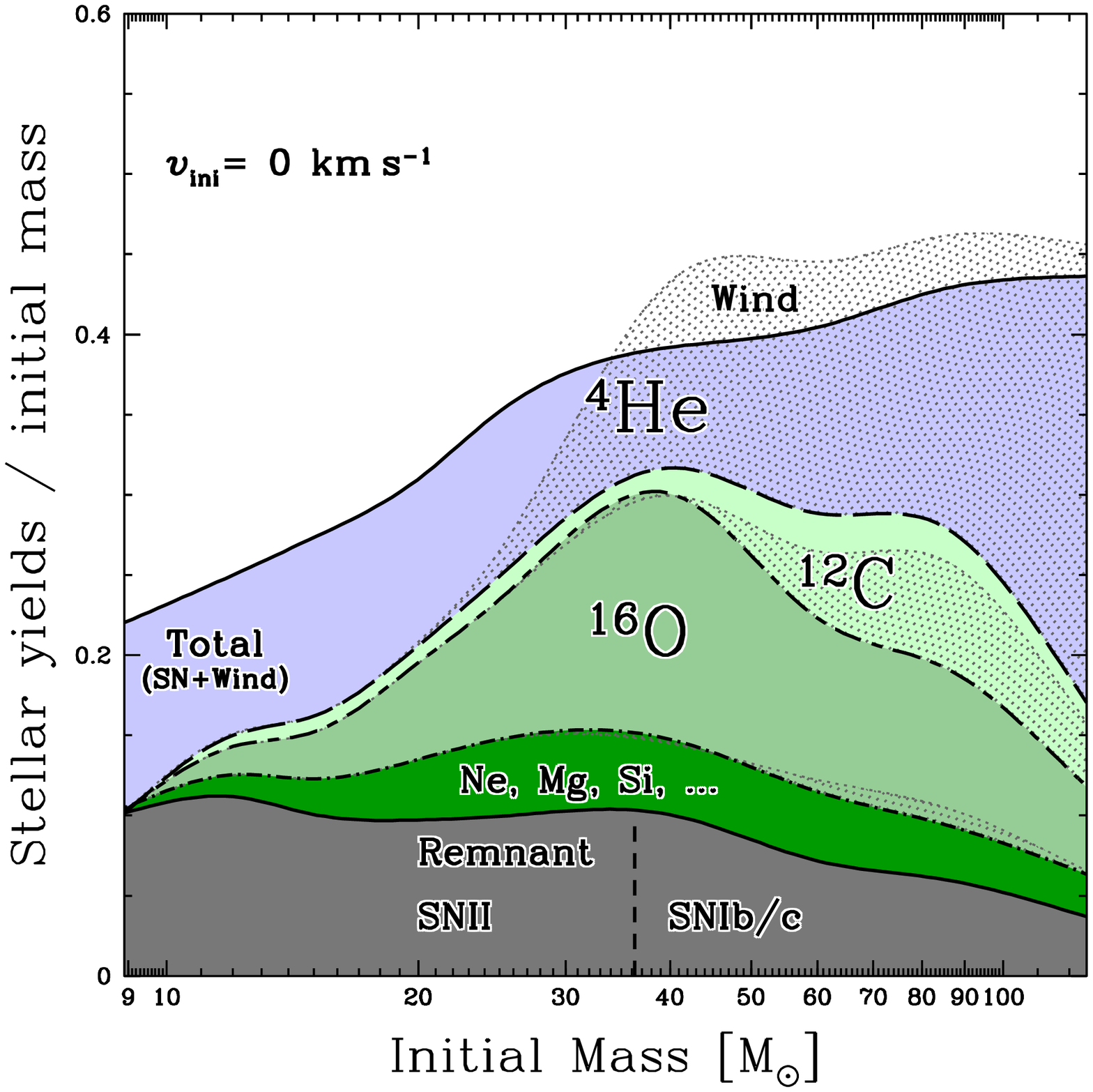}\includegraphics[width=8.8cm]{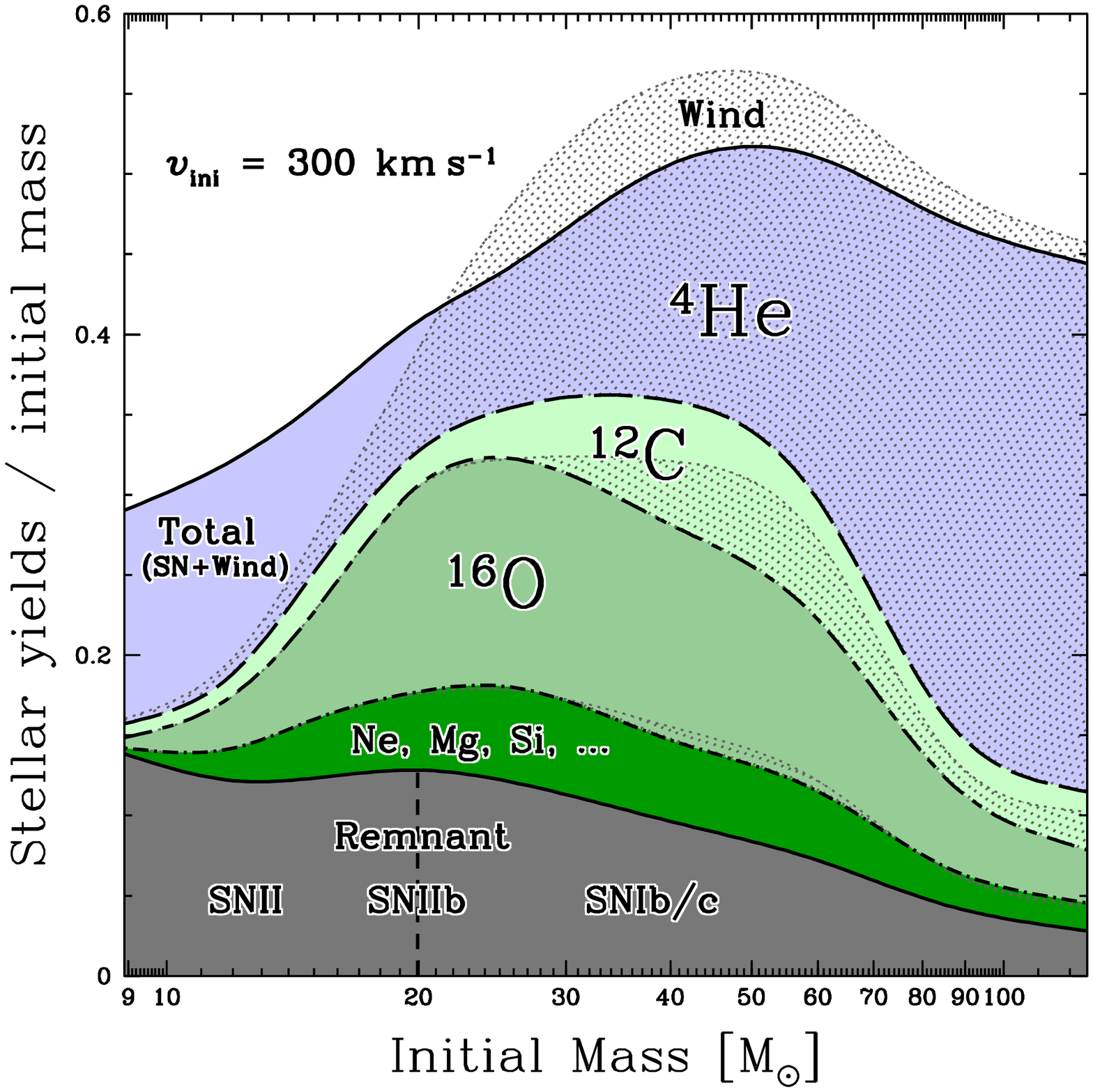}
\caption{Stellar yields divided by the initial mass, 
$p^{\rm{tot}}_{im}$, as a function of the initial mass
for the non--rotating 
 (left) and rotating  (right) 
models at solar metallicity. 
The different total yields (divided by $m$) are shown as 
piled up on top of 
each other and are not overlapping. $^4$He
yields are delimited by the top solid and long dashed lines (top
shaded area),
$^{12}$C yields by the long dashed and short--long dashed lines, 
$^{16}$O  yields by the short--long dashed and dotted--dashed lines and 
the rest of metals by the dotted--dashed and bottom solid lines. The
bottom solid line also represents the mass of the remnant 
($M^{\rm{int}}_{\rm{rem}}/m$). 
The corresponding SN explosion type is also given.
The wind contributions  are superimposed on these total yields for the
same elements between their bottom limit and the dotted line above
it. Dotted areas help quantify the fraction of the total yields
due to winds. Note that for
$^{4}$He, the total yields is smaller than the wind yields due to
negative SN yields. Figure taken from \citet{Hirschi2005b}.}
\label{yres}
\end{figure*}

\begin{figure*}[!tbp]
\centering
\includegraphics[width=8.8cm]{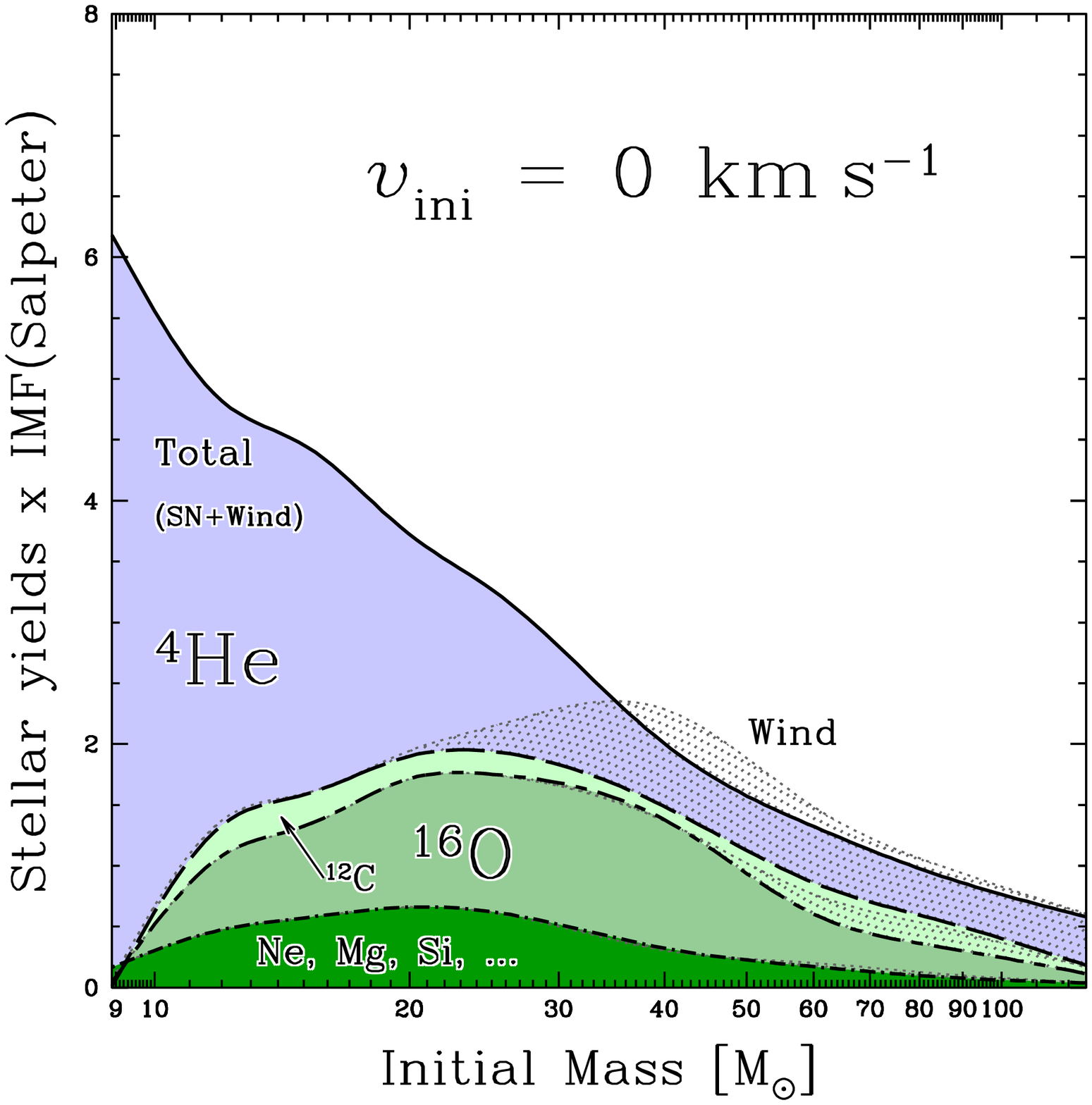}\includegraphics[width=8.8cm]{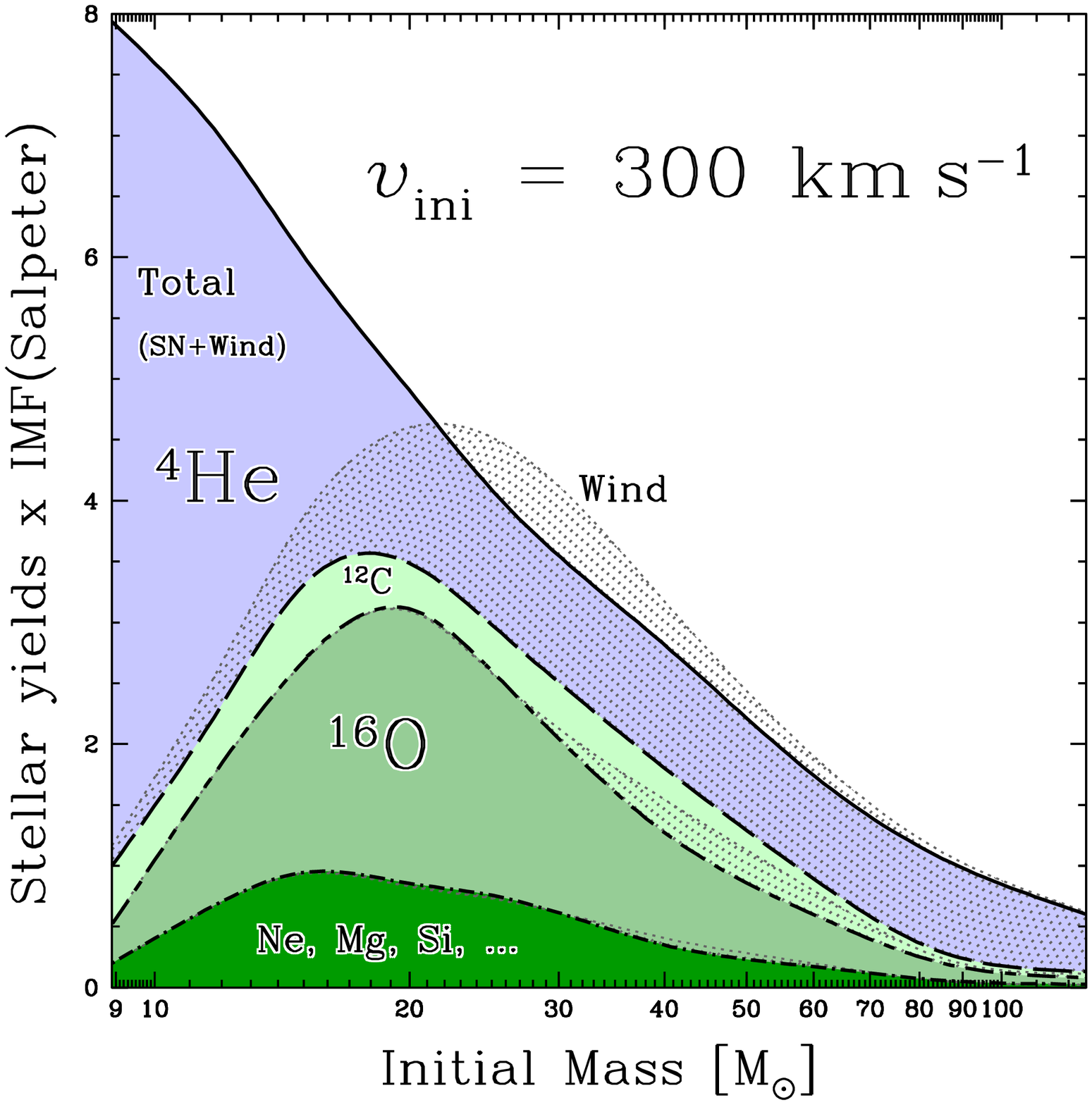}
\caption{Product of the stellar yields, $mp^{\rm{tot}}_{im}$ by 
Salpeter's
IMF (multiplied by a arbitrary constant: 1000 x $M^{-2.35}$), 
as a function of the initial mass
for the non--rotating (left) and rotating 
 (right) 
 models at solar metallicity. The different shaded areas correspond 
 from top
to bottom to $mp^{\rm{tot}}_{im}$ x 1000 x $M^{-2.35}$ for $^{4}$He, $^{12}$C, $^{16}$O and
the rest of the heavy elements. The dotted areas show for 
$^{4}$He, $^{12}$C and  $^{16}$O the wind contribution. Figure taken from \citet{Hirschi2005b}.}
\label{yimf}
\end{figure*}

\subsection{Impact on the early chemical evolution of galaxies}

Very metal poor halo stars
 have formed (at least in part) from matter  ejected by very metal poor massive stars
 \citep[see e.g.][]{Chiappini2006, Chiappini2008}. 
 By very metal poor stars we mean here Pop III stars and/or stars with an initial metal abundance
 lower or equal to that of the metal poor halo considered.
 Their surface
 compositions thus reflect
 the nucleosynthesis occurring in the first generations of massive stars
 (provided of course that no other processes as accretion or in-situ mixing
  mechanism has changed their surface composition).

 Interestingly, many observations of these stars show puzzling features.
 Among them let us cite the two following ones shown in Fig.~\ref{NOCO}: spectroscopic observations 
 \citep[e.g.][]{Spite2005} indicate a primary production
 of nitrogen over a large metallicity range;
 Halo stars with log(O/H)+12 inferior to about 6.5 present higher C/O ratios than
 halo stars with log(O/H)+12 between 6.5 and 8.2 \citep{Akerman2004, Spite2005}.

 Fast rotating
 massive stars are very interesting candidates
 for producing primary nitrogen at low metallicity and in a very short time delay as is required by
 the observations of the N/O plateau shown in the upper panel of Fig.~\ref{NOCO}. 
Primary nitrogen production occurs in both high mass and
intermediate mass stars provided they rotate and  have a metallicity below about
0.001 \citep{NprimL2002, paperVIII2002}. The initial rotation needed to obtain sufficient primary nitrogen production corresponds to values between
50 and 70\% of the critical rotation on the ZAMS. These values are 
above the usual ratio adopted for
solar metallicity models, which is 40\%, but are still far from extreme values near the critical limit. Let us note that
recent models of Pop III star formation seem to support very high initial rotational velocities \citep{Stacy2010}.

For a given initial velocity, primary nitrogen production is more efficient when
a small amount of metals is present than in stars made of primordial material. Indeed,
although rotational mixing in Pop III stars
is by far not a negligible effect, it remains at a relatively modest level due to the absence of
strong contraction at the end of the core H-burning phase. On the contrary, when $Z \ge \sim 10^{-10}$, the physical
conditions during the core H-burning phase and the core He-burning phase are so different that
a strong contraction occurs at the end of the core H-burning phase leading to steep $\Omega$-gradient, strong mixing and
important primary nitrogen production.
For metallicities higher than about 0.001, rotational mixing is not efficient enough for triggering
important primary nitrogen production (at least for the rotational velocities corresponding
to the observed ones at this metallicity) and thus rotational mixing, although still
important for explaining the surface enrichments, does not  change the stellar yields as much as at very low metallicity.

A comparison of various yields for CNO elements for Pop III stars and very metal poor stars is shown in Fig.~\ref{Yieldcompar}.
For $^{12}$C (\textit{left}) and $^{16}$O (\textit{right}), our non-rotating 15 and 25 $M_\odot$ yields compare within a factor 2 or 3 with those of \citet{Chieffi2004}. Their 25 $M_\odot$ produces also a huge amount of $^{14}$N (\textit{centre}), even a factor $\sim$ 10 higher than ours. Recent non-rotating models by \citet{Heger2010}  predict also important quantities of primary
nitrogen in some mass range, indicating that primary nitrogen production in $Z$=0 star models may not be linked to
rotation. However, according to chemical evolution model of \citet{Chiappini2006}, the production of primary nitrogen only by Pop III stars may not be sufficient to account for the observed N/O plateau shown by halo stars. Primary nitrogen production
over a metallicity range is needed. At the moment, rotating models can provide a mechanism which operates over
a sufficient extended metallicity range to account for the observations.

\begin{figure}[t]
\begin{center}
 \includegraphics[width=15cm,angle=0]{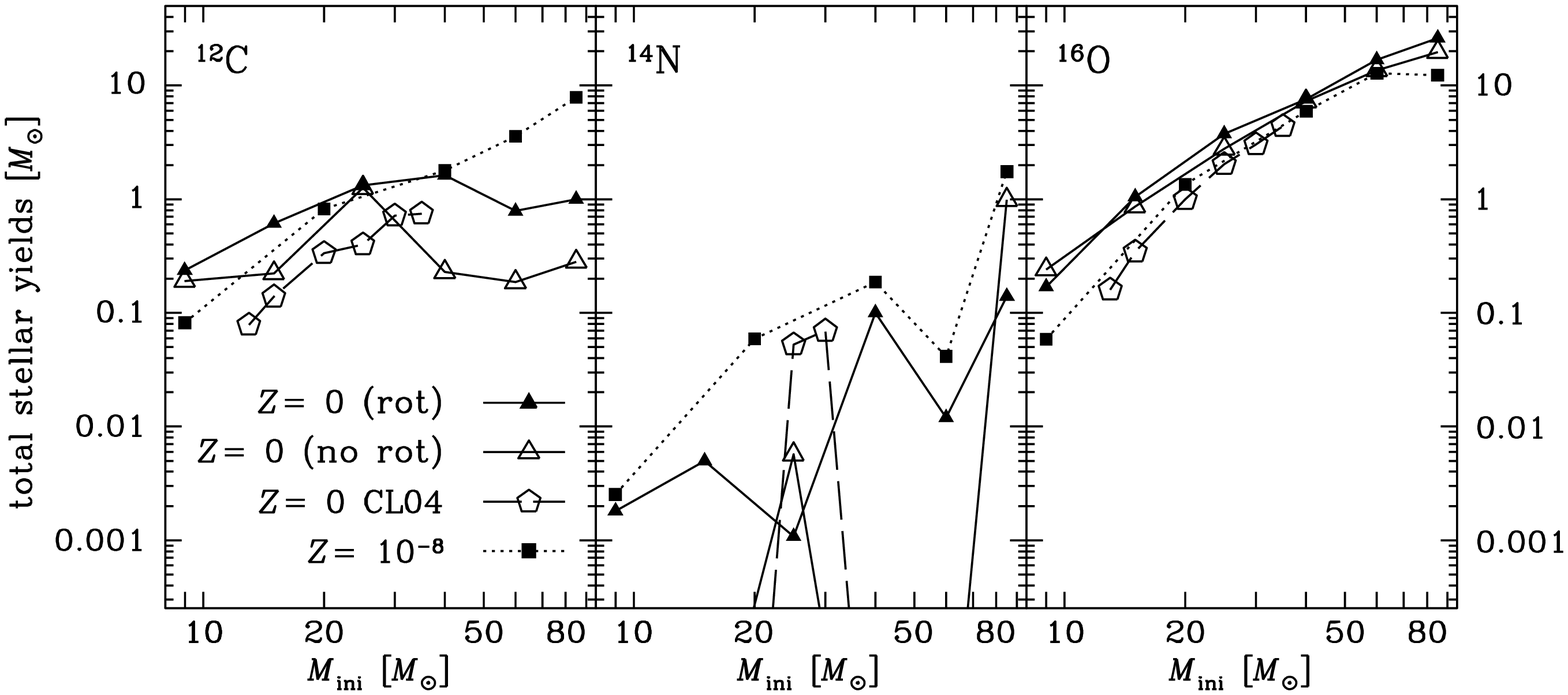} 
 \hfill
 \caption{Yields comparison between the non-rotating $Z=0$ models from \citet[CL04, open pentagons]{Chieffi2004}, the rotating $Z=10^{-8}$ models from \citet[filled squares]{Hirschi2007} and the rotating (filled triangles) and non-rotating (open triangles) $Z=0$ models from \citet{Ekstrom2008a}. \textit{Left:} $^{12}$C; \textit{centre:} $^{14}$N; \textit{right:} $^{16}$O. Figure taken from \citet{Ekstrom2008a}.}
   \label{Yieldcompar}
\end{center}
\end{figure}

When the yields of rotating massive star models are included in chemical evolution model
for the halo, nice agreement can be obtained with the observations.
It is possible to 
account for the high observed N/O ratio at very low log(O/H)+12
 values and to  reproduce the observed 
 C/O upturn mentioned just above. This is illustrated in Fig.~\ref{NOCO}, where
 predictions for the evolution of N/O and C/O of chemical evolution models using different sets of yields are compared (Chiappini et al. 2006a\footnote{The details of the chemical evolution models can be found in Chiappini et al. (2006b), where they show that such a model reproduces nicely the metallicity distribution of the Galactic halo. This means that the timescale for the enrichment of the medium is well fitted.}). We see that the observed N/O ratio is much higher than what is predicted by a chemical evolution model using the yields of the slow-rotating $Z=10^{-5}$ models from
 \citet{paperVIII2002} down to $Z=0$. When adding the yields of the fast-rotating $Z=10^{-8}$ models from 
 \citet{Meynet2006, Hirschi2007}\footnote{These models use the same physics as the models of \citet{paperVIII2002}, simply the initial rotation velocities considered are higher.}
the fit is much improved. The same improvement is found for the C/O ratio, which presents an upturn at low metallicity. These comparisons support fast rotating massive stars as the sources
of primary nitrogen in the most metal poor galactic halo stars.
 
High N/O and the C/O upturn of the low-metallicity stars are also observed in low-metallicity DLAs  
\citep[][see the crosses in Fig.~\ref{NOCO}]{Pettini2008}. We note that the observed points are
below the points for the halo stars in the N/O versus O/H plane. This may be attributed to two causes: either the observed N/O ratios observed in halo stars are somewhat overestimated or
the difference is real and might be due to different star formation histories in the halo and in DLAs. Let us just discuss these two possibilities. 

Measures of nitrogen abundances at the surface of very metal poor stars is quite challenging, much more than the measure of nitrogen in the interstellar medium as is done for the DLAs, therefore one expects that the data for DLAs suffer much smaller uncertainties than those for halo stars. 
In that respect 
the observed N/O ratios
in DLAs should be preferred to compare with the chemical evolution models. In case both DLAs and the galactic halo had the same star formation history, {\it i.e.} are the result of an intense and rapid star formation epsiode, then the chemical evolution models presented in Fig.~\ref{NOCO} can also apply to DLAs. We see that the DLA data still require models with fast rotation to be fitted
(the shift of the DLA data towards lower N/O ratios is by far not sufficient to discard fast rotating models).

Probably the star formation history in DLAs is not the same as in the halo. While in the halo we see the result of strong and rapid star formation episodes, in DLAs one might see the result of  much slower and weaker star formation episodes. In that case, both massive stars and intermediate mass stars contributed to the build up of the chemical abundances and the chemical evolution models presented in Fig.~\ref{NOCO} do no long apply to theses systems. It will be very interesting to study the results of chemical evolution models adapted to this situation and accounting for stellar yields from both massive and intermediate mass stars. Let us just mention at this stage that primary nitrogen production in metal poor intermediate mass stars is also strongly
favoured when rotational mixing is accounted for \citep{NprimL2002, paperVIII2002}\footnote{In that paper, the evolution was stopped in the
E-AGB phase and was not pursued in the TP-AGB stars, however the quantity of primary nitrogen obtained at that stage will
be a lower limit of what will be found at the end of the TP-AGB phase and thus represents some lower limit value.}.
Thus also in that case, rotation may play a key role.

\begin{figure}[t]
\begin{center}
 \includegraphics[width=3.5in,angle=0]{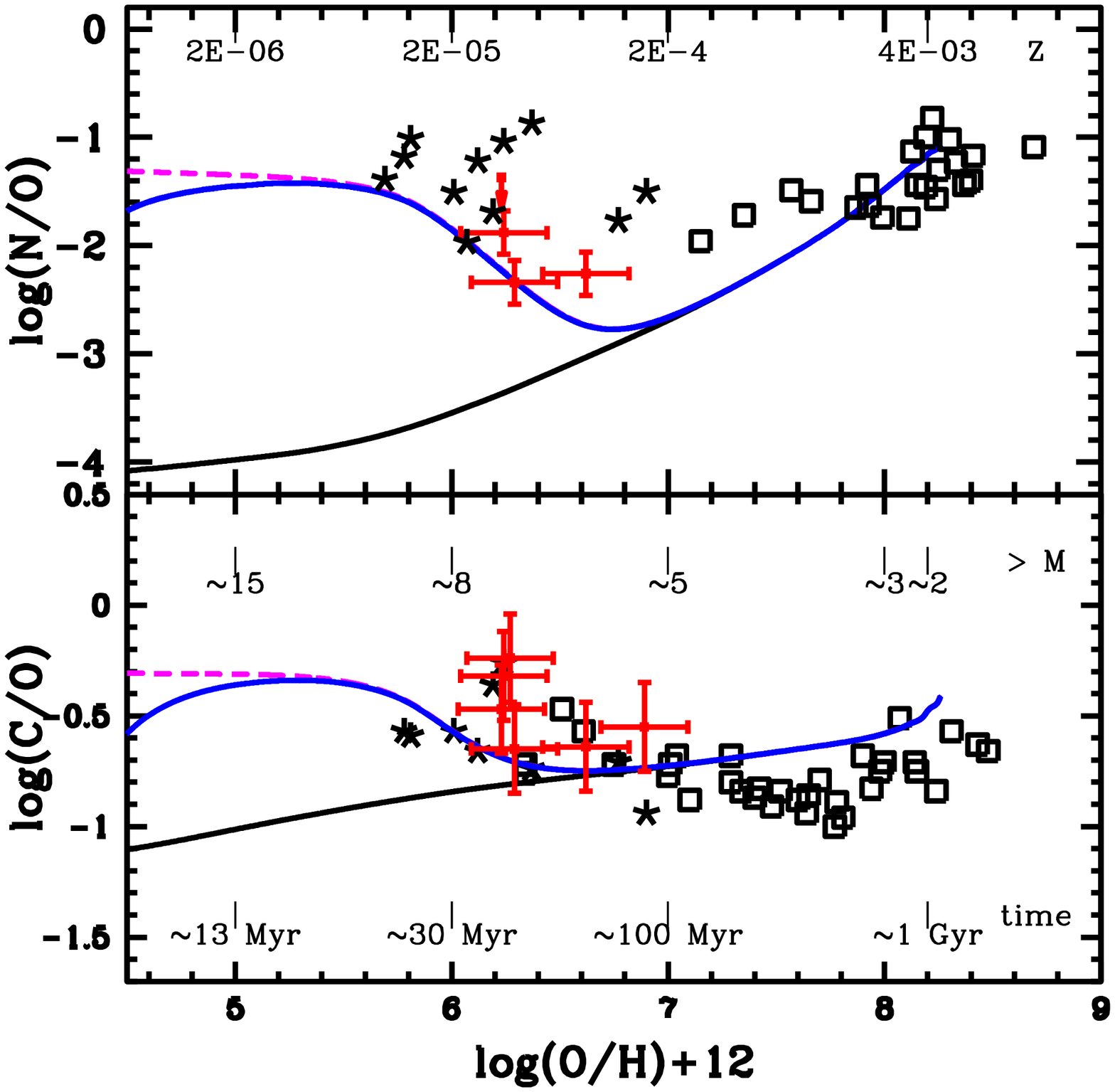} 
 \hfill
 \caption{Evolution of the N/O and C/O ratios. Data points for halo stars are from
 \citet[][open squares]{Israelian2004}
  and of \citet[][stars]{Spite2005}. The points with error bars are for DLA systems from
 \citet{Pettini2008}. The (black) continuous curve is the chemical evolution model obtained with the stellar yields of slow rotating $Z=10^{-5}$ models from \citet{paperVIII2002} and \citet{Hirschi2005b}. The (magenta) dashed line includes the yields of fast rotating $Z=10^{-8}$ models from \citet{Hirschi2007} at very low metallicity. The (blue) dotted curve is obtained using the yields of the $Z=0$ models presented in \citet{Ekstrom2008a} up to $Z=10^{-10}$. The chemical evolution models are from
\citet{Chiappini2006}.}
   \label{NOCO}
\end{center}
\end{figure}

The primary nitrogen production is accompanied by other interesting features such as the production
of primary $^{13}$C  \citep[see][]{Chiappini2008}, and of primary $^{22}$Ne. Primary $^{22}$Ne is produced by diffusion of primary nitrogen from the H-burning shell into the core He-burning zone, or by the engulfment of part of the H-burning shell by the growing He-buning core. These processes
occur in rotating massive star models \citep{paperVIII2002, Hirschi2007}. In the He-burning zone, $^{14}$N
is transformed into $^{22}$Ne through the classical reaction chain
$^{14}$N($\alpha$,$\gamma$)$^{18}$F($\beta^+$ $\nu$)$^{18}$O($\alpha$,$\gamma$)$^{22}$Ne.

In the He-burning zones (either in the core at the end of the core He-burning phase or in the
He-burning shell during the
core C-burning phase), neutrons are released through the reaction
$^{22}$Ne($\alpha$,n)$^{25}$Mg. 
These neutrons then can either be captured by iron seeds (or elements
with atomic numbers near the one of the iron called iron peak elements) and produce s-process elements or be captured by neutron poisons and
thus removed from the flux of neutrons which is useful for s-process element nucleosynthesis.
From what precedes one can easily conceive that the final outputs of s-process elements will
depend on at least three factors: the amounts of 1.- $^{22}$Ne, 2.- neutron poisons and 3.- iron seeds. In standard models, when the metallicity decreases, the amount of $^{22}$Ne decreases
(less neutrons produced), the amount of neutron poison becomes relatively more important because
for [Fe/H]
\footnote{[Fe/H]=$\lg(n(Fe)_*/n(H)_*)-\lg(n(Fe)_\odot/n(H)_\odot)$, where $n(Fe)_*$ is the number density of iron nuclei in the considered star and $n(Fe)_\odot$ in the Sun,
$n(H)_*$ and $n(H)_\odot$ the number of hydrogen nuclei in the star and in the Sun.}
$\le$ -2 main neutron poisons are primary elements and the amount of iron seeds decreases also. Thus very small quantities of s-process elements are expected (see the triangles in Fig.~\ref{spro}). 
When primary nitrogen and therefore primary $^{22}$Ne is present as given by rotating models
which can reproduce the observed trends for the N/O and C/O ratios observed in the
halo stars, then a very different output is obtained. One can see that globaly the abundances of the s-process elements are increased by many order of magnitudes, and that the elements produced in the greatest quantities are no longer in the iron-nickel region as is the case in standard models but
cover an extended region from strontium up to baryum \citep{Pignatari2008}. 

These first results need to be extended for other masses, rotation and metallicities. However they
already show that some heavy s-process elements, not produced in standard models, might be produced
in significant quantities in metal poor rotating stellar models. It will be very interesting in the future to find some non ambiguous signature of the occurrence of this process in the 
abundance pattern of very metal poor halo stars.

\begin{figure}[t]
\begin{center}
 \includegraphics[width=3.5in,angle=-90]{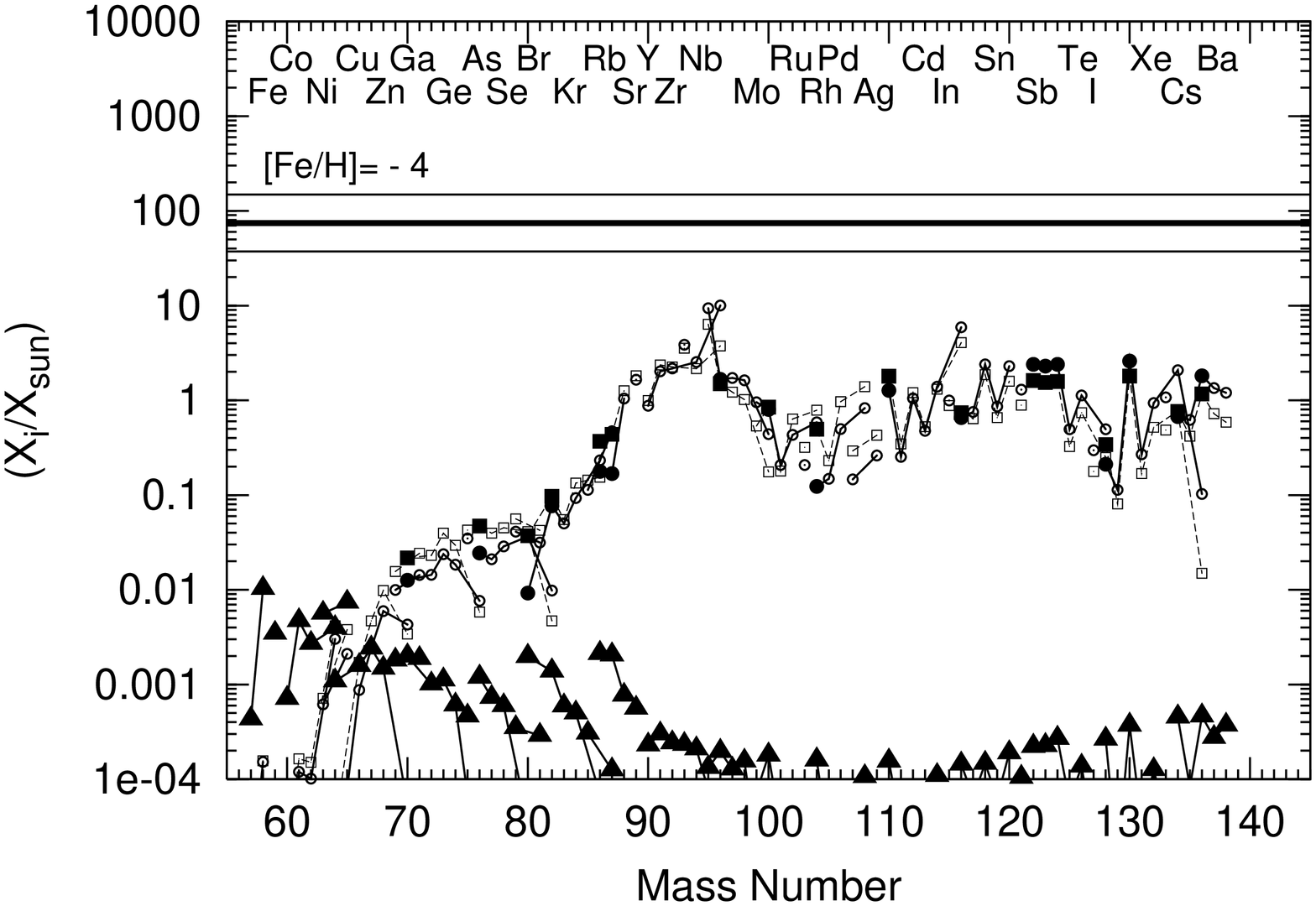} 
 \caption{s-process distributions between $^{57}$Fe and $^{138}$Ba normalized to solar for the 25 M$_\odot$ and [Fe/H]=-4 at the end of the convective C-burning shell. The horizontal line corresponds to the $^{16}$O overabundance in the C shell (thick line), multiplied and divided by two (thin lines). Isotopes of the same element are connected by a line. The cases presented are the following: {\it i)} case without additional $^{22}$Ne (black triangles) and X($^{22}$Ne)$_{\rm ini}=5.21 \times 10^{-5}$; {\it ii)} case with additional $^{22}$Ne (open squares, full squares for the s-only isotopes) and X($^{22}$Ne)$_{\rm ini}=5.0 \times 10^{-3}$; {\it iii)} case with additional $^{22}$Ne (open circles, full circles for the s-only isotopes) and X($^{22}$Ne)$_{\rm ini}=1.0 \times 10^{-2}$. Figure from \citet{Pignatari2008}.}
   \label{spro}
\end{center}
\end{figure}

\subsection{The Carbon-Enhanced- Metal- Poor stars}

At least 1 star out of 4, with [Fe/H] $\le$ -2.5, exhibits [C/Fe] $\ge$ +1.0 \citep[see the review by][]{Beers2005}. These stars are known as C-enhanced metal poor stars (CEMP). The two most iron poor stars presently known ([Fe/H] equal to -5.96 and -5.4) are C-rich, they are called C-rich Ultra-Metal Poor Stars (CRUMPS). The origin of their peculiar surface abundances is not known.
The high nitrogen abundances observed in some CEMP stars implies that the material which is responsible for
their peculiar abundance pattern must be heavily loaded in primary nitrogen. As seen above, rotating stars (both massive
and intermediate mass stars) can produce large amounts of primary nitrogen. Rotating models can account also for the observed enhancements in C, O, Na, Mg and Al, thus \citet{Meynet2006} and \citet{Hirschi2007} have proposed that CEMP stars
may be formed from material ejected by fast rotating stars (massive or intermediate mass stars).
Different chemical signatures are expected depending on the fact that the CEMP stars are formed
from wind ejecta, from wind ejecta together with supernova ejecta or from the envelope of an AGB star. For instance
CEMP stars formed from wind material of massive rotating stars mixed with small
amount of pristine interstellar medium produce stars which are He-rich, Li-depleted and with low  $^{12}$C/$^{13}$C ratios \citet{Meynet2010}.
By He-rich we mean stars that, at very low [Fe/H], present a helium mass fraction between 0.30 and 0.60.

Rotation is probably a key ingredient to explain the abundance patterns of CEMP stars. Similar non-rotating models, without any other extra-mixing, do not succeed to simultaneously
explain the enhancements in the three CNO elements.

\subsection{The chemical anomalies in globular clusters}

The finding of a double sequence in the globular cluster $\omega$ Cen, as well as
in many other globulars,
and the  further interpretation of the bluer sequence by a
strong excess of helium  constitutes a major enigma  for stellar and
galactic evolution \citep{Norris2004}. 
The bluer sequence with a metallicity
[Fe/H]= -1.2  or $Z =  2 \cdot 10^{-3}$ implies an He--content $Y=0.38$ (0.35-0.45), 
i.e. an helium enrichment $\Delta Y = 0.14$ \citep[see also][]{Norris2004}.
In turn, this demands a relative helium 
to metal enrichment  $\Delta Y/\Delta Z$ of the order of   70
\citep{Piotto2005}. 
This is enormous and more than one order of magnitude larger than the current value of $\Delta Y/\Delta Z= 4-5$ \citep{Pagel1992} obtained from  
  extragalactic  HII regions.  A value of 4--5  is  consistent with the chemical yields from supernovae
\citep{Maeder1992a} forming  black holes above about 20--25 M$_{\odot}$. \citet{Maeder2006} showed that the wind contributions of low $Z$ rotating stars  are able to produce the high $\Delta Y/\Delta Z$
 observed in the  blue Main Sequence in  $\omega$ Cen.

Recent observations show that He-rich stars in globular clusters might be a common feature.
For instance, clusters like NGC 2808, M13 and NGC 6441 \citep{Caloi2007} show a well-developed blue horizontal branch, and a strong slope upward from the red clump to the blue of the RR Lyrae regions. Both features could be explained if He-rich stars are present in these clusters.  
\citet{Kaviraj2007} studied the UV and optical properties of 38 massive globular clusters in M87
(elliptical galaxy in virgo cluster). A majority of these clusters appear extremely bright in the far-UV. Canonical models (i.e. models without any super He-rich stars) would imply ages for these clusters about 3 - 5 Gyr larger than the age of the Universe according to the WMAP data! This difficulty is removed when a  super-He-rich ($\Delta Y/\Delta Z > \sim 90$) stellar component is supposed to be present in these clusters.

It has also long been known that globular cluster stars present some
striking anomalies in their content in other light elements\footnote{On the contrary, the content in heavy
elements is fairly 
constant from star to star in any well-studied 
individual Galactic globular cluster (with the notable exception of $\omega$~Cen).}: while in all the
Galactic globular clusters studied so far one finds ``normal" stars with detailed chemical 
composition similar to that of field stars of same metallicity (i.e., same [Fe/H]), 
one also observes numerous ``anomalous" main sequence and red giant stars that are 
simultaneously deficient (to various degrees) in C, O, and Mg, and enriched in N, Na, and Al 
\citep[see the review by][]{Gratton2004}. 

It is clear now that these chemical peculiarities have been inherited at birth 
by the low-mass stars we observe today in globular clusters.
The observed abundance pattern indicates that the material from which these stars formed was
nuclearly processed by CNO burning and
the NeNa- and MgAl-chains
that occurred in an early generation of more massive and faster evolving 
globular cluster stars.
Therefore, there were at least two generations 
of stars in Galactic globular clusters. The first one 
corresponds to the bulk of ``normal'' stars born with the pristine composition 
of the protocluster gas; these objects are those with the highest O and Mg and 
the lowest He and Na and Al abundances also found in their field contemporaries. 
The second generation contains the stars born out of material
polluted to various degrees by the ejecta of more massive stars, and which present
lower O and Mg and higher He, Na and Al abundances than their first generation counterparts.

Any model aiming at explaining the chemical 
properties of globular cluster stars should at least give an answer to the following questions: 
(1) Which type of stars did produce the material enriched in H-burning products? 
(2) What is the physical mechanism responsible for selecting only material bearing the 
signatures of H-processing?
(3) Why does this process occur only in globular clusters?\footnote{Indeed up to now
the peculiar chemical patterns observed in globular clusters, i.e., the O-Na and Mg-Al 
anticorrelations, have not been found in field stars.}
Models for explaining the chemical anomalies in globular clusters invoke either TP-AGB stars \citep{Ventura2008}, massive binaries \citep{deMink2009} and massive rotating stars \citep{Decressin2007a}.
These last authors  propose that the matter from which the stars rich in H-burning products are formed, has been released by slow winds of fast rotating massive stars. Of course, part of the needed material can also be released by AGB stars. The massive star origin presents however some advantages: first a massive star can induce star formation in its surrounding, thus two effects, the enrichment and the star formation can be triggered by the same cause. Second, the massive star scenario allows to use a less flat IMF than the scenario invoking AGB stars \citep{Prantzos2006}. The slope of the IMF might be even a Salpeter's one in case the globular cluster lost a great part of its first generation stars by tidal stripping \citep{Decressin2007b}. Finally the material released through the mechanism invoked above presents
the property that the sum of the C+N+O elements remain constant. Interestingly, the sum
[(C+N+O)/Fe] for individual globular cluster stars presenting otherwise very different
chemical features is constant to within experimental errors in all the
  clusters studied so far with the exception of NGC 1851  \citep[see references in][]{Decressin2009}.
These last authors show also  that rotating AGB models lead to a large increase of the total C+N+O value, which disfavors
these models for explaning the origin of the chemical anomalies in most of the globular clusters.
  
\section{CONCLUSIONS}

The evolution of massive stars depends sensitively on the metallicity which has an impact on the
intensity of the line driven stellar winds and on rotational mixing. We can distinguish four
metallicity regimes: 1.- the very metal poor regime $0 \le Z < \sim 10^{-10}$; 2.-  The low metallicity regime $10^{-10} \le Z < 0.001$;  3.-  The near solar metallicity regime $0.001 \le Z < 0.020$; 4.-  The high metallicity regime $0.020 \le Z$. In each of these metallicity ranges, some specific physical processes occur.

In the (likely rare) Pop III regime, \citet{Ekstrom2008a} find that models rotate with an internal profile $\Omega(r)$ close to local angular momentum conservation, because of a very weak core-envelope coupling. 
Rotational mixing drives a H-shell boost due to a sudden onset of CNO cycle in the shell. This boost leads to a high $^{14}$N production, which can be as much as $10^6$ times higher than the production of the non-rotating models. 
Due to the low mass loss and the weak coupling, the core retains a high angular momentum at the end of the evolution. The high rotation rate at death probably leads to a much stronger explosion than previously expected, changing the fate of the models. 

In the low metallicity regime, rotating models produce large amounts of primary $^{13}$C, $^{14}$N and $^{22}$Ne. Rotation has also probably a strong impact on the s-process elements
nucleosynthesis. Rotational mixing, by increasing the CNO surface abundances, might trigger important mass losses through radiatively driven stellar winds. Stars may also lose some mass by mechanical mass loss when
rotating at the critical limit. For all these reasons, rotation may play 
an important role in explaining the CEMP stars and the chemical anomalies in globular clusters.

In the near solar metallicity regime, rotation and mass loss by stellar winds  are of
similar importance, however the balance of the two effects varies with mass. Neither of the two aspects can be neglected. This metallicity range
is also the one in which models can be checked and calibrated by comparisons with well observed features either of individual stars or of stellar populations. 
Among these observed features let us cite
\begin{itemize}
\item The observed changes of the surface abundances.
\item The observed changes of the surface velocities.
\item The width of the Main Sequence band.
\item The shape of fast rotating stars measured by interferometric technics, the shape
of the nebulosities around high mass losing stars, the anisotropies of the stellar winds.
\item In a near future, the size of the convective core as deduced from  asteroseismic analysis, as
well as the variation with depth of the angular velocity as obtained from asteroseismic observations.
Note that simultaneous observations in asteroseismology and interferometry \citep{Cunha2007}, together with estimate
of the surface magnetic field \citep{Wade2010} will provide new and important clues on massive star evolution.
\item The existence and variation with $Z$ of the populations of Be stars.
\item The variation with $Z$ of the blue to red supergiant ratios.
\item The variation with $Z$ of the Wolf-Rayet populations.
\item The rotation rates of young pulsars.
\item The variation with $Z$ of different core collapse supernova types.
\end{itemize}
As a general statement, it does appear that models including the effects of rotation provide a much better fit to most of the above observed features. Other aspects, not discussed in the present review as the synthesis of $^{26}$Al,
the origin of the isotopic anomalies in the galactic cosmic rays and in meteorites can also be affected
by rotation \citep{Palacios2005, Binns2005, Arnould2006}.

Above solar metallicities, radiative line driven winds become the dominant factor affecting the evolution of massive stars. In models with moderate rotation, we note however that the effects of rotational mixing are still important but their impact is less apparent being somewhat mixed with those of the stellar winds.
An interesting difference which occurs at high metallicity is the evolution of the angular momentum.  One expects that at high metallicity, angular momentum is more easily transported from the core to the surface, due to faster meridional currents, and more easily ejected at the surface by the stellar winds. Thus, everything being equal, one would expect that the angular momentum of the central regions will be lower when the metallicity is higher. Together with the fact that Black-Hole formation is probably more difficult at high $Z$ (also because of the strong stellar winds), this makes the formation of collapsars which are considered as a serious candidate for long soft gamma ray burst much less favorable at high metallicity. In case the rotation rate of young pulsar depends
to some extent to the rotation rate of the core at the presupernova stage, the above line of reasoning would lead to the conclusion that the rotation rate of young pulsars should be slower
in metal rich regions than in metal poor ones.

As we can see, rotation and mass loss are two key physical processes which have a strong impact on  the evolution of the massive stars during
the whole cosmic history. These two processes, which interact with each other, are responsible for a great diversity of outputs, going
from the shape of the nebulae, to the evolution of the abundances  in the interstellar medium, passing through
the origin of the Gamma Ray Bursts.
It is interesting to note that
the energy in form of rotational energy in stars is quite small, at most of the order of the percent of the gravitational energy. 
Similarly, metallicity or heavy elements, which shape the intensity of the line driven winds, represent only at most two percents of the
mass of all chemical elements. Nevertheless, the impacts of both rotation and metallicity are very important.
It reminds us that sometimes a pinch is sufficient for creating new worlds!

\bibliographystyle{aa}
\bibliography{MyBiblio}

\end{document}